\documentclass[12pt,a4paper,twoside,openright]{report}
\let\openright=\cleardoublepage

\usepackage[utf8]{inputenc}

\usepackage{graphicx}
\usepackage{amsthm}
\usepackage{datetime}
\usepackage{underscore}
\usepackage[group-separator={\,}]{siunitx}
\usepackage[usenames,dvipsnames,table]{xcolor}
\usepackage{fancyvrb}
\usepackage{listings}
\usepackage{caption}
\usepackage{enumitem}
\usepackage{verbatim}
\usepackage{pbox}

\definecolor{lstbgcolor}{rgb}{0.97, 0.97, 0.97}

\usepackage[unicode]{hyperref}
\hypersetup{pdftitle=Optimizing large applications}
\hypersetup{pdfauthor=Martin Liška}

\definecolor{Silver}{RGB}{230, 230, 230}

\definecolor{SpecGood}{RGB}{215, 244, 215}
\definecolor{SpecBetter}{RGB}{175, 233, 175}
\definecolor{SpecBad}{RGB}{244, 215, 215}
\definecolor{SpecWorse}{RGB}{233, 175, 175}

\makeatletter
\def\@makechapterhead#1{
  {\parindent \z@ \raggedright \normalfont
   \Huge\bfseries \thechapter. #1
   \par\nobreak
   \vskip 20\p@
}}
\def\@makeschapterhead#1{
  {\parindent \z@ \raggedright \normalfont
   \Huge\bfseries #1
   \par\nobreak
   \vskip 20\p@
}}
\makeatother

\begin{document}

\lefthyphenmin=2
\righthyphenmin=2

%%% Title page

\pagestyle{empty}
\begin{center}

\large

Charles University in Prague

\medskip

Faculty of Mathematics and Physics

\vfill

{\bf\Large MASTER THESIS}

\vfill

\centerline{\mbox{\includegraphics[width=60mm]{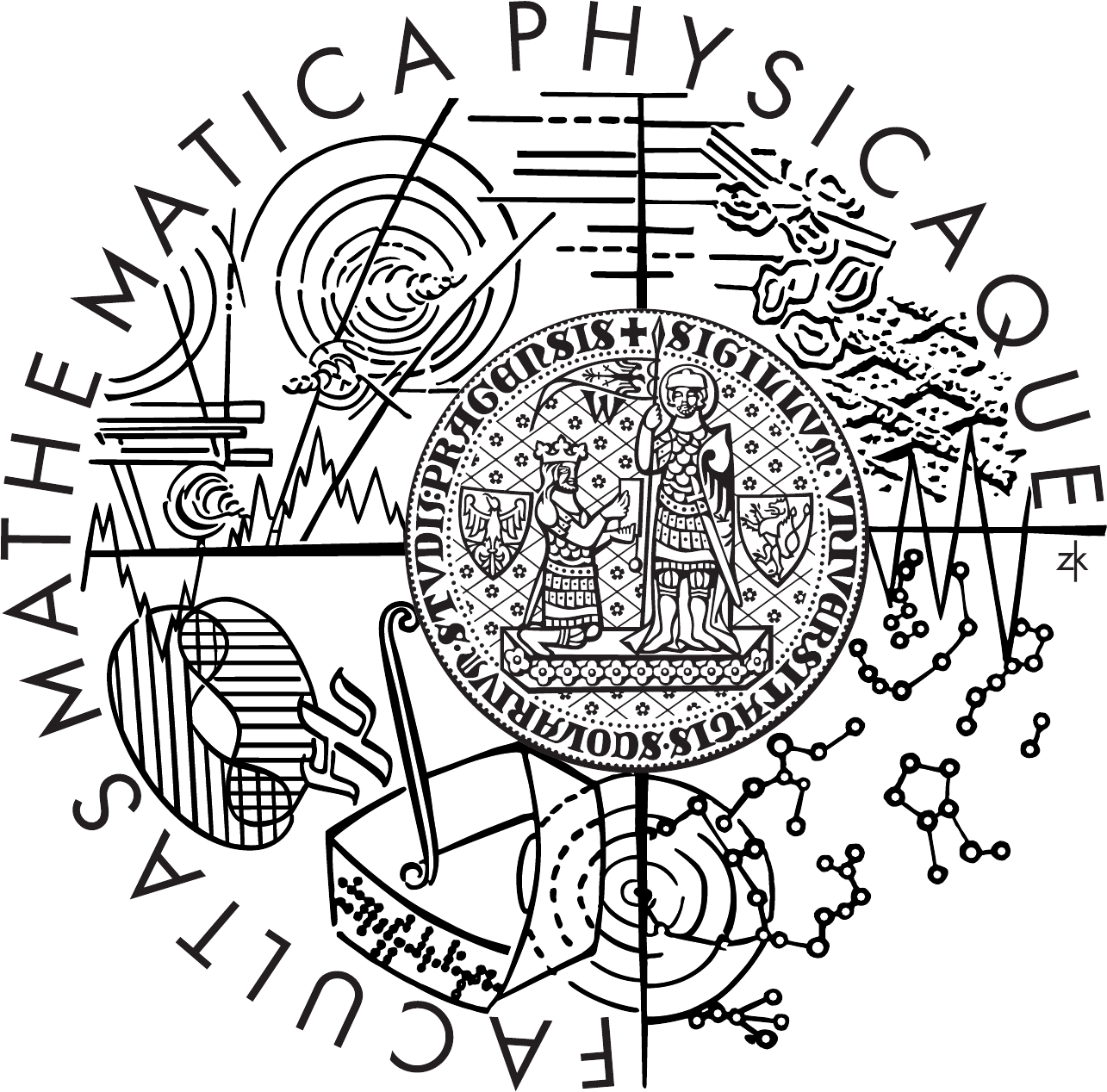}}}

\vfill
\vspace{5mm}

{\LARGE Martin Liška}

\vspace{15mm}

{\LARGE\bfseries Optimizing large applications}

\vfill

Department of Distributed and Dependable Systems

\vfill

\begin{tabular}{rl}

Supervisor of the master thesis: & Mgr. Jan Hubička, Ph.D. \\
\noalign{\vspace{2mm}}
Study programme: & Computer Science \\
\noalign{\vspace{2mm}}
Specialization: & Software Systems \\
\end{tabular}

\vfill

Prague \the\year

\end{center}

\newpage

%%% Následuje vevázaný list -- kopie podepsaného "Zadání diplomové práce".
%%% Toto zadání NENÍ součástí elektronické verze práce, nescanovat.

%%% Na tomto místě mohou být napsána případná poděkování (vedoucímu práce,
%%% konzultantovi, tomu, kdo zapůjčil software, literaturu apod.)

\openright

\noindent

I would hereby like to thank my supervisor Jan Hubička for his supervision, helpful advices and intensive cooperation. In addition, a thank to Lukáš Voráček, who helped me to do language correction.

\newpage

%%% Strana s čestným prohlášením k diplomové práci

\vglue 0pt plus 1fill

\noindent
I declare that I carried out this master thesis independently, and only with the cited
sources, literature and other professional sources.

\medskip\noindent
I understand that my work relates to the rights and obligations under the Act No.
121/2000 Coll., the Copyright Act, as amended, in particular the fact that the Charles
University in Prague has the right to conclude a license agreement on the use of this
work as a school work pursuant to Section 60 paragraph 1 of the Copyright Act.

\vspace{10mm}

\hbox{\hbox to 0.5\hsize{%
In Prague on \today
\hss}}

\vspace{20mm}
\newpage

\vbox to 0.5\vsize{
\setlength\parindent{0mm}
\setlength\parskip{5mm}

Název práce:
Optimalizace rozsáhlých aplikací

Autor:
Martin Liška

Katedra:
Katedra distribuovaných a spolehlivých systémů

Vedoucí diplomové práce:
Mgr. Jan Hubička, Ph.D., Informatický ústav Univerzity Karlovy

Abstrakt:
Oba hlavní open source překladače, GCC a LLVM, dnes dosahují stavu, kdy jsou schopny link-time optimalizovat velké aplikace. U rozsáhlých aplikací si nevystačíme jenom s klasickými měřítky výkonu jako je rychlost nebo paměťová náročnost. Zajímá nás typicky velikost kódu, doba studeného startu aplikace a podobně. Vývojáři těchto balíků tak často sahají k různým ad-hoc řešením, mezi které patří například utilita ElfHack, startování vlastních aplikaci pomocí předzavedené utility a dlopen, prelinking a různé nástroje pro přerovnání výsledného programu podle pořadí provádění funkcí. Práce si klade za cíl zmonitorovat dostupné techniky optimalizace, ocenit jejich účinnost a navrhnout nová řešení, postavená na platformě link-time optimalizací.

Klíčová slova:
překladač, optimalizace kódu, operační systémy

\vss}\nobreak\vbox to 0.49\vsize{
\setlength\parindent{0mm}
\setlength\parskip{5mm}

Title:
Optimizing large applications

Author:
Martin Liška

Department:
Department of Distributed and Dependable Systems

Supervisor:
Mgr. Jan Hubička, Ph.D., Computer Science Institute of Charles University

Abstract:
Both uppermost open source compilers, GCC and LLVM, are mature enough to link-time optimize large applications. In case of large applications, we must take into account, except standard speed efficiency and memory consumption, different aspects. We focus on size of the code, cold start-up time, etc. Developers of applications often come up with ad-hoc solutions such as Elfhack utility, start-up of an application via a pre-loading utility and dlopen; prelinking and variety of different tools that reorder functions to fit the order of execution. The goal of the thesis is to analyse all existing techniques of optimization, evaluate their efficiency and design new solutions based on the link-time optimization platform.


\vss}

\newpage

%%% TOC

\openright
\pagestyle{plain}
\setcounter{page}{1}
\tableofcontents

\chapter{Introduction}
\label{chapt:Introduction}

This thesis brings an review to problematic of large applications and we are focused on UNIX-like operating systems, primary on Linux. We introduce whole development toolchain (focusing on GNU toochain), classic build system procedure and state-of-the-art techniques built on the top of the toolchain. Moreover, we map the problematic of executable formats, which were primary designed to small- to medium-sized programs (by today standards).

Many topics discussed in this work are tightly connected to design of the executable file formats. Nowadays, \textit{Executable and Linkable Format} (ELF), the standard Linux executable format, plays a key role in many issues connected to large applications. The format is very flexible and, during the time, there were applied many fundamental improvements. These enhancements are very often written by the maintainers of the large applications and thus often these newly invented tools are ad-hoc project specific. On the other hand, in many situations a start-up of an application is slower and more complex.

Majority of production open-source compilers typically test performance on small benchmarks and only a small part of the compiler maintainers is interested in large applications. Idea of inter-procedural and inter-modular optimization frameworks came in 1990s and has been mainly developed in last 10 years (\cite{CompaqAlphaLTO}, \cite{gcc-lto}, \cite{GlekHubickaLTO}). The infrastructure is shown to be very powerful technique that can come up with appreciable benchmark results. Nevertheless, the build system is getting more complex and does not fit to every day compile/edit/debug cycle. Moreover, memory and CPU utilization is enormous and e.g. it is almost impossible to build a large application with link-time optimization on IA-32 architecture. Additionally, the approach requires a large set of modifications to the whole toolchain and will take few year the changes will establish.

The main goal of the thesis is to analyse performance bottlenecks of the large applications from toolchain developer point of view. Just few of large applications supply accessible statistics. Notable exception is Mozilla Firefox Telemetry~\cite{FirefoxTelemetry}, originally written by Taras Glek in 2011. Instead of any kind of benchmarking, it helps Mozilla's engineers to measure the application in the real world. The feature does monitoring of the browser and all usage statistics are sent to Mozilla. On the other hand, there are speed-up techniques for which it is very painful to do any kind of report. As an example, to collect information about read pages by a dynamic linker, we have to write quite a sophisticated script for the Linux kernel. For a purpose of benchmarking, we implemented many Python scripts that are attached on supplemented CD. Furthermore, many of experiments are very time consuming, e.g. complete build with enabled link-time optimization takes hours, even for a high-end CPU.

In the implementation part of the work, we come up with two new optimizations written for the most widely used open-source compiler, the GNU Compiler Collection (GCC). The idea of passes is not new, but we do optimizations on a different level of toolchain. The first one is a code placement pass which optimizes the order of functions in the ELF format. The layout primary reduces disk page faults and thus speed-ups cold start of applications. With the increasing usage of programming languages supporting classes and templates, many semantically equivalent functions and constructors (destructors) are generated. We try to find all these function patterns are merge as my function implementations as possible.

Important part of the effort was to make is possible to build major applications as Chromium and Libreoffice with link-time optimization. We are in contact with many maintainers of these large applications and we help them to integrate LTO to the build system. As the proposed patches will be applied to mainline source base of the projects, our workarounds, we were forced to do, will be necessary not longer. We had to do bug fixes on all sides and process was very long.

The rest of the thesis is organized as follows. Chapter \ref{chapt:ProgrammersGuide} is a guide for programmers of large applications. Chapter \ref{chap:Analysis} brings deeper analysis of existing techniques and rich variety of statistics is presented. Chapter \ref{chap:ProfileGuidedReorderingPass} is about implementation of function reordering in the GCC compiler. Chapter \ref{chap:SemanticFunctionEquality} talks about semantic equality pass that was also implemented as a part of thesis and Chapter \ref{chap:Conclusion} concludes this thesis.

\chapter{Programmer's Guide}
\label{chapt:ProgrammersGuide}

This part should be written as a guide to a programmer who wants to optimize his or her large application. We primary focus on advantages and disadvantages of individual techniques based on our analysis and statistics presented in Chapter~\ref{chap:Analysis}. We start the chapter with the description of terminology this thesis is built on. We introduce main compiler components and we examine state-of-the-art compiler techniques that are not limited to module-by-module compilation.

\section{Key Concepts}

The following list contains a short definition of all key terms used in the subsequent chapters, preventing any possible ambiguity. At the beginning, we explain all terms involved in a software build system. In general, toolchain is a collection of programming tools to create an executable file. These tools are often used in a chain, where the output of each tool becomes the input for the following one. In the Linux world, the main chain items, also called GNU programs, are following:

\begin{itemize}
	\item \textbf{GNU Automake} helps to create a portable \texttt{Makefiles}~\cite{Automake}. It transforms \texttt{*.am} input files and turns them to a corresponding \texttt{*.in} files that are used by the configure script.
	
	\item{\textbf{GNU Make} automatically builds executable programs by reading the content of so called \textit{makefiles}~\cite{GNUMake}. While integrated development environments can drive the build process, Automake is still very popular. Build of a program is driven by a set or rules, where the left side specifies a target and the right side list all its dependencies. Nowadays, as Symmetric Multiprocessing (SMP) became common even in embedded systems, make with introduced \texttt{-j} where we specify the number of jobs to run simultaneously.}

	\item \textbf{A compiler}, is the part of the toolchain we are interested the most. It transforms a source code written in a programming language to target machine specific assembly language. In detail, a compiler performs many consecutive operations: lexical analysis, preprocessing, parsing, semantic analysis, code optimization and generation. These phases can be separated to the following components: front-end/middle/back-end and the mostly used open-source compilers are:

	\begin{itemize}

		\item \textbf{GCC} (acronym for \textit{GNU compiler collection}) is an open-source compiler established in 1987, with rich front-end, supporting languages like C, C++, Fortran and Java, among others~\cite{GCC}. GCC is the most widespread compiler in UNIX world, thus being the best platform we can integrate any further optimizations and speed improvements.
	
		\item \textbf{LVVM}	(formerly \textit{Low Level Virtual Machine}) is a cross-platform compiler written in C++ language~\cite{LVVM}. The compiler covers a wide variety of front ends. Even though LVVM supports link-time optimization, there still many opened issued the maintainers must solve.
	
		\item \textbf{Open64} is an open-source compiler for x86-64 and Itanium architecture~\cite{Open64}. The project was adopted by University of Delaware and last stable build was released in November 2011.
	
	\end{itemize}	
	
	\item \textbf{Linker} is a program that loads multiple object files generated by a compiler and combines them into a single executable file. For the given list of objects, the linker resolves sections and symbols defined in these sections.
	
	\item \textbf{System runtime} provides essential support to an application and consists of following components:
	
	\begin{itemize}
	
		\item \textbf{Operating system} offers a variety of necessary system calls.
	
		\item \textbf{Dynamic loader}, also called dynamic linker, copies the content of the executable to memory, loads all shared libraries and performs dynamic relocations.
	
		\item \textbf{Standard library}	implements fundamental functions needed by almost every application.
	
	\end{itemize}

\end{itemize}

Because we are primary focused on the GCC compiler, the following enumeration of key concepts encompasses terms that are abundantly used in sequential chapters of the thesis:

\begin{itemize}
	\item \textbf{LTO} (acronym for \textit{Link Time Optimization}) is a compiler technique extending compilation unit boundaries. Unlike in a file-by-file build system, LTO can load all necessary source codes and optimize them as a single module. That encompasses deeper analysis and more optimization produced by inter-procedural passes.
	
	\item \textbf{PGO} (acronym for \textit{Profile Guided Optimization}), also known as \textbf{FDO} (\textit{Feedback-Directed Optimizations}), is a compiler optimization technique benefiting from running an instrumented program across a representative input set. All compiler optimizations are forced to make decisions based on heuristics. An application-collected profile  is deemed be more efficient and precise. Data indicates which parts of the program are hot (executed more often), and which are cold (less executed). However, due to caveat due-compilation model, PGO is not widely adopted by software projects.
	
	\item \textbf{LIPO} (acronym for \textit{Profile Feedback Based Lightweight IPO}) is a compiler cross-module optimization technique combining both IPO (Inter-Procedural Optimization) and PGO. LIPO sensitively enhances code visibility boundaries for optimization passes, where enlargement decisions are gained from the profile collected by the application test run. On the contrary, build system does not suffer from losing the degree of parallelism.
	
	\item \textbf{SPEC} (\textit{acronym for Standard Performance Evaluation Corporation}) is a well-known benchmark suite focused on different computer systems. CPU related benchmark is called SPEC CPU, latest version named as SPEC CPU 2006, consisting of two main components: compute-intensive integer performance (SPECint) and compute-intensive floating pointer performance (SPECfp). This well-established suite became de facto the standard in the world of compilers and computer performance analysis.	
\end{itemize}

\section{Software Build Process}

Build process of software, especially in UNIX-like systems, encompasses a suite of programming tools like Autotools, Make, a compiler and Binutils (all parts of GNU toolchain).

Starting point is a set of source code files and a makefile, driving the tree build process. Even though traditional compilation-linking build process is obvious for most readers, let me show a simplified scenario for the matter of subsequent comparison.

\subsection{Traditional Build System Scenario}

\begin{figure}[htbp]
	\begin{center}
		\includegraphics[width=9cm]{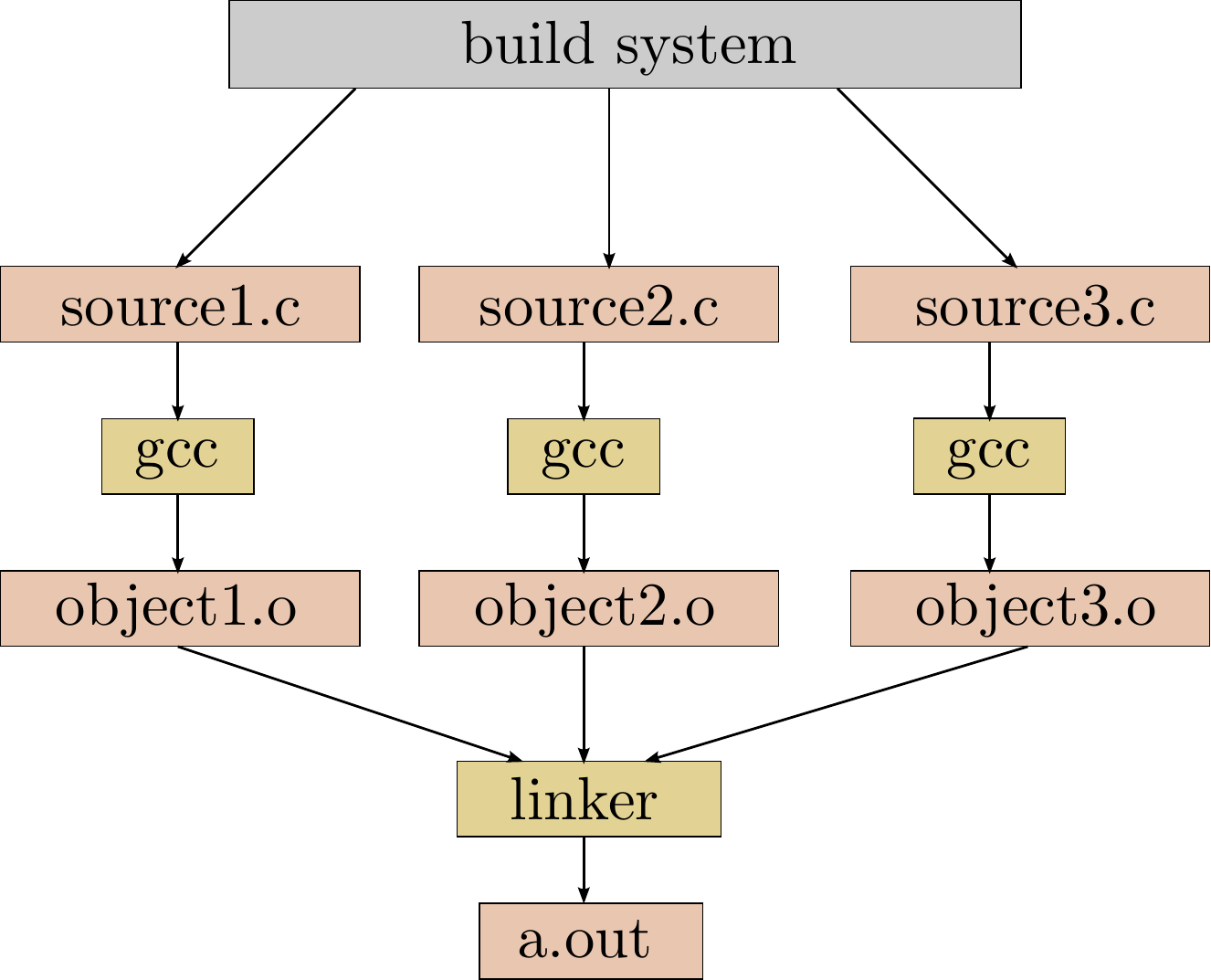}
		\caption{Traditional build system scenario.}
		\label{fig:TraditionalBuildSystem}
		\vspace{-10pt}
	\end{center}
\end{figure}

File-by-file oriented build system executes in parallel a compiler on all source files depends on. Compiler performs pre-processing, front-end parsing, code optimization and target machine assembler generation. Basic scenario is shown in Figure \ref{fig:TraditionalBuildSystem}.

Pre-processing step copies the content of all included header files, generates macros and expands all symbolic constants annotated by \texttt{\#define} keyword. Front-end parser transforms the text representation to a compiler internal representation, where all compiler optimizations take place. Finally, target machine architecture assembler is generated.

After that, a collection of object files is passed to a linker that produces the executable file.

\subsection{Profile Guided Optimization Build System}

\begin{figure}[htbp]
	\begin{center}
		\includegraphics[width=11cm]{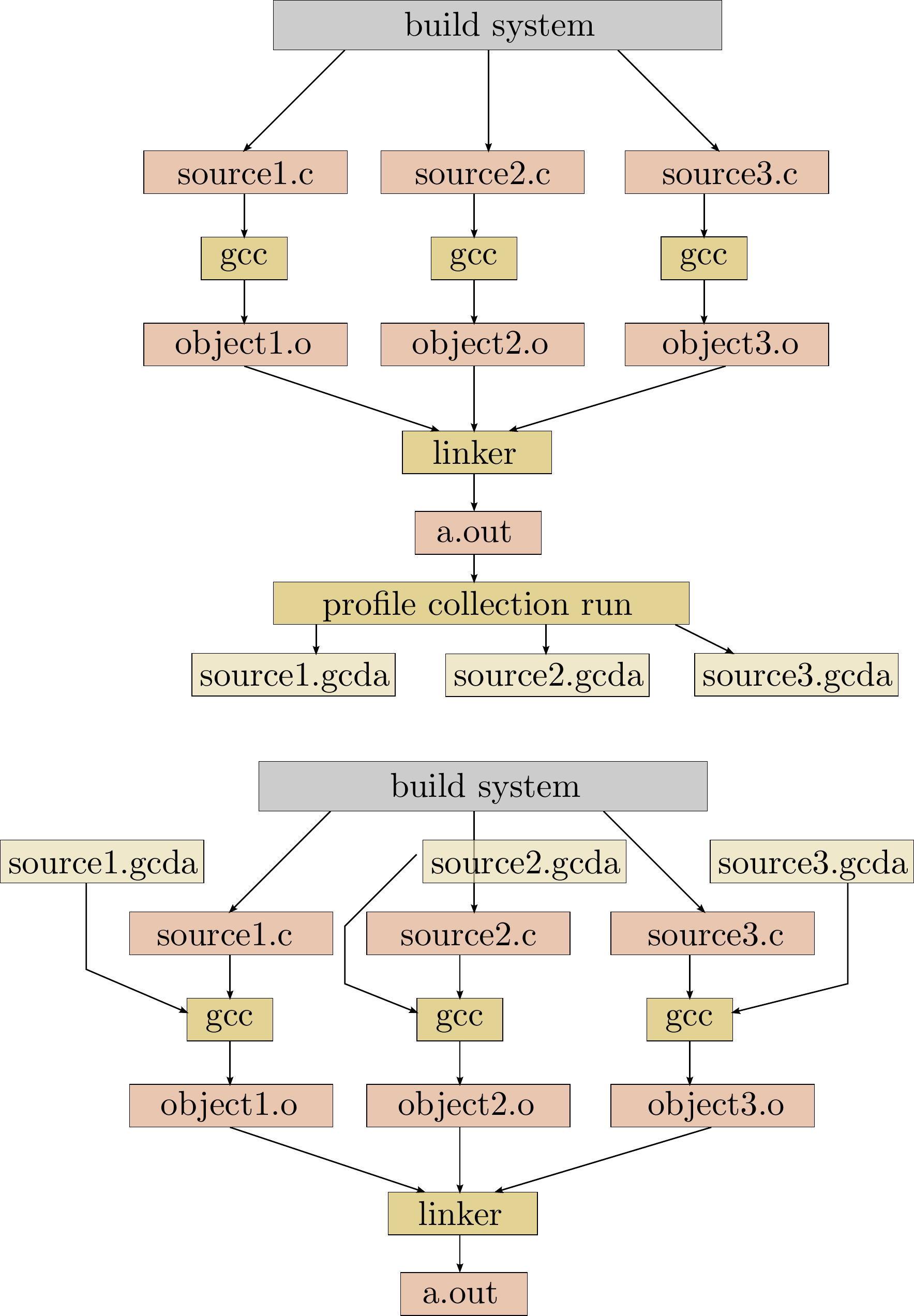}
		\caption{PGO build system scenario.}
		\label{fig:PGOBuildSystem}
		\vspace{-10pt}
	\end{center}
\end{figure}

The build system, introduced in Figure \ref{fig:PGOBuildSystem}, looks more complex than the previous one. Nevertheless, the system is just enhanced to collect profile information during train run with a representative set of input data and save the profile for later use (GCC uses \texttt{.gcda} files with names corresponding to the object files). To create profile data, we need to add \texttt{-fprofile-generate} to the compiler options. After the binary is instrumented, run\footnote{use test suite or run the application as usually} the application. Before compiling the project for the second time, remove all objects (typically by running \texttt{make clean}) and switch the compiler options to \texttt{-fprofile-use}. Final run loads the data from the profile files and we can quite easily reach significant speed-up for the program. 

Profile feedback is important hint for the compiler, enabling:

\begin{itemize}
	\item better inlining decisions
	\item better code placement
	\item cold code is optimized for size
	\item hot code is optimized more aggressively
	\item for SPEC benchmarks, we gain about 4\% of performance and the binary is more than 10\% smaller
\end{itemize}

\subsection{Link Time Optimization (LTO)}

\textit{Link Time Optimization} (LTO)~\cite{gcc-lto} extends the scope of inter-procedural optimizations to the whole binary. Instead of final machine code, the intermediate source code representation (in GCC case the GIMPLE) is written to an object file and the second optimization phase is invoked by a linker. File-by-file boundaries are merged and the whole program is visible for inter-procedural passes.

Main benefits of the technique are:

\begin{itemize}
	\item optimizer can process cross-module inlining
	\item dead code is eliminated
	\item whole program propagation
	\item for SPEC benchmarks, LTO with the same level of optimization brings another 4-5\% and the binary about 15-20\% smaller
	\item we can combine the optimization with PGO and this leads for SPEC to be the fastest compilation configuration
\end{itemize}

\begin{figure}[htbp]
	\begin{center}
		\includegraphics[width=11cm]{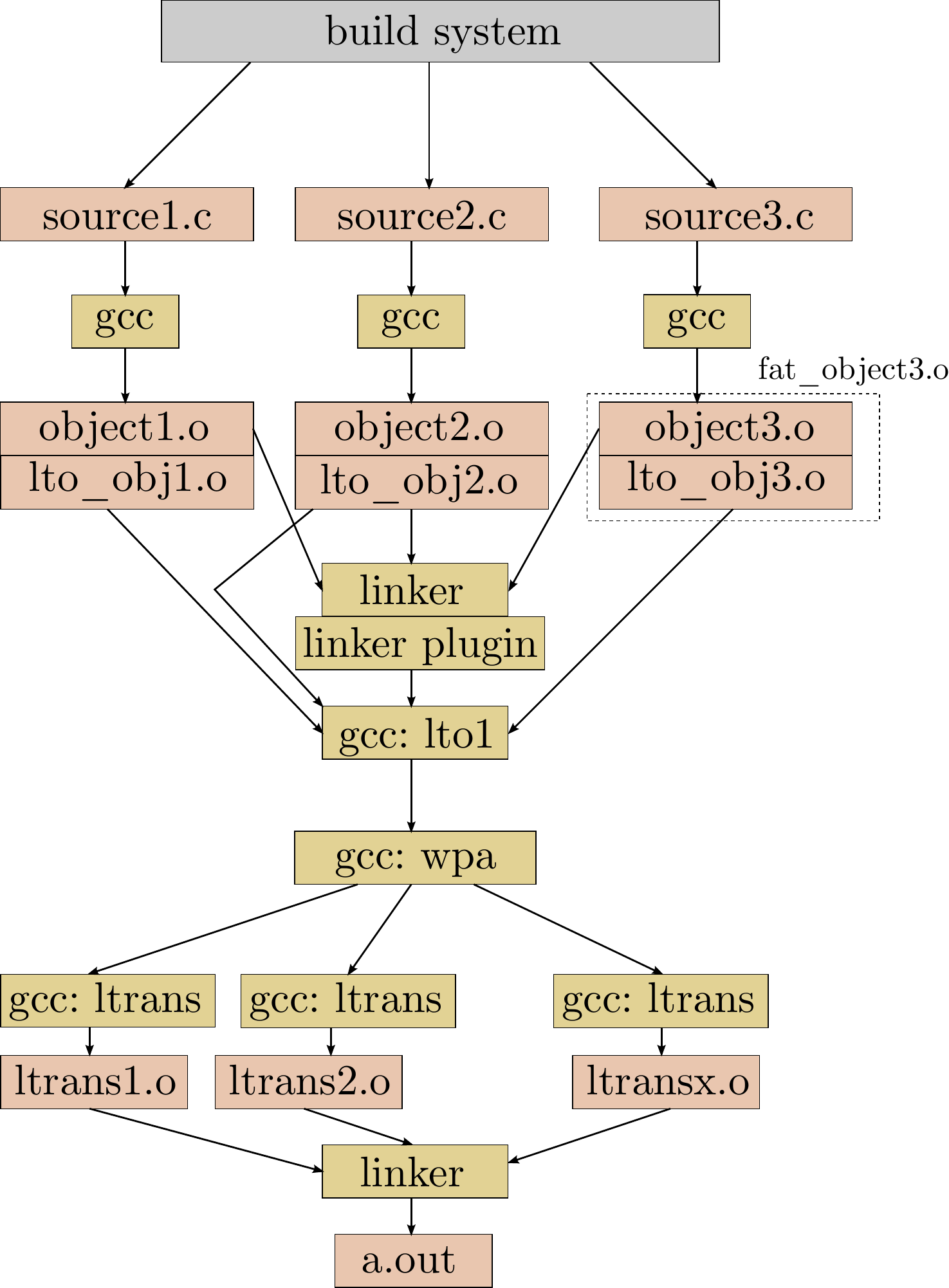}
		\caption{LTO build system scenario.}
		\label{fig:LTOBuildSystem}
		\vspace{-10pt}
	\end{center}
\end{figure}

LTO infrastructure development started in 2005 and GCC version 4.5.0 (released in 2009) became the first version where LTO was included. Beyond the compiler, it was necessary to apply patches to the whole toolchain.

As we can see in Figure \ref{fig:LTOBuildSystem}, first stage of the build system transforms all source files to \textit{fat} objects. The plug-in infrastructure designed by \texttt{gold} linker and adopted by \texttt{GNU linker} invokes GCC linker plug-in that is a start pointer for the entire link-time optimization. LTO front-end \texttt{lto1} loads all sections needed by link-time optimization from a hard drive. GCC LTO was designed for large applications and implements so-called WHOPR~\cite{WHOPR} mode that parallelizes link-time and conserves memory.

The GCC program can basically operate in two compilation modes:

\begin{itemize}
	\item \textbf{WHOPR} (acronym for WHOle Program optimizer) utilizes multiple CPUs for a faster distributed compilation.
	\item \textbf{LTO mode} loads entire program to memory and behaves similar to a single large compilation unit.
\end{itemize}

\pagebreak

Since GCC 4.6.0 the WHOPR mode is used by default. The mode tries to do the maximum of work in compile time and except one serial phase, the compilation is parallel. Compilation is done in following steps:

\begin{itemize}
	\item \textbf{Local generation phase (LGEN)} compiles all source files to intermediate representation.
	\item \textbf{Whole program analysis (WPA)} phase builds the program call graph.
	\item The compiler does all optimization decisions based on intermediate representation from the first phase. Important to notice, function bodies of the symbols are not approachable during the stage.
	\item Transformation decisions are computed, the call graph is split and streamed to local transformation (LTRANS) object files.
	\item \textbf{Local transformation phase (LTRANS)} materialized all decisions done in previous stages.
	\item Furthermore, all passes unable to make transformation, built on results coming from WPA, can operate on a LTRANS partition as in single compilation unit mode.
\end{itemize}

Big challenge for the infrastructure is to deal with pass cooperation. Conversely to LTO operation mode, all passes process analysis, propagation and transformation in sequential manner. Every modification applied to a function is seen correctly by a consecutive pass without any necessary interaction. However, in WHOPR mode, all inter-procedural optimizers performs analysis on the call graph snapshot established in WPA. Future modification are serialized to a corresponding LTRANS object file. When a pass makes a decision to modify call graph, that pass should somehow propagate the information to the rest of optimizers. Nevertheless, a concept of pass communication was simplified to creation of \textit{virtual clones}. Each newly created function became a clone of the function and all calls are redirected to a newly created symbol. Preservation of the original call graph node tends to guarantee that all succeeding passes can operate on the original function seen in WPA. Let's assume the inliner wants to inline function \texttt{foo} into function \texttt{bar} and function \texttt{foo} is declared as external. The inliner creates a new virtual clone \texttt{bar.clone1}, where the body of \texttt{foo} is included. All internal calls are redirected to the newly created \texttt{bar.clone1} and the primary function \texttt{foo} stays untouched. See~\cite{GlekHubickaLTO} for details.

To enable link-time optimization, \texttt{-flto} should be added to the compiler options and to the link flags. Moreover, \texttt{binutils} supporting the linker plug-in must be installed on your system. If we want to be sure that the LTO intermediate language was used, we add \texttt{-fno-fat-lto-objects} to our compiler options. Any problem of the toolchain set-up of will cause an error and the object files will be smaller. Fat object files were implemented to provide backward compatibility and will be in the future replaced with slim variant.

Developers of large application often want to reduce final binary. With O3 level of optimizations, the infrastructure provides a binary that is approximately of the same size as optimization focused on size reduction (-Os). Moreover, unit growth in LTO is more sensitive. If there is a very hot function, the optimizer can allow relatively large unit growth for function. But from the global perspective, the unit growth factor is preserved. As an example, \texttt{xalancbmk} SPEC benchmark runs about 50\% faster with LTO.

In the following list, we define GCC-specific terms and used in consecutive chapters:

\begin{itemize}
		\item \textbf{IPA pass} (acronym for \textit{Inter-procedural analysis pass}) is an optimization pass performed on GIMPLE code representation, benefiting from knowledge of more high-level informations. All these passes play main role in LTO, where inter-procedural passes are executed.

	\item \textbf{GIMPLE} is a tree-address intermediate source code representation targeted by all language front-ends that GCC supports. To transform the source code to GIMPLE, temporary variables are created to comply with the tree-address code notation. The idea of representation comes from the SIMPLE representation, chosen in the McCAT compiler.
	
	\item \textbf{RTL} (acronym for \textit{Register Transfer Language}) is the last low-level intermediate representation seen by GCC, where a bundle of optimizations are performed.	
\end{itemize}

\subsection{Profile Feedback Based Lightweight IPO (LIPO)}

\textit{LIPO}, a compiler technique for application performance boosting, was established as an alternative to GCC LTO in 2009~\cite{lipo}. The technique is built on conjunction of cross module optimization (\cite{HPCMO1}, \cite{HPCMO2}) and PGO. Among all transformations done in IPO, cross-module inlining is one of the most efficient. Due to procedure boundary elimination, larger optimization region and context sensitivity produce faster code. In fact, no additional hard drive storage is needed (no intermediate representation is stored). While most of time is spent in complex analysis algorithms, increased compile time is not observed and the method does not suffer from loss of degree of parallelism. Moreover, existence of profile comes with better decisions in inlining, loop unrolling, register allocation and value profiling.

\begin{figure}[htbp]
	\begin{center}
		\includegraphics[width=14cm]{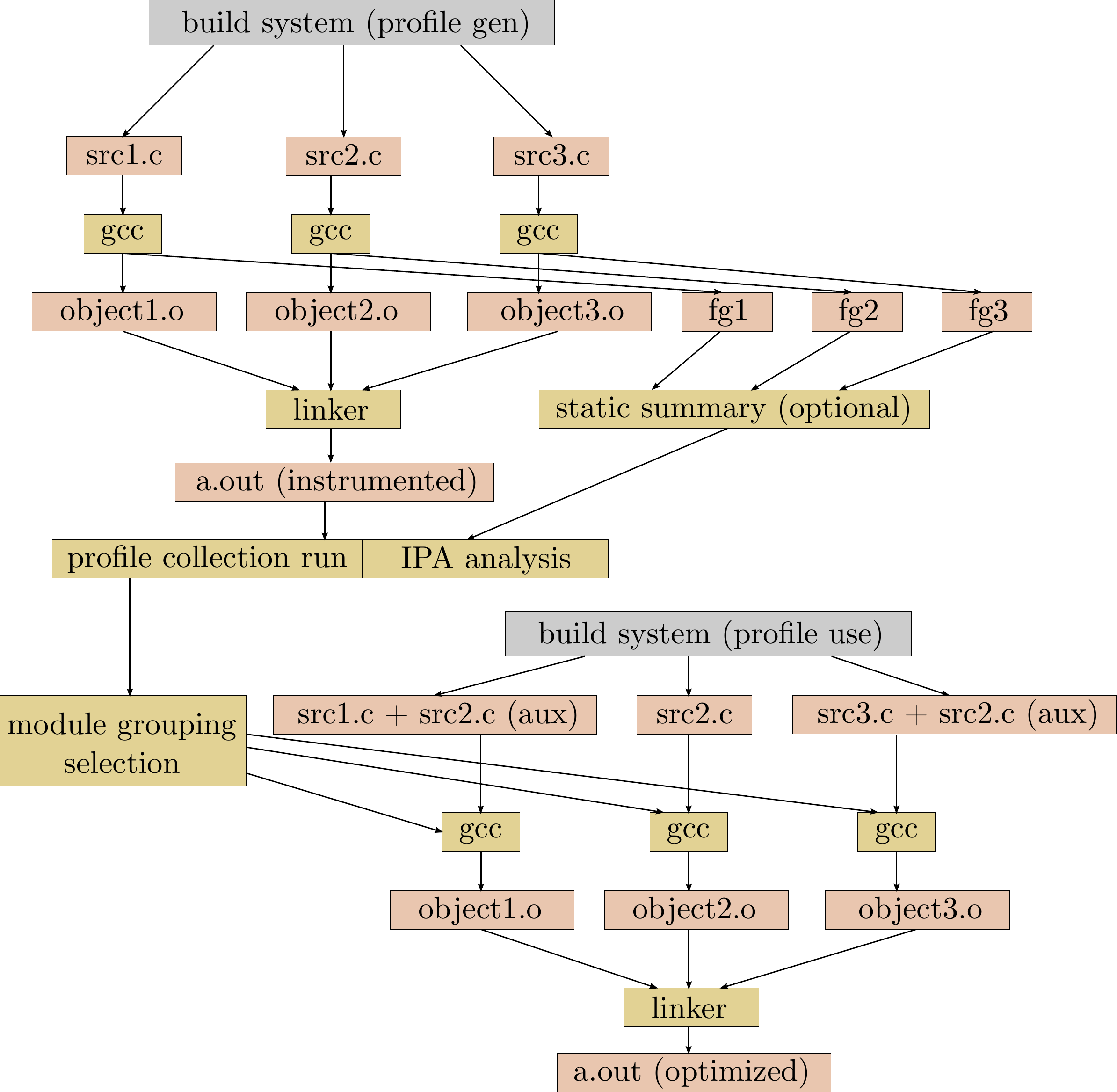}
		\caption{LIPO build system scenario.}
		\label{fig:LIPOBuildSystem}
		\vspace{-10pt}
	\end{center}
\end{figure}

For all users who want to be given the benefits of inter-procedural and PGO, LIPO brings mechanism to achieve this in a seamless fashion. The technique shifts all important decisions from compile time to training execution time, where the result is saved into corresponding grouping information program database. Moreover, all build systems that are familiar with profile-guided optimization paradigm do not need any modifications (except additional compile flag). The scenario is shown in Figure \ref{fig:LIPOBuildSystem}.

Compared to traditional build system, LIPO profile generation phase creates an additional information file by the LIPO runtime that stores the flow-graph (file with \texttt{fgX} extension). Deeper analysis is performed after execution of an instrumented binary. Module inlining builds a dynamic call graph with annotated call edges, and groups file modules by hot call edges. Profile database is enhanced by grouping decisions that are deemed to play a key role in the second build run (profile use run). In the profile use compilation, grouping decisions are read from a file. To every primary file, an auxiliary source file is parsed. Each primary module can encompass zero or more auxiliary source code files. From the conception point of view, all auxiliary files are included as header files with all functions defined as \textit{externally visible}. That opens door for more efficient inlining decisions.

\begin{minipage}{\linewidth}
\footnotesize
\begin{lstlisting}[label={lst:LIPOSample},caption=LIPO example,basicstyle=\ttfamily]
$ cat a.c
int foo(char *d, unsigned int i)
{
  return d[i] + d[i + 1];
}

$ cat b.c
char tokens[] = {1,3,5,7,11,13};

int extern foo(char *d, unsigned int i);

int bar(void)
{
  return foo(tokens, 3);
}

$ cat main.c
#include <stdio.h>
#include <string.h>

extern int sum(char *d, unsigned int i);
extern int bar(void);

int main(int argc, char **argv)
{
  unsigned int i, s = 0;

  if(argc != 2)
    return;

  for (i = 0; i < strlen(argv[1]) - 1; i++)
    s += foo(argv[1], i);

  printf ("sum: %d\n", s);
  printf ("tokens: %d\n", bar());
}
\end{lstlisting}
\normalsize
\end{minipage}

In Listing \ref{lst:LIPOSample} we prepared a trivial example. Source file \texttt{main.c} calls in a hot loop function \texttt{foo}, defined in source file \texttt{a.c}. Conversely, function \texttt{bar}, defined in source file \texttt{b.c}, contains a cold call to the same function. Let's assume that inlining \texttt{foo} into \texttt{main} is beneficial, group module for the main file is \texttt{\{main.c,~a.c\}}. Obviously, group module for \texttt{a.c} is \texttt{\{a.c\}} and, finally, group module for \texttt{b.c} contains also no auxiliary source files.

After a dynamic call graph is built, a greedy algorithm is used for module affinity analysis. If the edge is hot enough, all functions are put into the same group. More precisely, if a threshold is reached, all functions originating from different modules will become an auxiliary. As mentioned above, LIPO includes all functions placed in auxiliary files as external inline functions. Even though the idea looks simple, name lookup rules in f.e. C++ are very complex and complicated. Each compilation module in GCC maintains its own symbol table for declarations, types and name bindings, preventing cross module name lookup between modules. To achieve correct parsing of an auxiliary module, LIPO driver restores all tables to the state before primary module was loaded.

LIPO is maintained by a Google team and has a separate branch in source version control system: \url{svn://gcc.gnu.org/svn/gcc/google}.

To build a project with the infrastructure, we add \texttt{-fprofile-generate/use} and \texttt{-fripa} to our compiler and linker options. Like in PGO, we have to do two-phase compilation process.

\section{Representation of Executables and Shared Libraries}

Executable and Linkable Format (ELF) is a commonly used file format for executables (including object files, shared libraries etc.). As the idea of sharing code was growing, early Unix-days file formats as \textit{a.out} and \textit{COFF} were replaced. Many processes in an operating system share a large amount of code via shared libraries. Furthermore, all read-only parts of a shared library could be mapped to many virtual spaces. At the same time, memory storage for the rest of virtual spaces can be improved by copy-on-write mechanism.

Legacy formats designers did certain decisions leading to the fact that all relocations must be performed in link time, which does not allow load time relocations. Load address of all shared library must be fixed. Central authority for memory ranges must maintain all existing shared libraries with respect to library future growth and code refactoring. To make the matters even worse, virtual address application space gets fragmented, which brings about dynamic memory allocation restrictions. Moreover, address space on 32-bit architectures might cause space starvation even for the central authority.
Even though calling functions in old systems were efficient, every slightest range collision could cause catastrophic results. Therefore, Unix-like operating systems migrated to ELF at the turn of the millennium.

From the implementation point of view, every ELF file format simply consists of mandatory ELF Program Header. The header respects the same binary layout on every platform, defines basic information about binary and, especially, determines a distribution of ELF sections. As modern processors and operating systems can protect memory regions, each ELF sections can have a different subset of read, write and execute flags.

Following example shows that \texttt{.text} section, holding all of the generated assembler, has an executable flag and the executable data is split between writeable \texttt{.data} and read-only \texttt{.rodata} sections.

\begin{minipage}{\linewidth}
\scriptsize
\begin{lstlisting}[basicstyle=\ttfamily]

readelf -S `which echo`
There are 27 section headers, starting at offset 0x72d8:

Section Headers:
  [Nr] Name              Type             Address           Offset
       Size              EntSize          Flags  Link  Info  Align
... skipped sections ...
  [13] .text             PROGBITS         00000000004015c0  000015c0
       0000000000003028  0000000000000000  AX       0     0     16
... skipped sections ...
  [15] .rodata           PROGBITS         0000000000404600  00004600
       000000000000124b  0000000000000000   A       0     0     32
... skipped sections ...
  [24] .data             PROGBITS         0000000000607180  00007180
       0000000000000080  0000000000000000  WA       0     0     32
... skipped sections ...
\end{lstlisting}
\normalsize
\end{minipage}

The virtual space of the process is filled by \texttt{mmap} that accepts all aforementioned memory protection flags. Unlike in legacy formats, the binary is not complete. The kernel has has to execute the dynamic linker (\texttt{ld.so}) that is responsible for the rest of preparations. Chosen dynamic linker is not hard-coded, but ELF header defines path and dynamic linker name.

Every ELF executable file contains a set of shared libraries. \textit{Hello world} program written in C language has also a few of them:

\footnotesize
\begin{lstlisting}[basicstyle=\ttfamily]
ldd hello_world
    linux-vdso.so.1 (0x00007ffff9399000)
    libc.so.6 => /lib64/libc.so.6 (0x00007f2b995a4000)
    /lib64/ld-linux-x86-64.so.2 (0x00007f2b9994c000)
\end{lstlisting}
\normalsize

Virtual address space of every single process cannot live without mapped kernel segment. As we can see in the exhibit above, \texttt{linux-vdso.so.1} is a virtual ELF library which is contained in kernel itself (explains why no file path is specified) and provides system call API. The second one is a standard C library. The last library is a dynamic linker that plays role during the whole life of the executable.

Unlike in the example of simple \textit{hello world} program, large application can depend on tens or even hundreds of shared libraries. Examples of chosen applications are presented in Table \ref{fig:DiskSeekGraphFirefox25Elfhack}.

\begin{table}[htbp]
	\centering
	\begin{tabular}{|l|r|r|r|}
	\hline 
	& \textbf{Firefox 17.0.3} & \textbf{Libreoffice 3.6} & \textbf{Chromium 27} \\ \hline
	\textbf{Entry point binary} & \texttt{libxul.so} & \texttt{soffice.bin} & \texttt{chrome} \\ \hline
	\textbf{Binary size} & 34.0 MB & 6.0 KB & 74.0 MB \\ \hline
	\textbf{Shared lib. count} & 73 & 61 & 93 \\ \hline
	\textbf{Shared lib. size} & 59.1 MB & 122.0 MB & 90.7 MB \\ \hline
	\textbf{Aver. library size} & 818.3 KB & 2.0 MB	& 998.3 KB \\ \hline
	\textbf{Total size} & 93.1 MB & 122.0 MB & 164.7 MB \\ \hline
	\end{tabular}
	\caption{Large application shared library dependencies.}
	\label{fig:SharedLibraryDependencies}
\end{table}

\section{Relocations}

\subsection{Linkers}

\texttt{linker} is a program in GNU tool-kit that takes multiple object file, archives and combines them into a single executable file (either a shared library or an executable binary). Nowadays, \texttt{binutils}, a GNU package where the linker is included, contains a new linker called \texttt{gold}. The name is an acronym for \textit{Google Linker} and was developed by Ian Lance Taylor~\cite{Gold} and his small team at Google. The linker typically runs at the end of compilation process and is run serially. Even if we change a single source file, corresponding object is changes as well and the linker is invoked for linking. In a complex application, made of hundreds of objects files, the linker rapidly delays fast edit/debug development cycle.

Older \texttt{GNU ld} allows user to read, combine and write object file in many different formats, even legacy \texttt{a.out} or \texttt{COFF} formats are still supported. Nevertheless, most of the open-source operating system uses just \texttt{ELF} as its executable file format. Moreover, \texttt{GNU ld} accepts \textit{Linker Command Language} to offer total control over linker process and linker procedure itself is driven by a built in script. The source code of \texttt{GNU ld} is split into multiple components which communicate by various hooks. In fact, it showed that is would be easier to write a new linker from scratch instead of implementing any speed up to the existing code.

\subsection{Dynamic Linkers}

Dynamic relocations are one the mechanisms which enable transformation of position independent code (PIC) in dynamic linker. Every time we load a shared library to memory, we are given by the kernel an address where the library will be placed. It depends on a set of libraries and (as a security feature) address space layout randomization applied by kernel. Therefore, the dynamic linker has to resolve address information for each dynamically referenced symbol and all code and data must cope with that constraint. Fortunately, while the base address is variable, all symbol and data offsets are still constant. Knowing the base offset of the library, library code can easily reach offset of the data (used in the IA-32 ABI) or it knows where the current instruction is placed (used in the x86-64 ABI).

The relocation is the most expensive job done by a dynamic linker (we ignore Prelink tool which can significantly reduces relocation costs). Asymptotic computational complexity is according to~\cite{Drepper} at least O(R + \textit{nr}), where R is the number of relative relocations, \textit{r} the number of named relocations and \textit{n} is used for the number of Dynamically Shared Objects (DSO). Wrong hash table design even increases the complexity to O(r + \textit{nr\,log\,s}), where \textit{s} comprises the symbol count.

In general there are two types of dependencies:

\begin{itemize}
	\item Dependencies not tied to any symbol name that live in their own objects. Since the dynamic linker knows the relative position, the dependency can be easily computed. Conversely, application binaries are clean of these relocations since the static linker is knowledgeable to perform the relocation.
	\item Symbol-based dependencies, where symbol lookup process, among a set of different objects comes into play.
\end{itemize}

For all symbols used at run-time and coming from a different object, relocation mechanism by the linker must be applied. Many architectures allow deferral of symbol resolution, until the symbol is really needed. Final executable could contain many references to the same symbol, but a different lookup scope makes it more complex to cache resolution results. In general, every single object file builds its own list of lookup scopes and the length of the lists directly depends on loaded objects belonging to an executable file. Obviously, our motivation is to possibly reduce amount of such objects.

For more detail about speeding up dynamic linking, please visit Section \ref{sec:PerformanceProblemsOfDL}.

\section{Symbol Visibility}

Link-time optimization benefits from assumptions that could be applied to the biggest possible part of the application. To make matters worse, every function or variable which is \textit{externally visible} disables more aggressive optimization. It is very common that large applications are organised as a set of libraries, e.g. core functionality for all Libreoffice application is located in \texttt{libmergedlo.so} library. As described in~\cite{Drepper}, developers are often lax and do not declare functions with the \texttt{static} keyword. Solution for such a library can be compiler option \texttt{-fwhole-program}, that marks all functions except \texttt{main} as static and exceptions can be provided via \texttt{externally_visible} keyword. Historically, \texttt{-fwhole-program} option was used just for the compilation of a single source file and in context of LTO, the option is misused. 

Without the linker plug-in, LTO works with limited set of assumptions and must be more conservative in optimization decisions. The information about symbol usage plays a key role. Almost all new versions of Linux distribution have pre-installed version of binutils with enabled linker plug-in option.

\section{Faulty.lib}

Applications written for Android are distributed as a ZIP archive with \texttt{.apk} extension. These packages contain Java bytecode targeted for Dalvik virtual machine with just-in-time compilation. Apart from that, Android has capability to run a native code written in languages like C or C++, e.g. existing shared libraries are good candidates for the Native Development Kit (NDK). Nowadays, as hardware platforms the Android can run are growing, every time we write an native application, we have to deliver binaries for following application binary interfaces (ABI): \texttt{armeabi}, \texttt{armeabi-v7a}, \texttt{mips} and \texttt{x64}. Developers in Mozilla noticed that the native libraries are extracted by the system and installed to a separate location. Thus, the binary is effectively stored twice, in compressed and uncompressed version.

\texttt{faulty.lib} substitutes dynamic loader (on Android named \texttt{linker}) in the Android operating system with own implementation of dynamic loader related functions: \texttt{dlopen}, \texttt{dlerror}, \texttt{dlsym}, \texttt{dlclose} and \texttt{dladdr}. Moreover, all library access operations go through on-demand decompression layer. The layer touches a zip archive, compressed by blocks, each time a few bytes are required by the loader. The goal of the linker is to uncompress as few pages as possible, that primarily depends on how the binary layout is organised.

\section{Prelink}

\texttt{prelink} is an architecture independent ELF tool, originally written by Jakub Jelínek~\cite{Prelink}, significantly accelerating dynamic relocations during start-up of an application. Flexible ELF format, replacing legacy \textit{a.out} format, does not need a central authority for a distribution of virtual address space slots any more. At link time, due to relocation mechanism, dynamic linker can place any subset of existing shared libraries in a virtual memory. On the other hand, program loading, shared libraries dependency resolution and relative relocations are more complex and thus time consuming.

By adding new program's shared library dependency, symbol resolution process in runtime can result in a different symbol. Alternatively, one of the essential binaries can introduce a new dependency by defining the same symbol. First speed up method, originally part of \texttt{prelink}, caches symbol lookup results processed by the linker. When we look for a symbol multiple times, sorting all relocations to the same symbol by ascending order, relocation layout will improve. Moreover, such layout tends to optimize paging and caching. In case of Libreoffice 3.6, relocation cache hits about 86\% of all symbols relocation during start-up and totally the number of relocations resolved by the cache is close to three-quarters.

Primary goal of the tool is to allocate system unique virtual address slots, relocate them to the chosen load offset and cooperate with a dynamic linker in way that the dynamic linker will admit the offset and will skip relocation process as much as possible. Offset registry is local-machine-specific, by default saved in \texttt{/etc/prelink.cache}, and default lookup folders are located in \texttt{/etc/prelink.conf}. When using all mode, the tool collects all binaries and shared libraries located in those folders, transitively following dependencies in breadth first search order. Logic connected to symbol resolution process is quite complex and shared among the linker and \texttt{prelink}. All performed lookups are printed to a dump file, read by \texttt{prelink}. All conflicts seen by the prelink program are stored in a special ELF section and the list will go through a standard lookup mechanism processed by the linker in runtime.

When we execute a binary, the dynamic linker first checks if it is prelinked. If yes, the linker verifies that lookup search scope is unchanged and all libraries seen in the first phase are not modified. Moreover, no additional library can be added. Accomplishing all rules, prelink mode is launched, in which classic relocation handling is skipped and just conflicting symbols are needed to by resolved. On the other hand, if any of these conditions is not satisfied, the dynamic linker will fall back to normal operations, without any knowledge prepared in the first phase.

\pagebreak

Let us compare results of Libreoffice before and after \texttt{prelink}:

\footnotesize
\begin{lstlisting}[basicstyle=\ttfamily]
$ LD_DEBUG=statistics ./soffice.bin --writer

runtime linker statistics:
  total startup time in dynamic loader: 36659467 clock cycles
	    time needed for relocation: 33061559 clock c. (90.1%)
                 number of relocations: 10504
      number of relocations from cache: 67112
        number of relative relocations: 204730
	   time needed to load objects: 3076100 clock c. (8.3%)

$ LD_DEBUG=statistics ./soffice.bin --writer
runtime linker statistics:
  total startup time in dynamic loader: 4036111 clock cycles
	    time needed for relocation: 672804 clock c. (16.6%)
                 number of relocations: 15
      number of relocations from cache: 946
        number of relative relocations: 0
	   time needed to load objects: 2701730 clock c. (66.9%)
\end{lstlisting}
\normalsize

Before we applied the prelink program, there were almost \num{80000} symbol relocations complemented by more than \num{200000} relative relocations that were applied by the linker. Loading of objects takes a minor role in the time spent. Unlike in the first example, prelinked binary consumes most of time waiting for object files to be loaded from the hard drive. With almost no relocation, total start-up time spent in the linker shrinks about nine times. Aforementioned facts make a powerful tool from \texttt{prelink}, but for a modern CPU, these accelerations can save just hundreds of seconds.

Even though \texttt{prelink} can accelerate start-up of an application on slow machine significantly, the tools is not common in Linux distributions and is mainly installed as a dependency of e.g. Libreoffice package. Moreover, to prelink all system libraries we must have root access and the shared libraries are modified. With any system update, user should rerun the tool to keep the database up-to-date and even a small change to a common library can devaluate entire work done by \texttt{prelink}.

\section{Function Reordering}

During the start-up of an application, the dynamic linker, by default, does not preload entire binary to a memory. Conversely, on-demand loading of hard drive pages is utilized and the kernel loads just pages that will be really used by the dynamic linker and the application.

In general, poor code locality degrades performance of an application, either by a slow start-up, or by slower calls of hot functions that are not close enough to each other. In case of large application written in just-in-time compilation languages, it could lead to instruction and cache misses. In~\cite{DynamicCodeManagement} we can find fast on-line algorithm dealing with dynamically generated code. We are more interested in functions participating in an application cold start. In this chapter, we demonstrate different ways of tracing functions that are usually spread over the entire \texttt{.text} section of an ELF binary file. In fact, all existing techniques collaborate with a dynamic linker, either by linker script or ordering file (\texttt{gold}). We created a new approach built on top of link-time optimization and profile-guided optimization.

\subsection{Function Count Statistics}

Probably the easiest tool for use is \texttt{valgrind}, a generic framework for creating dynamic analysis. It essentially simulates a CPU. Tracing dump provides useful data that can be utilized for sorting-out of the executed functions:

\scriptsize
\begin{lstlisting}[basicstyle=\ttfamily]
$ valgrind --tool=none --trace-flags=10000000 --demangle=no firefox

0x88006f6 NS_InitXPCOM2_P+2450 libxul.so+0x15346f6
0x77e1740 UNKNOWN_FUNCTION UNKNOWN_OBJECT+0x0
0x8800729 NS_InitXPCOM2_P+2501 libxul.so+0x1534729
0x880073d NS_InitXPCOM2_P+2521 libxul.so+0x153473d
0x8800759 NS_InitXPCOM2_P+2549 libxul.so+0x1534759
0x7db2430 _ZN10nsRunnable6AddRefEv libxul.so+0xae6430
0x880077f NS_InitXPCOM2_P+2587 libxul.so+0x153477f
0x7db244e _Z14NS_NewThread_PPP9nsIThreadP11nsIRunnablej+14 libxul.so+0xae644e
0x7db246b _Z14NS_NewThread_PPP9nsIThreadP11nsIRunnablej+43 libxul.so+0xae646b
0x7db248f _Z14NS_NewThread_PPP9nsIThreadP11nsIRunnablej+79 libxul.so+0xae648f
0x7db249c _Z14NS_NewThread_PPP9nsIThreadP11nsIRunnablej+92 libxul.so+0xae649c
0x77e4780 UNKNOWN_FUNCTION UNKNOWN_OBJECT+0x0
0x7db24eb _Z14NS_NewThread_PPP9nsIThreadP11nsIRunnablej+171 libxul.so+0xae64eb
0x7db24f8 _Z14NS_NewThread_PPP9nsIThreadP11nsIRunnablej+184 libxul.so+0xae64f8
\end{lstlisting}
\normalsize

Tracing approach periodically monitors application's stack and prints currently executed function (mangled function name) in a corresponding shared library or application. This method gives just an approximation as some functions are skipped because the tracing frequency is not quick enough to catch them. Additionally, there are entries in the dump file pointing to unknown location: \texttt{UNKNOWN_FUNCTION}.

We have noticed that Taras Glek developed a more precise tool for \texttt{valgrind} called \texttt{icegrind}\footnote{\url{https://bugzilla.mozilla.org/show_bug.cgi?id=549749}}. Unfortunately, the patch was just a piece of experimental code and cannot be applied to current mainline of the \texttt{valgrind} tool. Because of that, the tool's repository was cloned and we added code which is called at the beginning of function \texttt{\mbox{CLG_(push_call_stack)}}. The function prints every first occurrence of a function to standard error output. Source code is available as a Github project: \url{https://github.com/marxin/valgrind}.

The result is similar to the first introduced technique:

\scriptsize
\begin{lstlisting}[basicstyle=\ttfamily]
INIT:_ZNKSt9basic_iosIcSt11char_traitsIcEE5rdbufEv
INIT:_ZNSt15basic_streambufIcSt11char_traitsIcEE7pubsyncEv
INIT:_ZN9__gnu_cxx18stdio_sync_filebufIcSt11char_traitsIcEE4syncEv
INIT:fflush
INIT:_IO_file_sync@@GLIBC_2.2.5
INIT:_ZNSt13basic_ostreamIwSt11char_traitsIwEE5flushEv
INIT:_ZNKSt9basic_iosIwSt11char_traitsIwEE5rdbufEv
INIT:_ZNSt15basic_streambufIwSt11char_traitsIwEE7pubsyncEv
INIT:_ZN9__gnu_cxx18stdio_sync_filebufIwSt11char_traitsIwEE4syncEv
\end{lstlisting}
\normalsize

\begin{figure}[htbp]
	\begin{center}
		\includegraphics[width=150mm]{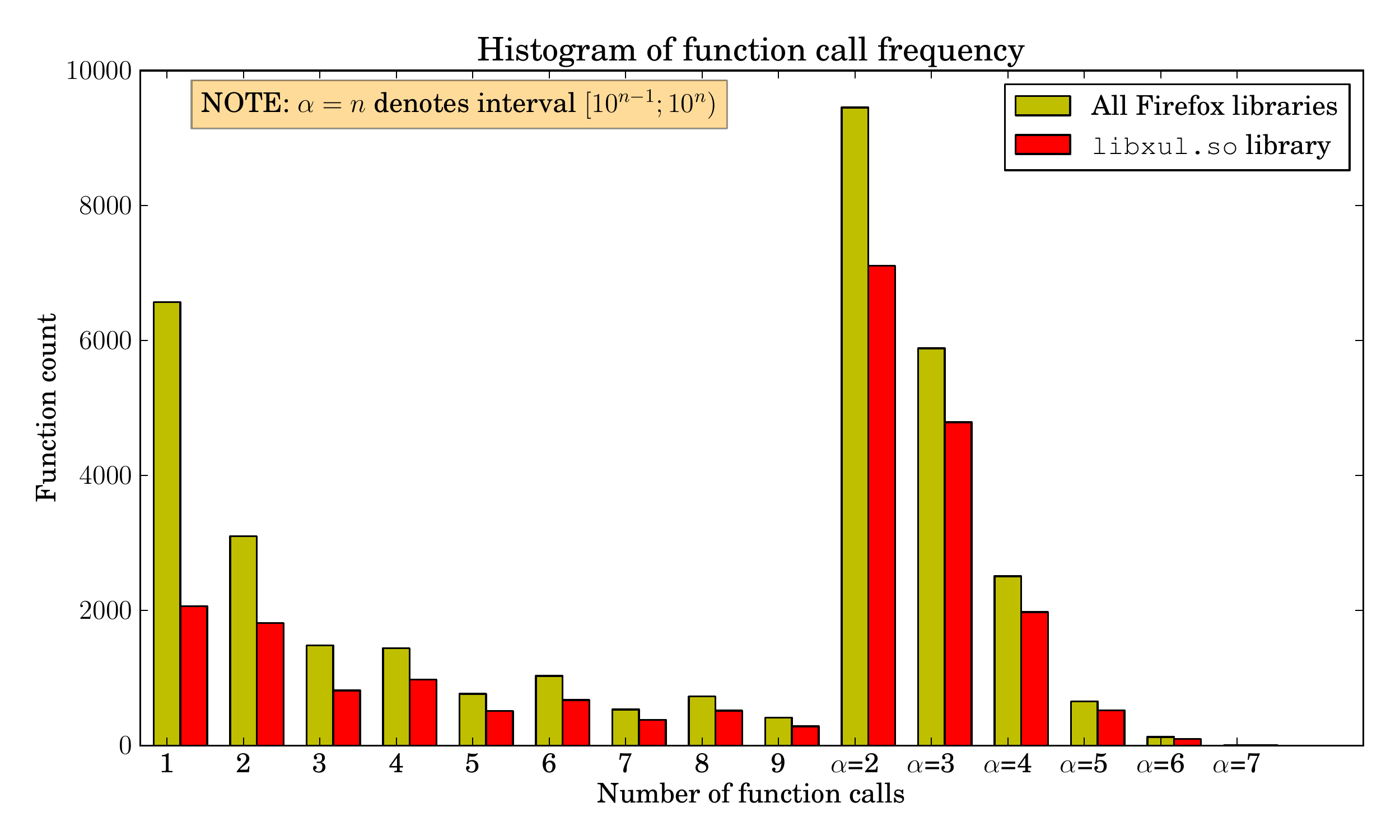}
		\vspace{-10pt}
	\end{center}
	\caption{Function call frequency for \texttt{libxul.so} during start-up.}
	\label{fig:FirefoxFunctionCallFrequency}	
\end{figure}

With a small modification of the code, one can reach total number of executed functions in start-up process. Function distribution, according to Figures \ref{fix:FirefoxFunctionCallFrequency} and \ref{fig:FirefoxFunctionCallFrequency}, can be approximated by a multiplicative inverse ($x^{-1}$). We present the data in intervals of power of ten with the exception in the first interval. Our prime motivation is to follow and observe functions which are called just once (or few times). We are specifically interested in internally defined symbols, which can be seen in the two columns on the right. As we can seen, more than \num{2000} functions, representing almost 10 percent, are called during the start-up process just once and should be placed to a similar location in a binary. What is more interesting, occurrences in the interval [2,4] cover additional \num{3600} functions, enhancing our ordering potential to a quadruple.

\begin{table}[htbp]
	\centering
	\begin{tabular}{|l|r|r||r|r|}
	\hline
	Number of calls & Function count & Portion & Function count & Portion \\ \hline
	& \multicolumn{2}{c||}{\cellcolor{Silver}All \texttt{Firefox} libraries} & \multicolumn{2}{c|}{\texttt{\cellcolor{Silver}libxul.so library}} \\ \hline
	1 & \textbf{\num[detect-weight]{6561}} & \textbf{18.91\,\%} & \textbf{2064} & \textbf{9.16\,\%} \\ \hline
	2 & \num{3102} & 8.93\,\% & 1817 & 8.06\,\% \\ \hline
	3 & \num{1484} & 4.27\,\% & 814 & 3.61\,\% \\ \hline
	4 & \num{1444} & 4.16\,\% & 978 & 4.34\,\% \\ \hline
	5 & 766 & 2.20\,\% & 511 & 2.27\,\% \\ \hline
	6 & \num{1033} & 2.97\,\% & 673 & 2.99\,\% \\ \hline
	7 & 533 & 1.53\,\% & 382 & 1.69\,\% \\ \hline
	8 & 727 & 2.09\,\% & 516 & 2.29\,\% \\ \hline
	9 & 413 & 1.18\,\% & 285 & 1.26\,\% \\ \hline
	[10;100) & \num{9454} & 27.24\,\% & \num{7109} & 31.54\,\% \\ \hline
	[100;1000) & \num{5884} & 16.95\,\% & \num{4792} & 21.26\,\% \\ \hline
	[1000;\num{10000}) & \num{2510} & 7.23\,\% & \num{1976} & 8.77\,\% \\ \hline
	[$10^4$;$10^5$) & 653 & 1.88\,\% & \num{523} & 2.32\,\% \\ \hline
	[$10^5$;$10^6$) & 127 & 0.36\,\% & 97 & 0.43\,\% \\ \hline
	[$10^6$;$10^7$) & 7 & 0.02\,\% & 1 & 0.00\,\% \\ \hline
	\hline	
	\textbf{Total} & \textbf{\num[detect-weight]{34706}} & \textbf{100\,\%} & \textbf{\num[detect-weight]{22538}} & \textbf{100\,\%} \\ \hline
	\end{tabular}
	\caption{Function call frequency for \texttt{libxul.so} during start-up.}
	\label{fix:FirefoxFunctionCallFrequency}
\end{table}

To make data presentation complete, the number of function calls exceeds \num{71000000}, leaded by memory allocation functions like \texttt{malloc} and \texttt{free}, followed by thread synchronization functions (\texttt{pthread_mutex_lock}, \mbox{\texttt{pthread_mutex_unlock}}, \texttt{__pthread_mutex_unlock_usercnt}). \texttt{libxul.so} library itself utilizes mainly string functions, e.g. \texttt{js::frontend::TokenStream::getChar()} or \mbox{\texttt{longest_match}}.

\subsection{Reordering in Linker}
\label{subsec:ReorderingInLinker}

Both linkers (\texttt{GNU ld} and \texttt{gold}) have a facility to educate the linker about the symbol order. Unfortunately, neither \texttt{GNU ld} nor gold support to order functions by names, but only by section names. It is necessary to produce object files with enabled compiler option \texttt{-ffunction-sections}, which places each function to its separate section. Although, the feature is supported by most systems using \texttt{ELF}, it tends to generate larger object files. Naming convention just appends the function name to traditional \texttt{.text} section, i.e. \texttt{.text.main}. While the \texttt{gold} supports section ordering in natural manner with \texttt{--section-ordering-file}, there is a powerful script language in \texttt{GNU ld} that can help to do the same. An example of such a script can be find hereafter:

\begin{minipage}{\linewidth}
\footnotesize
\begin{lstlisting}[basicstyle=\ttfamily]
SECTIONS
{
  .text :
  {
    *(.text.function_name1)
    *(.text.function_name2)
    *(.text.function_name3)
    *(etc.)
  }
}
INSERT BEFORE .fini
\end{lstlisting}
\normalsize
\end{minipage}

In order to generate a section ordering file, we have to do a mapping between symbol and section names. Unfortunately, not all function names are marked just with the prefix \texttt{.text.[function_name]}. Compiler can prefix a function with prefixes such as \texttt{.startup}, \texttt{.hot} or \texttt{.unlikely}. Apart from that, a virtual thunk or a mangled symbol with a slightly different flags make mapping more complicated.

Google branch of the GCC repository contains a linker plug-in that annotates the call graph and repeatedly groups sections that are connected by hot edges. Original implementation was presented in this mailing post: \url{http://gcc.gnu.org/ml/gcc-patches/2011-09/msg01440.html}. Statistics are collected by PGO and streamed to a newly created ELF section. In the linker, after all symbols are loaded, the plug-in performs reordering. Optimization technique is motivated to speed-up extremely large applications and according to the author of the plug-in, it brings about 2\% of performance.

\section{Conclusion}

We presented in the chapter majority of techniques that can be adopted by a build system to any large project. Benefits coming from usage of these methods are described in following chapter. Moreover, we add to the chapter all proprietary solutions that are distributed independently with the software source code.

If you are interested in optimizations introduced in the thesis, visit Chapters \ref{chap:ProfileGuidedReorderingPass} and \ref{chap:SemanticFunctionEquality}. Moreover, in the last chapter we compare our pass for semantic function equality with Identical Code Folding. For Mozilla Firefox, one can easily reduce the binary by about 4\% with just adding single linker option.

\chapter{Analysis}
\label{chap:Analysis}

This part explores major issues, limitations and techniques tightly connected to performance of large real-world applications. Beside small applications, large programs suffer from a special kind of problems. We will mainly focus on time spent during start-up of a program and size of a binary created by the toolchain. Even though Moore's law has been still valid, huge expansion of less powerful hardware, as e.g. modern smart phones, tablets or book readers, makes more sense for any further optimizations. Despite all the limitations, these platforms are running modern operating system, as e.g. Linux kernel, and are equipped with rich variety of applications (web browsers, simplified office suites, video player etc.). An integral part of this chapter is a comparison of all aforementioned build systems.

This comparison is performed on SPEC CPU2006 benchmark suite. Important to notice that the biggest benchmark from the suite is about 20$\times$ smaller than Mozilla Firefox.

\section{Complexity of Large Applications}

Nowadays, the number of applications that can be considered being large is still growing. In the thesis, we primarily focus on two modern web browsers: Mozilla Firefox and Chromium; an office suite Libreoffice and, finally, the Linux kernel. Apart from these applications, we also chose a set of medium-size applications for testing purpose: the GNU Image Manipulator and Inkscape. All these applications are open-source, so that any kind of code modification can easily be done.

\begin{figure}[htbp]
	\begin{center}
		\includegraphics[width=140mm]{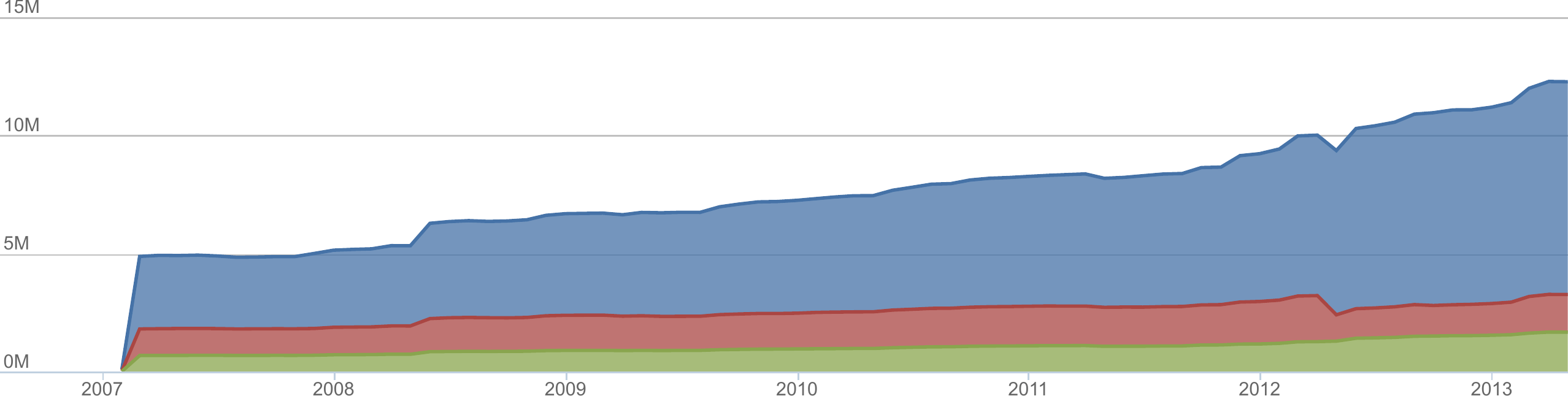}
		\caption{LOC of Firefox (blue: code; red: comments; green: blanks), taken from~\cite{OhlohFirefox}.}
		\label{fig:FirefoxLOCGrowth}
		\vspace{-10pt}
	\end{center}
\end{figure}

In Figure \ref{fig:FirefoxLOCGrowth}, we can see source code growth of Mozilla Firefox. More complex data can been seen in Table \ref{fix:LOCStatistics}. According to data taken from the source code statistics server Ohloh~\cite{OhlohFirefox}, the source code of Mozilla Firefox grows by nearly one third annually. As for the Chromium browser, established just 5 years ago, contributors add an additional quarter of source code every year. Even though, the Linux kernel code base is huge as well, a source code for device drivers consumes majority of the project. We typically build just a very small subset of drivers fitting to the target machine. Code base for Libreoffice looks stabilized, standing somewhere in the middle between the aforementioned web browsers and the Linux kernel.

\begin{table}[htbp]
	\centering
	\begin{tabular}{|l|r|r|r|r|}
	\hline
	& \textbf{Firefox} & \textbf{Chromium} & \textbf{Libreoffice} & \textbf{Linux} \\ \hline
	\textbf{Lines of code} & \num{9134485} & \num{7423679} & \num{13273239} & \num{15746046} \\ \hline
	\textbf{Monthly LOC growth} & \num{149120} & \num{104791} & \num{-13287} & \num{62520} \\ \hline
	\textbf{Annual LOC growth} & \num{1789440} & \num{1257492} & \num{-159444} & \num{750240} \\ \hline
	\textbf{Monthly perc. growth} & 2.68\% & 2.13\% & -0.1\% & 0.44\% \\ \hline
	\textbf{Annual perc. growth} & 32.2\% & 25.62\% & -1.2\% & 5.27\% \\ \hline
	\end{tabular}
	\caption{Large applications source code statistics.}
	\label{fix:LOCStatistics}
\end{table}

For our experiment, we have chosen Firefox 8.0.1 (released~on~November~21,~2011), having about 5.24\,M lines of code, and latest Firefox 25.0a1 (June 1, 2013), consisting of about 7.42\,M lines of code. As we can see in Table \ref{fix:FirefoxCompilationStatistics}, compilation time is not linear to the size of the source code. While the code has grown by about 40 percent, real compile time has extended more than twice. We utilized 8-core CPU without Hyper-threading technology (AMD FX-8350), which is used by degree of parallelism equal to 4.64 (on average) for Firefox 8.0.1. Nevertheless, utilizing factor increases in the latest Firefox release to 6.37 and it helps to significantly minimize total compile time.

\begin{table}[htbp]
	\centering
	\begin{tabular}{|l|r|r|r|}
	\hline
	& \textbf{Firefox 8.0.1} & \textbf{Firefox 25.0a1} & \textbf{Proportion} \\ \hline
	\textbf{Lines of code} & 5.24M & 7.42M & 142\% \\ \hline	
	\textbf{Real compilation time} & 37.2 min. & 86.7 min. & 233\% \\ \hline
	\textbf{User compilation time} & 172.6 min. & 551.9 min. & 320\% \\ \hline
	\textbf{CPU utilization factor} & 4.64 & 6.37 & 137\% \\ \hline
	\end{tabular}
	\caption{Compilation statistics for Firefox.}
	\label{fix:FirefoxCompilationStatistics}
\end{table}

\section{Real world usage of Link-Time Optimization}

In this section, we present statistical informations and CPU and memory utilization. 

\subsection{Feasibility of LTO to Compile Large Applications}

Having a large application, LTO can hardly run on IA-32 architecture, where the usage of memory is limited to 4\,GB. As seen at the beginning chapter, the application are still growing and it would be painful to utilize link-time optimization on average person computer.

In the following graph, we can easily distinguish both WPA and LTRANS phase of the compilation. As WPA reads all streamed object files to memory, the utilization increases to about 12\,GB and just a single CPU code is used. Memory usage deflection is a divide, where WPA decisions are streamed to LTRANS object files. Finally, all parallel LTRANS phases load the content of the files and run in parallel. LTRANS partitions with the biggest number of symbols are run first, thus memory utilization decreases during the time.

\begin{figure}[htbp]
	\begin{center}
		\includegraphics[width=140mm]{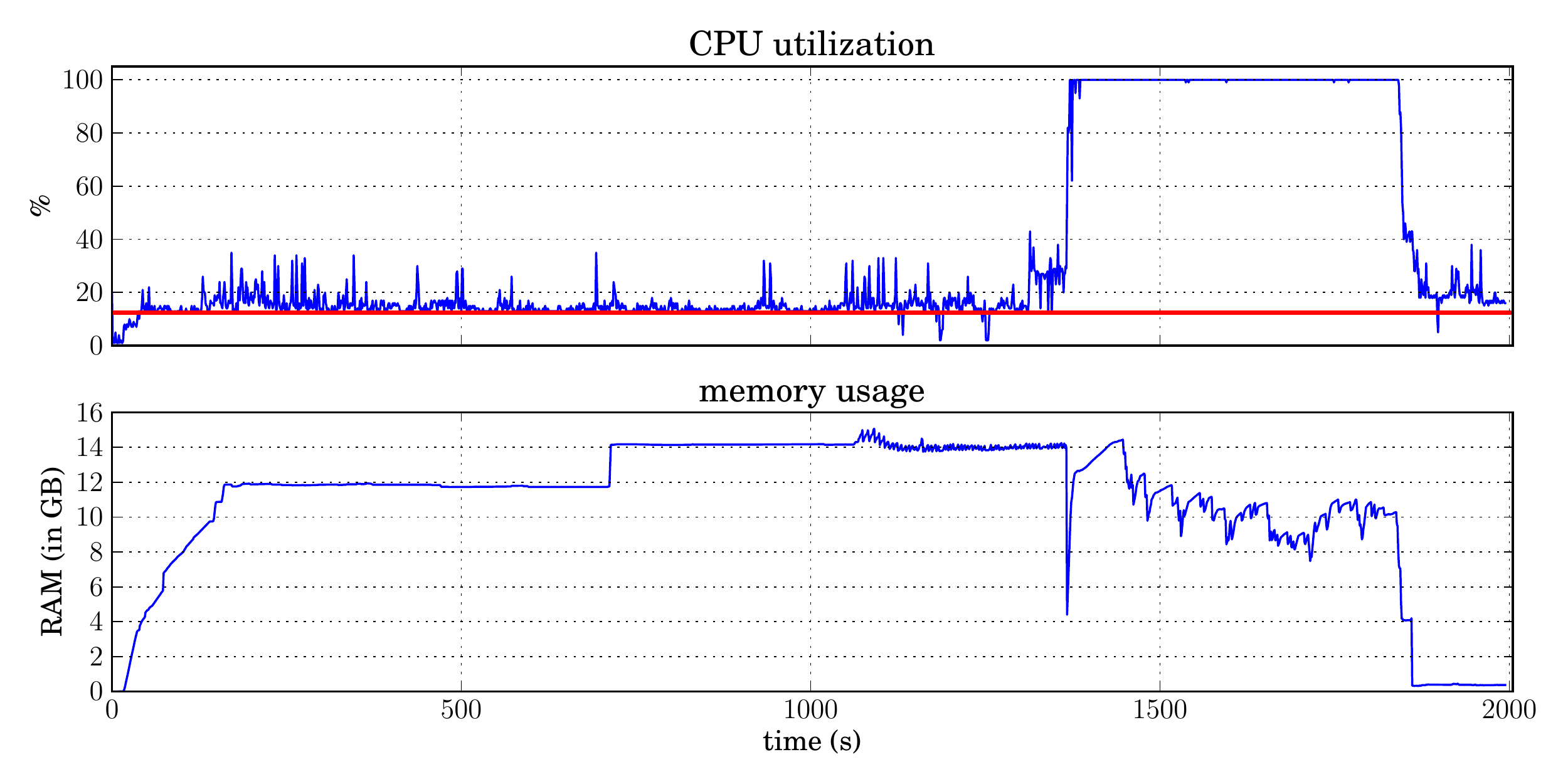}
		\caption{System utilization for Chromium.}
		\label{fig:VmstatChrome}
		\vspace{-10pt}
	\end{center}
\end{figure}

From perspective of time, please follow Table \ref{fix:FirefoxLTOCompilationStatistics}. The compilation is three times and complete rebuild of Firefox takes more than 100 minutes. Important to notice, the compilation was built with a high-end CPU (AMD FX-8350). These numbers document that LTO is going to be technique targeted at build servers.

\begin{figure}[htbp]
	\begin{center}
		\includegraphics[width=140mm]{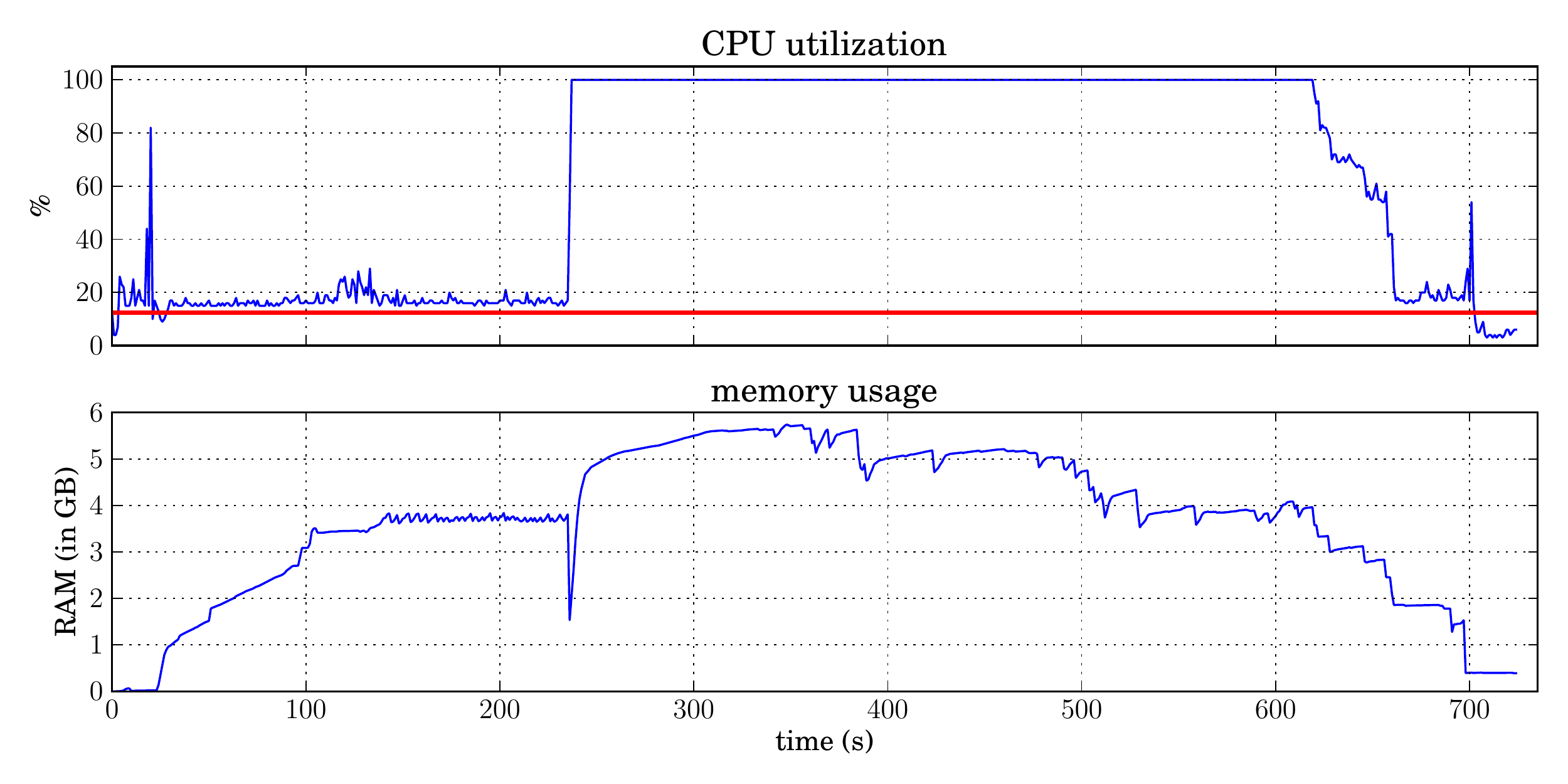}
		\caption{System utilization for \texttt{libxul.so} (Firefox 8.0.1).}
		\label{fig:VmstatFirefox8}
		\vspace{-10pt}
	\end{center}
\end{figure}

\begin{figure}[htbp]
	\begin{center}
		\includegraphics[width=140mm]{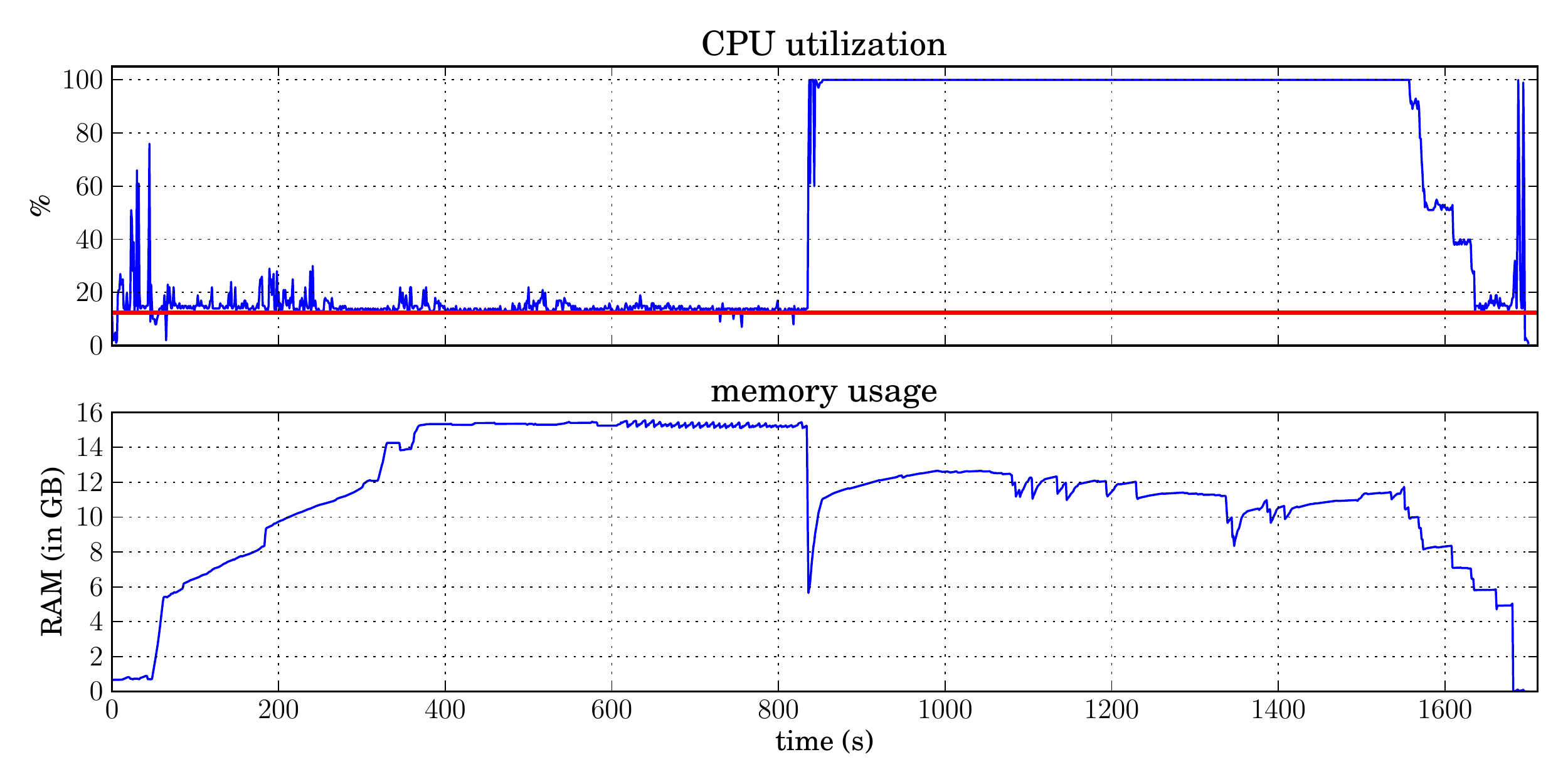}
		\caption{System utilization for \texttt{libxul.so} (Firefox 25.0a1).}
		\label{fig:VmstatFirefox8}
		\vspace{-10pt}
	\end{center}
\end{figure}

\begin{table}[htbp]
	\centering
	\begin{tabular}{|l|r|r|r|}
	\hline
	& \textbf{Firefox 8.0.1} & \textbf{Firefox 25.0a1} & \textbf{Proportion} \\ \hline
	\textbf{Real compilation time} & 41 min. & 107.9 min. & 263\% \\ \hline
	\textbf{User compilation time} & 202.3 min. & 648.3 min. & 320\% \\ \hline
	\textbf{Real CT for \texttt{libxul.so}} & 11.39 min. & 28.3 min. & 248\% \\ \hline
	\textbf{User CT for \texttt{libxul.so}} & 56.5 min. & 112.4 min. & 199\% \\ \hline
	\end{tabular}
	\caption{Compilation statistics for Firefox with LTO.}
	\label{fix:FirefoxLTOCompilationStatistics}
\end{table}

On the other hand, during the work on the thesis, there were applied many patches to GCC that significantly reduced memory utilization. Most beneficial was a collection of patches submitted by Richard Günther which improved type merging. According to the maintainers, there is still space for improvement in call graph area.

\subsection{Growth of Compilation Unit}
\label{subs:GrowthOfCompilationUnit}

Function inlining, as a inter-procedural optimization, is probably the most beneficial optimization technique. Unit growth boundaries are set relatively to the size of an application and the default value configuration allow growth of 30\%. To show the speed-up progress influenced by the inlining, we take the largest SPEC benchmark called \texttt{xalancbmk}. The speed-up curve is stepwise and O3 profile is more than 10\% faster. The experiment shows that further unit growth expansion bring a noticeable benefit. On the other hand, there are SPEC benchmarks like \texttt{bzip2} that are small enough to fit the CPU caches and the growth of the binary leads to speed degradation.

We propose a patch, attached on the supplemented CD, which makes a boundary between small and large applications. If the application is recognized as small, we enhance growth threshold that can take place in link-time optimization. Different compilers use similar techniques, where the factor can be a linear function to the size of the application.

\begin{figure}[htbp]
	\begin{center}
		\includegraphics[width=140mm]{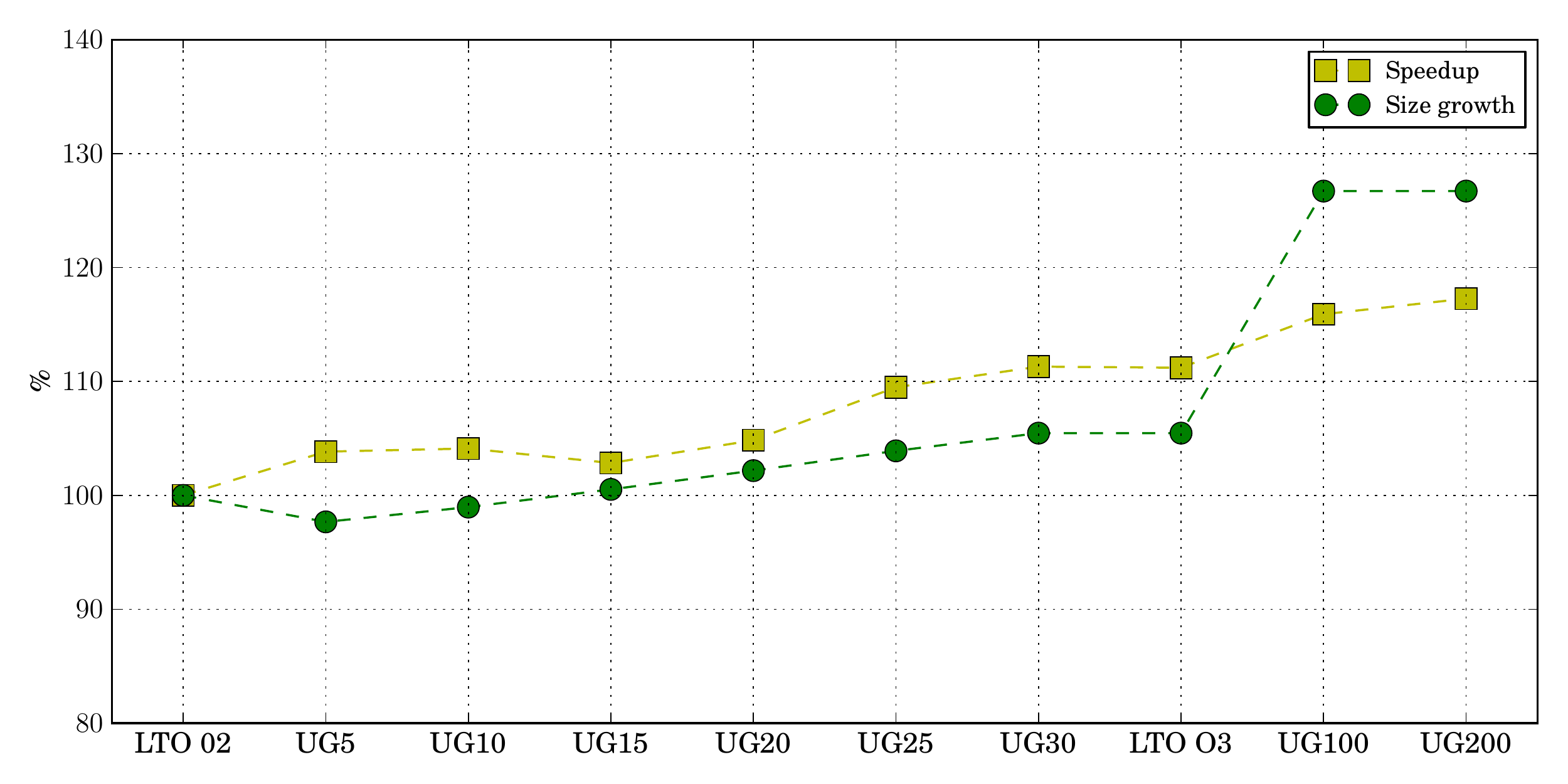}
		\caption{Unit growth statistics for \texttt{xalancbmk} SPEC benchmark.}
		\label{fig:UnitGrowthXalancbk}
		\vspace{-10pt}
	\end{center}
\end{figure}

\subsection{Compilation Troubleshooting}

As we mentioned earlier, despite the link-time optimization infrastructure looks mature enough to deal with large application. We have met many issues related to ecosystem toolchain as e.g. compiler, linker or build system. On the top of that, reporting such issues encompass many dependencies. Reporting can dependent on application repository version, version of compiler, selected linker, correct compiler and linker flags. They are passed either by the system environment or they are added to the configuration script. Additionally, dumps coming from LTO are extremely large, e.g. inter-procedural and whole program analysis dumps of \texttt{libxul.so} achieve 16\,GB.

When we started to examine LTO, Firefox and the Linux kernel were capable of utilizing the infrastructure. Unlike in SPEC, where one can just add the option and compilation works, problems connected with Libreoffice and Chromium were more complex. We have to overcome many consecutive issues and in my opinion, nowadays, it would be easier to port any new large application.

Chosen LTO related issues:

\begin{itemize}
	\item \textbf{GCC bug 56312} (\url{http://gcc.gnu.org/PR56312})
	
	The bug actually consists of source code modification for Firefox and wrong toolchain configuration. Similarly to other large applications, GCC optimizes out a function or a variable that is intended to be externally visible as API and the compiler finds out that the function is unused. In the following situations, we ought to decorate such a function with \mbox{\texttt{__attribute__((used))}} keyword.
	
	The second problem related to toolchain was caused by using the system version of \texttt{ar}, \texttt{nm} and \texttt{ranlib}. All these executables take advantage of linker plug-in. For example, if we create \textit{slim} object files, the output from \texttt{nm} differs in the following way:
	
\footnotesize
\begin{lstlisting}[basicstyle=\ttfamily]
	
$ nm-with-lto gimpselection.o
         U _gimp_selection_float
00000000 T gimp_selection_clear
00000000 T gimp_selection_float
         U gimp_selection_none
         
$ /usr/bin/nm gimpselection.o
0000000000000001 C __gnu_lto_slim
0000000000000001 C __gnu_lto_v1
	
\end{lstlisting}
\normalsize

As a result, the archive was mixed with slim object files in linking phase and some of compiler's assertion was reached. The patch for decoration of Firefox symbols was submitted to following bug issue: \url{https://bugzilla.mozilla.org/show_bug.cgi?id=826481}.
	
	\item \textbf{Libreoffice bug 61627} (\url{https://bugs.freedesktop.org/show_bug.cgi?id=61627})
	
	Configuration for Libreoffice build system was really complex. More precisely, working directory contained multiple header files of an identical definition. That lead to multiple definition of a symbol, where objects used a different definition according to priority of included header files. As a result, an error was returned by the compiler. Conflicting header files were placed in the following folder locations:
	
\footnotesize
\begin{lstlisting}[basicstyle=\ttfamily]

$ find -name Sequence.h
./cppu/inc/com/sun/star/uno/Sequence.h
./solver/unxlngx6.pro/inc/com/sun/star/uno/Sequence.h
./workdir/unxlngx6.pro/Zip/cppu_odk_headers/ \
include/com/sun/star/uno/Sequence.h

\end{lstlisting}
\normalsize

	These is still a couple of units tests that do not work properly with enabled LTO. Most of them have assumptions about symbols optimized-out by the compiler. In fact, Libreoffice testing suite encompasses tens of tests, successfully running with LTO and even the main standalone applications like \textit{Writer}, \textit{Calc} or \textit{Impress} were deeply tested and no problem has been encountered.

	\item \textbf{Binutils bug 15516} (\url{http://sourceware.org/bugzilla/show_bug.cgi?id=15516})
	
	For the compilation, we used the system version of \texttt{GNU ld}. The linker was confused when the plug-in turned COMDAT symbol into the static without renaming it. Mainline \texttt{binutils} works fine and even in the system version of \text{gold} the linker did not suffer from the issue.
	
	\item \textbf{GCC bug 57703} (\url{http://gcc.gnu.org/PR57703})
	
	Source code tree of Chromium contains a top-level assembler function that is placed to a different LTRANS partition, resulting in undefined symbol error. This problem is known as a missed feature of the top-level assembler and will be discussed more. Workaround could be  done by simply disabling partitioning for that file, either by setting \texttt{--lto-partition=none}, or by setting \texttt{--param lto-partitions=1}. Important to notice, content of LTO partitions is very volatile. Every slightest change in the configuration, in the compiler release or in the optimization flags, causes dissimilar partitioning result. Obviously, this fact makes any bug reproduction difficult.
	
	\item \textbf{GCC bug 57208} (\url{http://gcc.gnu.org/PR57208})
	
	This bug, similarly as the Firefox bug, aggregates all issues connected to the Chromium browser. First thing to notice, the Chromium tree contains binary of \texttt{gold} compiler for x64 Linux architecture. Moreover, some component require to be dynamically linked with \texttt{GNU ld}, which is searched on the system by \texttt{ld_bfd.py} script. To enable system gold linker, we have to comment on the line in \texttt{build/common.gypi}. This line is responsible for usage of the linker: \texttt{"-B\textless(PRODUCT_DIR)/../../third_party/gold"}.
	
	The second issue is related to serialization of LTRANS files. There is a LTRANS object with more than \num{65280} sections and the gcc compiler, using its own ELF file generator, does not count with that limitation. Fortunately, we were able to extend the number of partitions by \texttt{--param lto-partitions=64}. Moreover, there is a patch (\url{http://gcc.gnu.org/bugzilla/attachment.cgi?id=30323}) that removes the limitation. The patch implements extended numbering scheme for the ELF format that removes any section limitation, so that LTRANS export phase has no further restraint.
	
	Finally, last bug that blocked the final linking phase was caused by a missing reference, overlooked by the inter-procedural constant propagation pass (IPA-CP). In fact, Libreoffice was also affected by the same bug and the mainline version of GCC was fixed by the following patch: \url{http://gcc.gnu.org/viewcvs/gcc?view=revision&revision=200468}.
	
	\item \textbf{GCC bug 57698} (\url{http://gcc.gnu.org/PR57698})
	
	A couple of functions in \texttt{RegExp.cpp}, in the Firefox tree, are decorated with GCC's keyword \texttt{__attribute__((always_inline))}. Developers mark these performance critical functions to force the compiler to always inline. Thanks to this, the call overhead will not happen. The error is always reported after the early inliner is finished. In the previous version of the compiler, on the other hand, this kind of errors was reported only if another inlining happened. Submitted patch (\url{http://gcc.gnu.org/viewcvs/gcc?view=revision&revision=201039}) removed the described issue.	
	
\end{itemize}

The aforementioned list proves that LTO is much easier to be adopted by a small program than by large and complex application. Even though there will be still a couple of issues, LTO as a infrastructure has big potential.

\subsection{The Linux Kernel}

The Linux kernel, as an ubiquitous piece of software, plays the key role in software industry from small embedded devices (cell phones, personal computers or even supercomputers). From the perspective of link-time optimization, Andi Kleen's~\cite{AndiKleenKernelLTO} extensive set of patches allow the kernel to be built with LTO. According to the patch set applied to his branch, to make the kernel compilable, many \texttt{externally_visible} attributes were added to variables that tend to be optimized-out. Besides that, similar amount of \texttt{asmlinkage} was added to enable global visibility for top-level inline assembler and some files were decorated to only be ignored by LTO.

First look at statistics in Table \ref{fix:KernelLTOStatistics}, collected on 8-core CPU, looks unsatisfying. The compilation time for the configuration file fitting a common PC prolonged 4$\times$ (from 6.2 minutes to 24.3 minutes). The kernel is a specific software that executes linking procedure 2-4$\times$. Running WPA phase, the most time consuming operation in LTO, makes the build process very slow. Such complicated link phases are caused by Linux kernel symbols, called \texttt{kallsyms}. When developing kernel, we sometimes want to access a kernel symbol that is presented in \texttt{/proc/kallsyms} (PC kernel image comprises about \num{50000} symbols).

\begin{table}[htbp]
	\centering
	\begin{tabular}{|l|r|r|r|}
	\hline
	\multicolumn{4}{|c|}{\cellcolor{Silver}\textbf{PC config (linked: 4$\times$)}} \\ \hline
	& \textbf{O2} & \textbf{LTO,Os} & \textbf{LTO,O2} \\ \hline
	Real compilation time & 6.2\,min. & 19.3\,min. & 24.3\,min. \\ \hline
	Real CT comparison & 100\% & 311.29\% & 391.94\% \\ \hline
	User compilation time & 46.5\,min. & 116.8\,min. & 151.3\,min. \\ \hline
	Compilation parallelism & 7.5 & 6.05 & 6.23 \\ \hline
	Binary size & 5.10\,MB & 4.49\,MB & 5.14\,MB \\ \hline
	Binary proportion & 100\% & 87.35\% & 99.22\% \\ \hline	
	\multicolumn{4}{|c|}{\cellcolor{Silver}\textbf{allyes config (linked: 2$\times$)}} \\ \hline
	& \textbf{O2} & \textbf{LTO,Os} & \textbf{LTO,O2} \\ \hline
	Real compilation time & 39.5\,min. & 74.8\,min. & 79.03\,min. \\ \hline
	Real CT comparison & 100\% & 189.37\% & 171.85\% \\ \hline
	User compilation time & 301.2\,min. & 484.3\,min & 517.6\,min. \\ \hline
	Compilation parallelism & 7.63 & 6.47 & 6.55 \\ \hline
	Binary size & 32.27\,MB & 27.1\,MB & 31.02\,MB \\ \hline
	Binary proportion & 100\% & 83.98\% & 96.13\% \\ \hline
	\multicolumn{4}{|c|}{\cellcolor{Silver}\textbf{allno config (linked: 4$\times$)}} \\ \hline
	& \textbf{O2} & \textbf{LTO,Os} & \textbf{LTO,O2} \\ \hline
	Real compilation time & 0.547\,min. & 1.40min. & 1.44\,min. \\ \hline
	Real CT comparison & 100\% & 255.94\% & 263.25\% \\ \hline
	User compilation time & 3.29\,min. & 7.10\,min. & 7.77 \\ \hline
	Compilation parallelism & 6.01 & 5.07 & 5.40 \\ \hline
	Binary size & 845.40\,KB & 827.44\,KB & 742.56\,KB \\ \hline
	Binary proportion & 100\% & 97.88\% & 87.84\% \\ \hline
	\end{tabular}
	\caption{Link-time compilation statistics for the Linux kernel.}
	\label{fix:KernelLTOStatistics}	
\end{table}

\begin{figure}[htbp]
	\begin{center}
		\includegraphics[width=140mm]{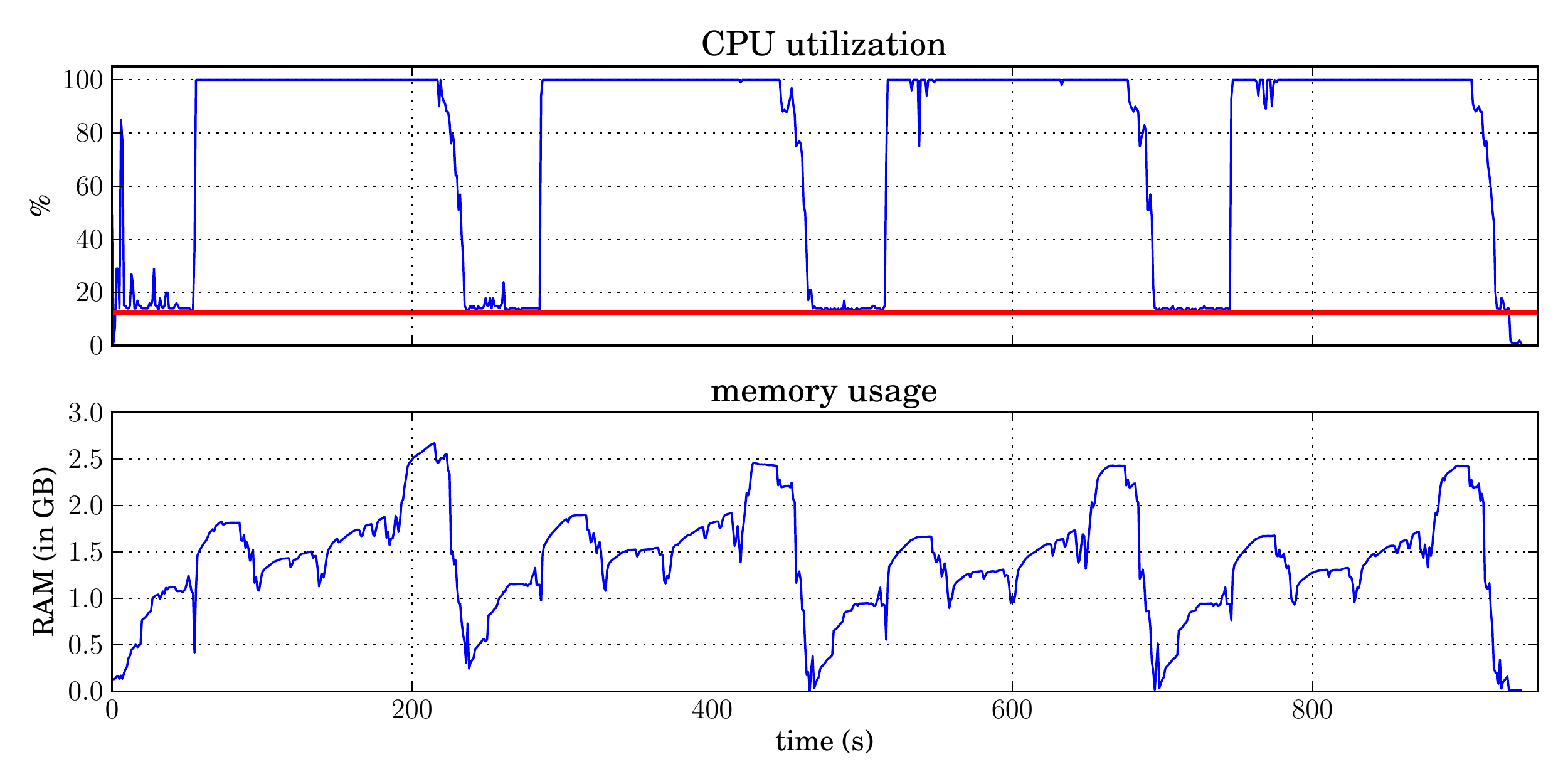}
		\caption{System utilization for the Linux kernel (PC configuration).}
		\label{fig:VmstatKernelMarxinbox}
		\vspace{-10pt}
	\end{center}
\end{figure}

Apart from the PC configuration, we also present data for special kind of kernel binaries: allyes configuration, where all options are set to \texttt{Y}, respectively \texttt{N} for allno configuration. While non-link-time compilation utilizes an 8-core CPU approximately by factor of 7.5, the whole program analysis (WPA) degrades the number to 6 for PC configuration (6.5 for allyes configuration). Figure \ref{fig:VmstatKernelMarxinbox} demonstrates CPU utilization, every single WPA phase touches the red line (single CPU utilization boundary), before the LTRANS stage utilizes all cores.

\begin{figure}[htbp]
	\begin{center}
		\includegraphics[width=140mm]{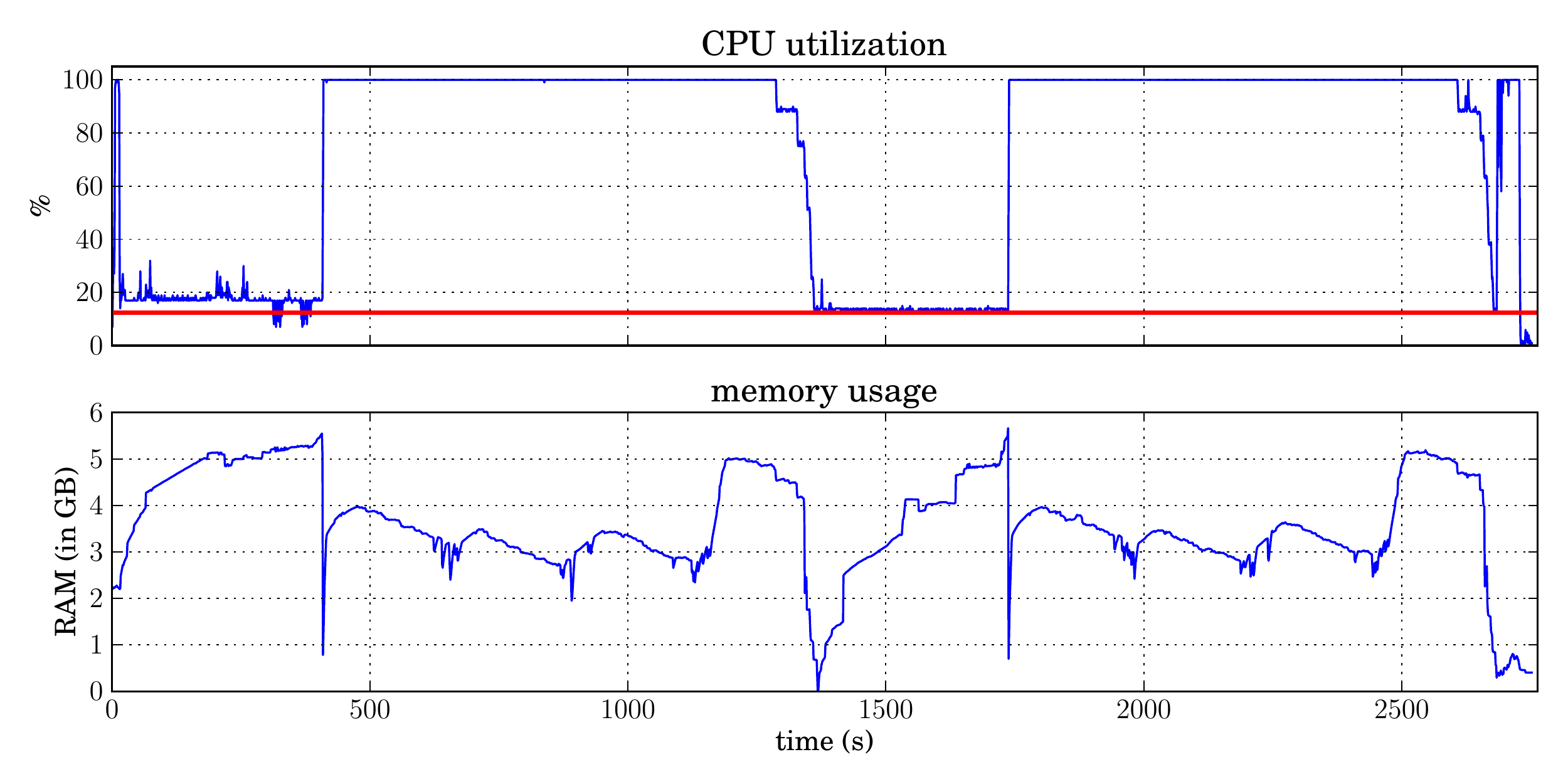}
		\caption{System utilization for the Linux kernel (allyes configuration).}
		\label{fig:VmstatKernelAllyes}
		\vspace{-10pt}
	\end{center}
\end{figure}

From memory perspective, according to Kleen~\cite{AndiKleenKernelLTO}, parallel build with \texttt{-j16} option increases GCC 4.8.0 to a 15\,GB peak. Nevertheless, since the complicated iterative hash, consisting of 4 hash tables, was replaced with a more memory efficient implementation. Memory usage of Allyes configuration utilising a half degree of parallelism shrinks to less than 6\,GB (Figure \ref{fig:VmstatKernelAllyes}).

Additionally, link-time optimization produces smaller binary - allyes image is about 4\% smaller. If we optimize for size, we can reach another 10-15\%. Obviously, the kernel speed aspect is the most important. Benchmarks performed on a 4-core Intel\textsuperscript{\textregistered}~ Core\texttrademark~ i5 mobile CPU and is presented in the following table.

\begin{table}[htbp]
	\centering
	\begin{tabular}{|l|r|r|r|}
	\hline
	& Kernel 3.9.0 & Kernel 3.9.0,LTO & System kernel 3.9.6 \\ \hline
	PyBench & 0\% & -0.80\% & -0.43\% \\ \hline
	PgBench & 0\% & +0.01\% & \cellcolor{SpecWorse}-15.18\% \\ \hline
	ApacheBench & 0\% & \cellcolor{SpecGood}+3.40\% & \cellcolor{SpecBad}-3.76\% \\ \hline
	AIMX9 benchmark & 0\% & \cellcolor{SpecBetter}+16.14\% & N/A \\ \hline
	\end{tabular}
	\caption{Runtime speed-up of kernel images.}
	\label{fix:KernelRuntimeSpeedup}
\end{table}

\section{Performance Problems of Dynamic Linker}
\label{sec:PerformanceProblemsOfDL}

Traditional ELF symbol lookup algorithm, often called SysV, does a symbol lookup in the following steps:

\begin{enumerate}
	\item Hash value for a symbol name is computed.
	\item For each lookup scope determined by the symbol, following steps are executed:
		\begin{enumerate}
		\item Find the hash table bucket for the symbol and use the name offset as a string terminated by NULL character.
		\item Compare relocation name with the symbol name.
		\item Is case of both the reference and the definition are versioned, compare versions. If the versions are equal, or no versions are propagated, the definition we found is exactly the one we are looking for.
		\item If not, iterate the rest of the hash bucket until match is found.
		\item If the hash bucket chain lookup is not successful, repeat aforementioned steps until we go through all object lookup scopes.
	\item If all lookup scopes do not contain the definition, lookup error arises.
	\end{enumerate}
\end{enumerate}

According to the description of algorithm, multiple definitions do not cause a failure, the first definition is used. The most expensive part of the algorithm are lookups in hash table chains and the number of objects in lookup scope. For both successful and unsuccessful lookup, ELF hash table construction primarily determines the average length of a chain. In reality, the key role is played by the unsuccessful lookups, outbalancing the successful ones due to the number of objects we search for. \texttt{readelf} utility is able to serve these data for the original SysV hash table format:

\begin{minipage}{\linewidth}
\footnotesize
\begin{lstlisting}[label={lst:ReadelfHashFirefox8},caption=Hash table bucket distribution for Firefox 8.0.1,basicstyle=\ttfamily]
$ eu-readelf -I /tmp/firefox-8.0.1/lib/firefox-8.0.1/libxul.so 

Hist. for bucket list length in section '.hash' (4099 buckets):
 Addr: 0x00000190  Offset: 0x000190  Link to section: '.dynsym'
 Length  Number  % of total  Coverage
      0    1424       34.7%
      1    1461       35.6%     33.2%
      2     818       20.0%     70.3%
      3     304        7.4%     91.0%
      4      67        1.6%     97.1%
      5      23        0.6%     99.7%
      6       1        0.0%     99.8%
      7       1        0.0%    100.0%
 Average number of tests:   successful lookup: 1.544381
			  unsuccessful lookup: 1.074652
\end{lstlisting}
\normalsize
\end{minipage}

The dynamic linker offers a rich variety of debugging outputs connected to dynamic relocation statistics (\texttt{LD_DEBUG=help cat}). We execute Firefox 8.0.1 with set environment variable \texttt{LD_DEBUG=symbols}\, for total number of \num{12669} symbol relocations performed by the linker. The linker searches on average 25.37 different DSOs with average symbol string length equal to 21 characters. Important to notice, all string comparisons must be processed until a difference character is found or the end of the string is reached. On the other hand, modern CPUs are equipped with vector instructions. In case of AMX FX-8350, the dynamic linker uses \texttt{__strcmp_sse42} that could speed up the string comparison up to 12$\times$. For each symbol, the linker has to perform on average $1.074652 \times 25.37 = 27.2639$ string comparisons. The normal symbol length is about 21.13 characters, assuming a representative number of comparisons is, according to ~\cite{Drepper}, only 20\%. These statistics would theoretically lead to almost 1.5 million characters to be loaded from memory and compared.

Let's compare it with the telemetry collected by the dynamic linker:

\footnotesize
\begin{lstlisting}[label={lst:ReadelfTelemetryHashFirefox8},caption=Dynamic linker statistics for Firefox,basicstyle=\ttfamily]
$ date +%s.%N && LD_DEBUG=statistics /tmp/firefox/bin/firefox 
1373470808.856270113

runtime linker statistics:
  total startup time in dynamic loader: 3485280 clock cycles
	    time needed for relocation: 2586494 clock c. (74.2%)
                 number of relocations: 1263
      number of relocations from cache: 1006
        number of relative relocations: 1811
	   time needed to load objects: 691319 clock c. (19.8%)
[XRE_main] 1373470808.881
     ...
runtime linker statistics:
           final number of relocations: 12669
final number of relocations from cache: 9634
\end{lstlisting}
\normalsize

For Firefox' major library \texttt{libxul.so}, the telemetry from the dynamic linker shows that total time spent by the dynamic linker is negligible, only about 25\,ms. This library encapsulates all core functionality, so the dynamic symbol resolution process is minimal.

GNU hash table style, introduced in 2006\footnote{\url{http://www.sourceware.org/ml/binutils/2006-06/msg00418.html}}, tries to improve the most painful problem in dynamic symbol algorithm, the symbol lookup. The new table format lives in a separate ELF section (\texttt{.gnu.hash}) to support back compatibility. That means the dynamic linker could combine the old and the new format in seamless fashion. Primary motivation for the new implementation was to significantly reduce comparisons.

Lookup process is a bit different:

\begin{enumerate}
	\item Hash value for a symbol name is computed.
	\item For each lookup scope determined by the symbol process following steps:
		\begin{enumerate}
		\item Hash value computed in the previous step is used for testing whether the symbol is present at all. 2-bit Bloom filter\footnote{http://en.wikipedia.org/wiki/Bloom_filter} is used for that test. The filter is a space-efficient probabilistic data structure giving fast results, if an item is presented in a set. False positives are prohibited, but false negatives are not. False answer (not probabilistic result) means that following object should be searched.
		\item Find the hash table bucket for the symbol and use it as a symbol index.
		\item Access a symbol at the index determined by the symbol index and compare these hash values, ignoring bit 0.
		\item If they match, compare corresponding string names.
		\item In case of the reference and the definition are versioned, perform version comparison. If the versions are equal, or no versions are propagated, the definition we found is exactly one we are looking for.		
		\item If not, iterate the rest of the hash bucket until match is found or bit 0 is set.
		\item If the hash bucket chain lookup is not successful, repeat aforementioned steps until we go through all object lookup scopes.
	\item If all lookup scopes do not contain the definition, lookup error arises.
	\end{enumerate}
\end{enumerate}

Thanks to the Bloom filter, amount of lookups needed by symbols decreases about 10$\times$, using just a single memory access. In addition, even if hash chain is degraded, expensive chain iteration is performed just occasionally.

Speed improvements are even noticeable for Inkscape:

\footnotesize
\begin{lstlisting}[basicstyle=\ttfamily]
$ LD_DEBUG=statistics /tmp/inkscape-telemetry-gnu/bin/inkscape 

runtime linker statistics:
  total startup time in dynamic loader: 110504530 clock cycles

$ LD_DEBUG=statistics /tmp/inkscape-telemetry/bin/inkscape 

runtime linker statistics:
  total startup time in dynamic loader: 126128305 clock cycles
     
\end{lstlisting}
\normalsize

Time spent by the linker before application is run is reduced by about 15\%. Space requirements are similar - in case of old implementation 155\,KB, 160\,KB is needed for the new (both occupy about a percent of the binary's size).

\section{Kdeinit}

KDE\footnote{http://kde.org/} is an open-source application collection providing basic desktop functionality as well as applications for daily usage. KDE serves as an umbrella for many standalone projects, sharing core functionality in shared KDE libraries like \texttt{libkdecore.so}, \texttt{libkhtml.so} or \texttt{libkdeui.so}. Among these, all KDE applications are based on  QT framework\footnote{https://qt-project.org/}, a graphical user-interface toolkit. \texttt{kdeinit} is a process launcher similar to \texttt{init} capable of running all KDE applications and kdeinit loadable modules (KLMs). \texttt{kdeinit} is linked against all KDE shared libraries and provides a fast fork for any of KDE applications (through \texttt{kdeinit_wrapper}, \texttt{kshell} and \texttt{kwrapper commands}).

\section{ElfHack}
\label{sec:ElfHack}

In general, the ELF format introduces two classes of relocations: REL and RELA. Both of them aggregate several types of relocation, where some of them require an addend. The data representation is shown in the following listing:

\footnotesize
\begin{lstlisting}[basicstyle=\ttfamily]
typedef struct
{
	Elf64_Addr r_offset;
	Elf64_Xword r_info;
}
Elf64_Rel;

typedef struct
{
	Elf64_Addr r_offset;
	Elf64_Xword r_info;
	Elf64_Sxword r_addend;
}
Elf64_Rela;
\end{lstlisting}
\normalsize

\begin{itemize}
	\setlength{\itemsep}{0cm}
	\setlength{\parskip}{0cm}
	\item \textbf{r_offset} - indicates the location where the relocation should be applied
	\item \textbf{r_info} - contains both a symbol table index and relocation type. 64-bit systems use first 32-bit for the symbol reference and 32-bit for the relocation type.
	\item \textbf{r_added} - constant addend used to compute the relocation
\end{itemize}

As we can see, first (shorter) form loads the addend from the offset where the relocation takes place. Nevertheless, for all RELA entries, we waste space, as the addend is, in fact, stored twice. First time in the entry and second time at the offset. Obviously, one occurrence could be eliminated. To make thing more complicated, x86-64 uses only RELA relocations. Having words of 64 bits, each relocation costs us 8 bytes extra space. On the other hand, IA-32 uses just REL relocations, where each relocation takes 8 bytes (3 times less than x86-64).

ElfHack was originally introduced by Mike Hommey in his blog post~\cite{glandium-elfhack} in November 2010. He noticed that huge portion of \texttt{libxul.so} library is occupied by dynamic relocations. According to Table \ref{fig:FirefoxElfHackBenefit}, in time of writing the post, almost 20\,\% were taken by relocations. He did a set of observations about ELF relocation entries that could save space:

\begin{itemize}
	\setlength{\itemsep}{0cm}
	\setlength{\parskip}{0cm}
	\item \textbf{addend} - waste of space on x86-64 ABI in RELA relocations
	\item \textbf{relocation type} - currently no ABI uses more than 256 relocation types, meaning the rest of 3 bytes is never used
	\item \textbf{symbol reference} - following x86-64 ABI relocation types do not require symbol reference: R_X86_64_RELATIVE, R_X86_64_DTPMOD64; as we can see in Table \ref{fig:Firefox2010TableRelocations}, majority of relocation can spare symbol reference
	\item \textbf{r_offset} - all pointers on x86-64 ABI have 8 bytes, but binaries are far from 4GB, so that we can save other 4 bytes
\end{itemize}

First ElfHack version, presented in November 2012, took care of just R_X86_64_RELATIVE relocations. They generate majority of all relocations and, having just one type, none relocation type information is needed to be stored. A bump format of 8 bytes is chosen: first 4 bytes is used for \texttt{r_offset} (we suppose the binary is smaller than 4\,GB) and the second half stores the number of relocations applied consecutively. He noticed that a lot of relocation were coherent and using some kind of range is more efficient. Table \ref{fig:Firefox2010TableRelocations} presents distribution of relocation types and in Table \ref{fig:Firefox2010WasteOfSpaceTable}, we can see theoretically wasted space in ELF binary.

\begin{table}[htbp]
	\centering
	\begin{tabular}{|l|r|l|r|}
	\hline 
	\textbf{Relocation type} & \textbf{Number} & \textbf{ELF section} & \textbf{Size in B} \\ \hline
	R_X86_64_DTPMOD64 & \num{1} & .rela.dyn & \num{24} \\ \hline
	R_X86_64_GLOB_DAT & \num{238} & .rela.dyn & \num{5712} \\ \hline
	R_X86_64_64 & \num{27611} & .rela.dyn & \num{662664} \\ \hline
	R_X86_64_RELATIVE & \num{208043} & .rela.dyn & \num{4993032} \\ \hline
	R_X86_64_JUMP_SLOT & \num{3773} & .rela.plt & \num{90552} \\ \hline
	\textbf{Total} & \textbf{\num{239666}} & & \textbf{\num{5751984}} \\ \hline
	\end{tabular} 
	\caption{Dynamic relocations in mozilla-central, November 22, 2010.}
	\label{fig:Firefox2010TableRelocations}
\end{table}

\begin{table}[htbp]
	\centering
	\begin{tabular}{|l|r|l|r|l|}
	\hline 
	\textbf{Description} & \textbf{Entries} & \textbf{Space} & \textbf{Total} & \textbf{Portion} \\ \hline
	Duplicated addend & \num{239666} & 8 & \num{1917328} & 33.33\% \\ \hline
	Relocation type encoding & \num{239666} & 3 & \num{718998} & 12.49\% \\ \hline
	Relocation without sym. ref. & \num{208044} & 4 & \num{832176} & 14.46\% \\ \hline
	Relocation offset & \num{239666} & 4 & \num{958664} & 16.93\% \\ \hline
	\textbf{Total} & \multicolumn{2}{r|}{} & \textbf{\num{4427166}} & \textbf{78.20\%}  \\ \hline
	\end{tabular} 
	\caption{Wasted space in mozilla-central, June 1, 2013.}
	\label{fig:Firefox2010WasteOfSpaceTable}
\end{table}

Unlike in November 2010, ElhHack implementation became more complex, today supporting IA-32, x86-64 and ARM architectures. My observation is primarily focused on x86-64, where \texttt{R_X86_64_64} and \texttt{R_X86_64_RELATIVE} relocations are supported. Important to say that the tool packs just \texttt{R_X86_64_RELATIVE} relocation, which takes place in writeable sections. Unfortunately, injected code is not allowed to modify not-writeable memory segments. As the number of lines is growing during the time (let us see Figure \ref{fig:FirefoxLOCGrowth}), number of relocations has been increasing as well.

\begin{table}[htbp]
	\centering
	\begin{tabular}{|l|r|l|r|l|}
	\hline 
	\textbf{Description} & \textbf{Entries} & \textbf{Space} & \textbf{Total} & \textbf{Portion} \\ \hline
	Duplicated addend & \num{314369} & 8 & \num{2514952} & 	33.33\% \\ \hline
	Relocation type encoding & \num{314369} & 3 & \num{943107} & 12.5\% \\ \hline
	Relocation without sym. ref. & \num{295757} & 4 & \num{1183028} & 15.68\% \\ \hline
	Relocation offset & \num{314369} & 4 & \num{1257476} & 16.67\% \\ \hline
	\textbf{Total} & \multicolumn{2}{r|}{} & \textbf{\num{5898563}} & \textbf{78.18\%}  \\ \hline
	\end{tabular} 
	\caption{Wasted space in mozilla-central, November 22, 2010.}
	\label{fig:Firefox2013WasteOfSpaceTable}
\end{table}

Summary of Firefox ElfHack is presented in Table \ref{fig:FirefoxElfHackBenefit}. Data trend shows that the representation of relocations in comparison to the size of the binary is decreasing.

\begin{table}[htbp]
	\centering
	\begin{tabular}{|l|r|r|}
	\hline 
	\textbf{libxul.so} & \textbf{November 22, 2010} & \textbf{June 1, 2013} \\ \hline
	\multicolumn{3}{|c|}{\cellcolor{Silver}Before} \\ \hline
	Dynamic relocations in B & \num{5661432} & \num{7544856} \\ \hline
	Binary size in B & \num{29629040} & \num{54760917} \\ \hline
	Lines of code & \num{5455206} & \num{8996787} \\ \hline
	\multicolumn{3}{|c|}{\cellcolor{Silver}After} \\ \hline	
	Dynamic relocations in B & \num{668400} & \num{319296} \\ \hline
	Injected code & \num{84} & \num{76} \\ \hline
	Packed relocations in B & \num{228568} & \num{311928} \\ \hline
	Binary size in B & \num{24865520} & \num{47847429} \\ \hline
	Binary reduction in B & \num{4763520} & \num{6913488} \\ \hline
	Saved size & \textbf{16.07\%} & \textbf{12.62\%} \\ \hline
	\end{tabular} 
	\caption{Elfhack results.}
	\label{fig:FirefoxElfHackBenefit}
\end{table}

Primarily caused by relocation dominance in read-only ELF sections, Elfhack can save for Chromium about 5 percent in case by of default disabled linker option \texttt{-z relro}. From security reasons the browser protects virtual tables with the linker option. With enabled protection, Elfhack saves just only 0.59\%.

\begin{table}[htbp]
	\centering
	\begin{tabular}{|l|r|r|r|r|}
	\hline
	\textbf{Name} & \textbf{Size} & \textbf{ElfHack bin. size} & \textbf{Reduction} & \textbf{Portition} \\ \hline
	libmergedlo.so & \num{43418584}\,B & \num{42886296}\,B & \num{532288} & 1.24\% \\ \hline
	chrome & \num{154931887}\,B & \num{147743607}\,B & \num{458552} & 4.64\% \\ \hline
	\end{tabular} 
	\caption{ElfHack statistics.}
	\label{fig:LargeApplicationElfhackStatistics}
\end{table}

\section{ELF Format and Start-up Time Problems}

A big binary file layout tightly influences start-up time, more precisely total time needed for the kernel, a dynamic library loader and all initialization steps of the primary executable before the program is ready for user interaction. The biggest issue is to minimize the amount of hard drive page misses that are touched either by the dynamic loader or the program itself. Furthermore, a non-sequential reading makes a hard disk drive (HDD) seek which tends to hurt the performance. While solid-state drives are getting more cheaper and do not suffer from seek latency, a cell phone internal storage (ensured by e.g. Secure Digital cards) face even worse access curve than HDD. As an example, missed seek costs about 30\,ms (depending on manufacturer and the number of platters and rotation speed). On the other hand, latency of SSD is in order of magnitude lower (units of milliseconds).

Every block device in the Linux kernel has assigned read-ahead constant, telling how many 256-bytes sectors are pre-fetched during reading from the device. \texttt{blockdev} kernel utility calls a corresponding \texttt{ioctl} in the kernel:

\footnotesize
\begin{lstlisting}[basicstyle=\ttfamily]
$blockdev --getra /dev/sdb1
256
\end{lstlisting}
\normalsize

As we can see, each time a disk read operation is executed, the kernel pre-loads 16 pages ($256\times256 = \num{65536} = 16 \times \num{4096}$). As many modern computer subsystems, the disk drives are equipped with a buffer as well.

For a hard drive monitoring, we use SystemTap\footnote{http://sourceware.org/systemtap/}, a scripting language and tool for dynamic instrumentation of the Linux operating system. We are interested in all file system read operations and the script fits for \texttt{ext4} file system (inspired by~\cite{GlandiumSystemtap}):

\footnotesize
\begin{lstlisting}[basicstyle=\ttfamily]
probe begin {
	targetpid = target();
}

probe kernel.function("ext4_readpages") {
	if (targetpid == pid())
		file_path[tid()] = $file;
}

probe kernel.function("ext4_readpage") {
	if (targetpid == pid())
		file_path[tid()] = $file;
}

probe kernel.function("do_mpage_readpage") {
	if (targetpid == pid() && (tid() in file_path)) {
		now = gettimeofday_us();
		printf("%d %p %d\n", now, file_path[tid()],
			$page->index * 4096);
	}       
}
\end{lstlisting}
\normalsize

Every time the kernel is triggered for a file system read, either \texttt{ext4_readpages} or \texttt{ext4_readpage} sets global variable \texttt{targetpid} to the current process's ID and read file location is saved to \texttt{file_path} variable. After that, the kernel calls \texttt{_mpage_readpage}, the low-level function directly responsible for loading data to a kernel buffer.

The script produces the following output for the distribution installation of Inkscape. The first column presents a timestamp with microsecond precision. The second one is a pointer that could be used as a unique file identifier. The last one shows the offset in the read file:

\footnotesize
\begin{lstlisting}[basicstyle=\ttfamily]
$ stap stap_readpage_ptr.stp -c `which inkscape`

1373831122972881 0xffff8805ce149700 12869632
1373831122972900 0xffff8805ce149700 12873728
1373831122972906 0xffff8805ce149700 12877824
1373831122972912 0xffff8805ce149700 12881920
1373831122972917 0xffff8805ce149700 12886016
1373831122972933 0xffff8805ce149700 12890112
1373831122972939 0xffff8805ce149700 12894208
1373831122972944 0xffff8805ce149700 12898304
1373831122972949 0xffff8805ce149700 12902400
\end{lstlisting}
\normalsize

As seen in Figure \ref{fig:StapStatisticsFirefox25}, the dynamic loader reads during a start-up phase of the Firefox almost 80\% of the ELF binary, where sections \texttt{.rela.dyn}, \texttt{.data.rela.ro} and \textit{.data} are loaded entirely. Moreover, the biggest ELF section \texttt{.text} was loaded from more than 90\%, that gives a great opportunity for an improvement.

\begin{table}[htbp]
	\centering
	\begin{tabular}{|l|r|r|r|r|}	
		\hline
		\textbf{Section name} & \textbf{Size} & \textbf{Size portion} & \textbf{Disk read} & \textbf{Disk read portion} \\ \hline
		.rela.dyn & 7.1 MB & 14.98 \% & 7.1 MB & 99.97\% \\ \hline
		.text & 24.8 MB & 52.18 \% & 22.7 MB & 91.57\% \\ \hline
		.rodata & 3.5 MB & 7.28 \% & 3.1 MB & 88.31\% \\ \hline
		.eh\_frame\_hdr & 1.1 MB & 2.22 \% & 44.0 KB & 4.07\% \\ \hline
		.eh\_frame & 5.3 MB & 11.09 \% & 68.0 KB & 1.26\% \\ \hline
		.data.rel.ro & 2.8 MB & 5.98 \% & 2.8 MB & 99.94\% \\ \hline
		.data & 1.1 MB & 2.25 \% & 1.1 MB & 99.97\% \\ \hline
		.bss & 1.5 MB & 3.19 \% & 4.0 KB & 0.26\% \\ \hline
		\textbf{Total} & \textbf{47.6 MB} & & \textbf{37.4 MB} & \textbf{78.48\%}\\ \hline	
	\end{tabular}
	\caption{Disk read statistics for the ELF sections (covering more than a percent of the size) in \texttt{libxul.so}.}
	\label{fig:StapStatisticsFirefox25}
\end{table}

\begin{figure}[htbp]
	\includegraphics[width=150mm]{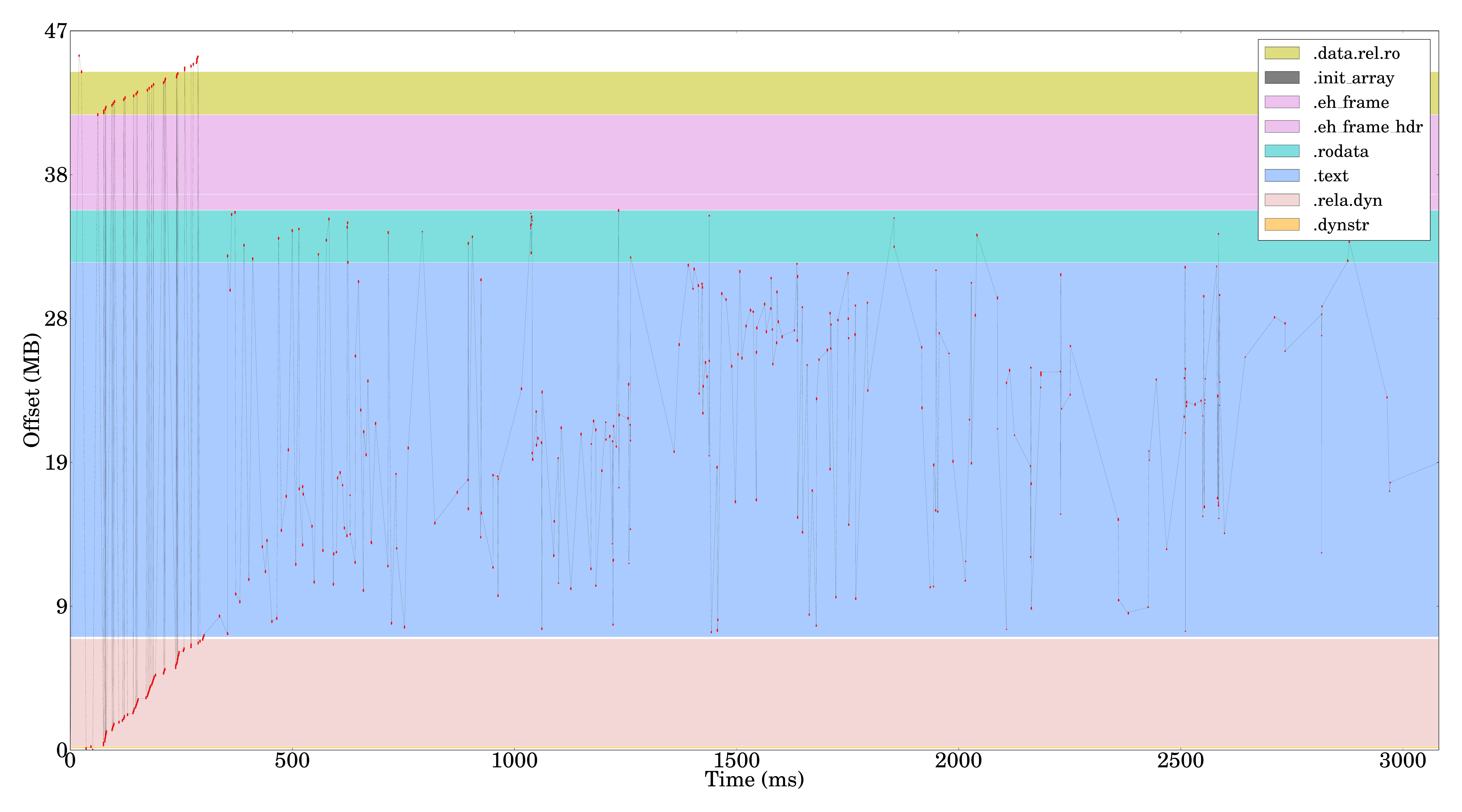}
	\caption{Disk seek graph for \texttt{libxul.so}, built with enabled LTO.}
	\label{fig:DiskSeekGraphFirefox25}
	\vspace{-10pt}
\end{figure}

In previous Figure \ref{fig:DiskSeekGraphFirefox25}, all starts with reading several disk pages at the beginning (ELF header and \texttt{.hash}), followed by tens (almost a hundred) of seeks forced by dynamic relocation process. \texttt{.rela.dyn} contains a set of rules, where we have to change the library depending on the base offset. As we can see, \texttt{ld.so} instead of reading the section at one time, the loader actually reads in small batches, as each single relocation is applied. In the meantime, all kind of sections like \texttt{.data.rel.ro}, \texttt{.data} and \texttt{.plt} are touched. Seeking to \texttt{.data.rel.ro}, placed almost at the end of the binary, forces the linker to skip the content of large sections \texttt{.text}, \texttt{.rodata} and \texttt{.eh_frame/.eh_frame_hdr}. This results in seek operations of more than 40\,MB.

Before LTO infrastructure was introduced, GCC created a static initializer function (and corresponding ELF section \texttt{.ctors}) for each compilation unit. After all the units were concatenated, the initializers were spread over the entire \texttt{.text} section and even run reversely! LTO's latest approach, in a similar way like the Microsoft C++ compiler and linker do, collects all occurrences of static constructors and destructors, sorts them by initialization priority and creates a single function calling all of them.

Once the dynamic loader executes \texttt{main} function, a random access pattern occurs over \texttt{.text} section with combination of \texttt{.rodata}. This random order will be observed in following sections. The function locality can be optimized for a size needed during start-up. Alternatively, hot functions, often calling each other, can reside very closely.

\begin{figure}[htbp]
	\includegraphics[width=150mm]{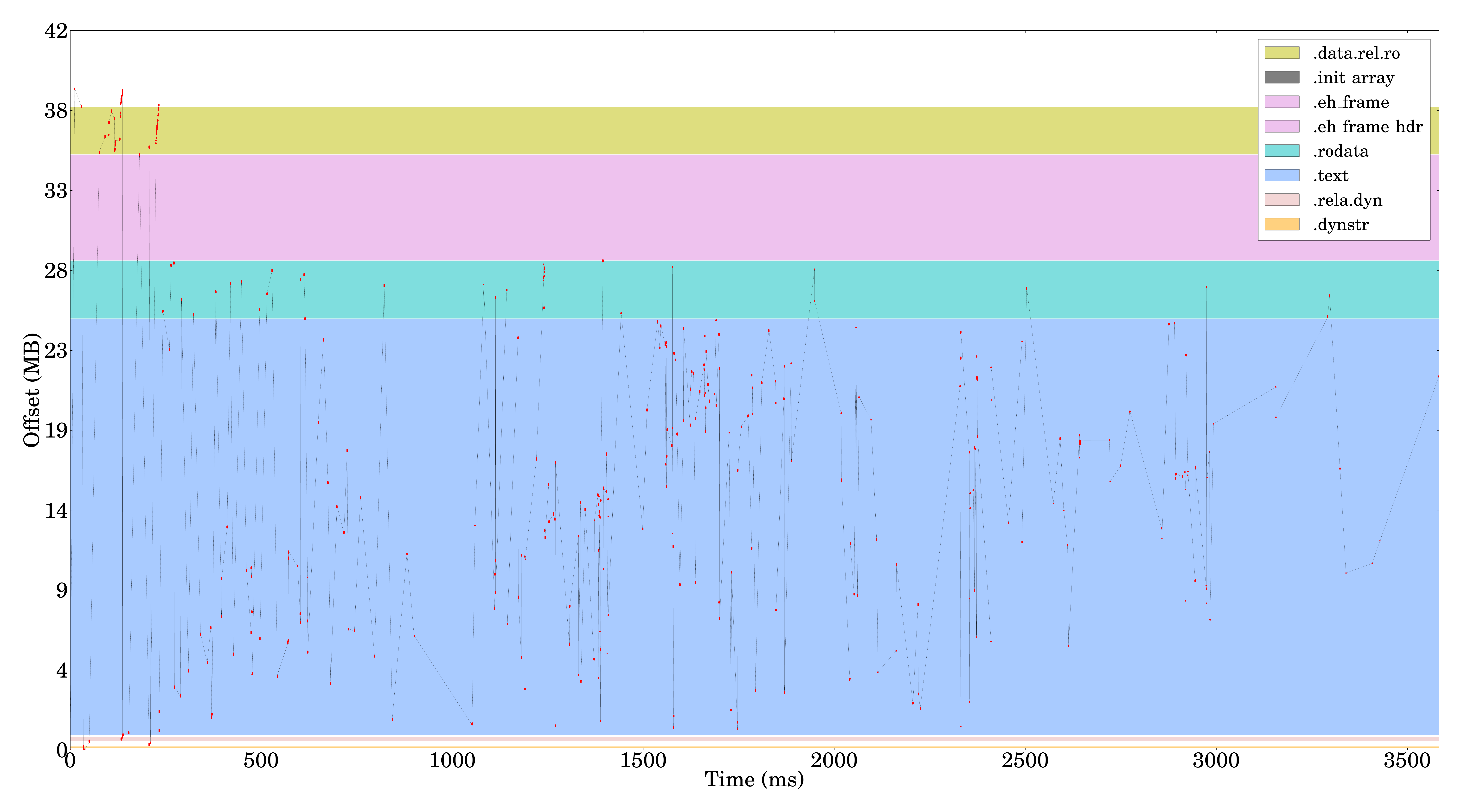}
	\caption{Disk seek graph for \texttt{libxul.so}, built with enabled LTO and Elfhack.}
	\label{fig:DiskSeekGraphFirefox25Elfhack}
	\vspace{-10pt}
\end{figure}

In Section \ref{sec:ElfHack}, the ElfHack technique rapidly reduces \texttt{.rela.dyn} section, compared to the previous graph, we save \num{1717} hard drive page reads (about 18\,\% of pages) and the final binary is about 15\,\% smaller. As seen in Figure \ref{fig:DiskSeekGraphFirefox25Elfhack}, majority of pages are read from \texttt{.text} ELF section.

\subsection{Binary preload}

Faster binary start-up could be reached by the prevention of the hard drive seeks. This especially applies for drives that suffer from random access I/O. As seen in the previous section, we read almost 80\,\% of the binary during the start-up. It is reasonable to do a preload in order to avoid the seeking process. Two years ago, Firefox project enabled preloading as a default option, according to~\cite{glandium-preload}, getting 20\,-\,30\% faster start-up times. Even if we have a fragmented file, it would be likely stored in chunks large enough so that it behaves approximately as a sequential scan. During the start-up, Firefox iterates all shared libraries. Every library is searched for \texttt{PT_LOAD} segment, which describes how the dynamic linker is going to map the file to memory. Having the information, the biggest offset from the binary is found. On the Linux system, \texttt{readahead} system call ensures the file will be read through page cache. We can demonstrate, how long it takes to read the entire \texttt{libxul.so} library in case of both fast so solid-state drives and slower hard-disk drives.

\begin{table}[htbp]
	\centering
	\begin{tabular}{|l|r|r|}
		\hline
		& \num{7200} RPM HDD (1\,TB) & OCZ 4 SSD (128\,GB) \\ \hline
		Size of \texttt{libxul.so} & \multicolumn{2}{r|}{\num{54370648}\,B} \\ \hline
		Read time & 0.693\,s & 0.150\,s \\ \hline
		Average read speed & 78.47 MB/s & 362.47 MB/s \\ \hline
	\end{tabular}
	\caption{Sequential read statistics for \texttt{libxul.so}.}
\end{table}

Mozilla Firefox maintainers spent much effort with a function reordering technique and invented ElfHack tool. But all these techniques were shown to be slower than simple preload of the binary in the kernel.

\section{Portable Executable vs. ELF Format}

The Portable Executable (PE) is a format used for executables and shared libraries in the Windows operating system. Among others, the format covers IA-32, x86-64 and IA-64 instruction set architectures. The goal of this section is to introduce the layout of the format and techniques dealing with the load mechanism for shared libraries.

The format is a variant of the Common Object File Format (COFF) that was used in early UNIX days before the ELF format. In addition, as you will see, many designs and principals are de facto equal to the ones mentioned in context of the ELF format. PE files comprise of section, e.g. \texttt{.text} or \texttt{.data}, that are mapped by memory pages to a virtual address space of an application with respect to the import attributes. These flags indicate a memory protection level (read-only, read/write, execute). The memory pages must be aligned for IA-32 architecture to 4\,KB (IA-64 uses 8\,KB aligned pages).

Even though the instrumented code is position-dependent to a base address, rebasing mechanism can be applied, if the operating system is unable to map a shared library to demanded base offset. The operating system calculates a delta value, a difference between assigned and base address, and the delta is added to all absolute virtual addresses. Such approach suffers from few things. First of all, rebasing is a process-specific operation and memory pages sharing mechanism is no longer possible. Second, loading of a shared library is significantly delayed. On the other hand, Microsoft ships all its libraries with precomputed base addresses, so that none of them overlap. Microsoft behaves as a central authority, actually distributing virtual address space slots, similarly to the \textit{Prelink} tool.

We are much more interested in function importing from a different shared library. Firstly, the resolution process for the main EXE file searches all shared libraries. Each dynamic library exports all API code and data in the Exports section, where all symbol names and ordinal numbers are stored. Ordinal number lookup is faster, but programmers are more familiar with function import by name. Export Address Table (EAT), the only mandatory array, is a container for pointers to all exported functions. Moreover, Export Names Table (ENT) holds a list of names for these functions, sorted for ensuring logarithmic search complexity. The index in the array is simply used to access function's address in the EAT table.

Assembler implementation of an API imported function call for IA-32 architecture can look like following sample:

\footnotesize
\begin{lstlisting}[basicstyle=\ttfamily]
CALL DWORD 0x005220AC
...
0x005220AC:
JMP DWORD [FunctionPointerAddr]
\end{lstlisting}
\normalsize

In situation of an API function, compiler generates a CALL instruction to the address, which will be filled by the linker. Last instruction call is realised via a short stub code that calls the API function through JMP instruction. One may notice that there is an opportunity for optimization. If we want to produce faster API calls, we can decorate exported function in a header file by \texttt{__declspec(dllimport)} attribute. In addition to generated assembler, the compiler also emits information about newly created function according to following name convention: \mbox{\texttt{__imp__[FunctionName]}}. From linker perspective, each of these functions will reside in the IAT table, rather than to stub code.

Data structures responsible for imported symbols are designed in similar manner. The linker assigns each imported library a structure called\\\mbox{\texttt{IMAGE_IMPORT_DESCRIPTOR}}. The structure stores name of a library together with two analogous arrays: Import Name Table (INT) and Import Address Table (IAT). An entry in the IAT table holds either the ordinal number of an API function or a pointer to the API function name. The goal of the linker is to overwrite all mentioned entries by actual address. The INT table duplicates the information from the IAT, but plays important role in Binding mechanism.

Microsoft toolchain can create a bound library, where an executable contains actual in-memory addresses of all API functions in the IAT table. As we can see, this technique is actually prelinking. Information about dependent shared libraries is placed in the executable and is quickly checked by the linker, whether all references are still valid. For an imported library, an entry with timestamp, library name and an array of imported symbols must be validated. If no change occurs, IAT table entries are valid. On the other hand, these entries are invalidated and a standard symbol resolution algorithm, using IAT table, comes to play.

Executable format in the Windows world does have the same predecessor, even though the format itself is different. Virtual address slot assignment and dynamic symbol resolution process behave similarly. Optimization approaches fully correspond with each other.

\section{SPEC CPU2006 Benchmark}

\subsection{Benchmark Configuration}

For reference, we include results of SPEC CPU2006 benchmark on a Linux machine with the Linux Kernel 3.3.8, Intel\textsuperscript{\textregistered} Core\texttrademark ~i5 CPU M 460 (2.53\,GHz) and equipped with 8\,GB RAM. For my benchmark, we chose three versions of GCC:

\begin{itemize}
	\item \textbf{gcc48} - checkout of \texttt{gcc-4_8-branch} cloned on July 4, 2013, compiled with enabled bootstrap
	\item \textbf{gcc49} - checkout of \texttt{trunk} cloned on July 4, 2013, compiled with enabled bootstrap
	\item \textbf{gcc-lipo} - checkout of \texttt{google} branch, where LIPO infrastructure is included, cloned on July 4, 2013 and also compiled with enabled bootstrap
\end{itemize}

For all compiler profiles, we chose the following set of compiler options:\\\texttt{-fno-strict-aliasing -fpeel-loops -ffast-math -match=native} and succeeding Table \ref{fix:SpecCPUProfileOptions} presents additional compiler (linker) flags. 

\begin{table}[htbp]
	\centering
	\begin{tabular}{|l|l|}
	\hline
	\textbf{Profile name} & \textbf{Compile/Linker options} \\ \hline
	gcc48-O2 & -O2 \\ \hline		
	gcc48-O3 & -O3 \\ \hline
	gcc49-O2 & -O2 \\ \hline
	gcc49-O3 & -O3 \\ \hline
	gcc49-O2-LTO & -O2 -flto=5 -fno-fat-lto-objects -fwhole-program \\ \hline
	gcc49-O3-LTO & -O3 -flto=5 -fno-fat-lto-objects -fwhole-program \\ \hline
	gcc49-O3-LTO-UG5 &  \pbox{20cm}{-O2 -flto=5 -fno-fat-lto-objects -fwhole-program \\ --param inline-unit-growth=5} \\ \hline
	gcc49-O3-PGO & -O3 -fprofile-generate/-fprofile-use \\ \hline
	gcc49-O3-LTO-PGO & \pbox{20cm}{-O3 -flto=5 -fno-fat-lto-objects -fwhole-program \\ -fprofile-generate/-fprofile/use} \\ \hline
	gcc49-O2-LIPO & -O2 -fripa -fprofile-generate/-fprofile-use \\ \hline
	gcc49-O3-LIPO & -O3 -fripa -fprofile-generate/-fprofile-use \\ \hline
	\end{tabular}
	\caption{SPEC CPU2006 compiler and linker profile options.}
	\label{fix:SpecCPUProfileOptions}	
\end{table}

\subsection{Benchmark Results}

\begin{figure}[htbp]
	\begin{center}
		\includegraphics[width=140mm]{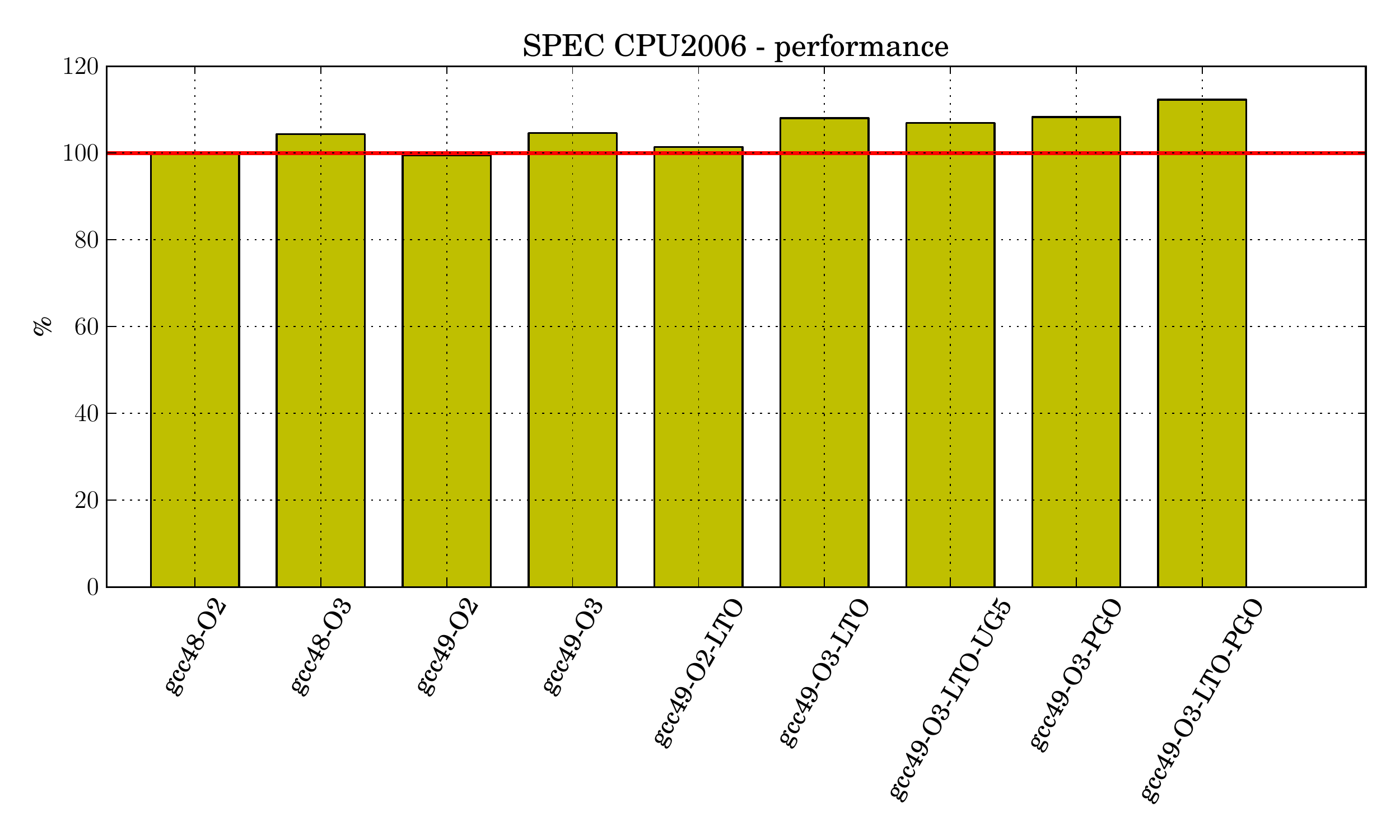}
		\vspace{-10pt}
	\end{center}
	\caption{SPEC CPU2006 speed-up graph.}
	\label{fig:Spec2006SpeedupGraph}
\end{figure}

In order to have representative results, we chose a branch of GCC 4.8 to cover officially released compiler. More precisely, this branch is going to be formal 4.8.2. As we can see from Table \ref{fix:Spec2006SpeedupSummary} and correlative Figure \ref{fig:Spec2006SpeedupGraph}, difference between GCC 4.8 and GCC 4.9 is within the noise level, both for \texttt{-O2} and \texttt{-O3}. In general, level O3 runs approximately 4-5\,\% faster than O2, producing 15\,\% larger binaries. While level O2 with enabled link-time optimization produces slightly faster code and level O3 adds another 4\,\% compared to non-LTO level O3. Moreover, resulting binaries are of the same average size as level O2 with enabled LTO. Last profile built without profile-guided optimization is based on the previous profile configuration, but the unit growth is limited to only 5\,\%. Profile generates slightly slower code with huge reduction in amount of 20\% compared to level O3 with enabled LTO and about half size to the same level without LTO. More about inlining can be found in Subsection \ref{subs:GrowthOfCompilationUnit}.

Last four profiles are feedback-driven. First run (with \texttt{-fprofile-generate}) produces an instrumented binary that was performed on the same testing data. Even though the final binary was in fact seen precisely on the same input data, PGO technique rapidly overcomes all compiler internal heuristics. Therefore, even with a worse set of input data, the final application speed is improved significantly. As a demonstration, level O3 profile is even faster than link-time optimization with the same level of optimization. If we combine both these techniques, we can gain additional 12\% of speed compared to O2 profile. In other words, the benefit of LTO coming from the collected profile adds one third of the speed-up. Finally, we tried to evaluate the speed-up benefit brought by LIPO infrastructure. Unfortunately, LIPO supports just C and C++ languages and SPEC contains quite a lot of Fortran based benchmarks. Moreover, even though there are benchmarks which can be built in an appropriate way, a runtime failure was encountered. Thus, the profile was removed from overall statistics.

Important to notice, in case of speed-up, a positive number indicates improvement and if the benefit is noticeable, we decorate the cell with green color. On the hand hand, negative numbers mean slow-down and call marked with read color. Conversely, binary size numbers with a negative value are marked with shades of green color and mean binary file reduction. To make the enumeration complete, we use red color for tests, where the final executable file dilates. These SPEC data convention is valid through the thesis.

Complete benchmark results are located in Appendix \ref{app:SpecCPU2006Results}.

\begin{table}[htbp]
	\centering
	\begin{tabular}{|l|r|r|r|}
	\hline
\textbf{} & \textbf{Speed-up} & \textbf{Speed-up (FP)} & \textbf{Speed-up (INT)} \\ \hline
\textbf{gcc48-O2} & 0.00\% & 0.00\% & 0.00\% \\ \hline
\textbf{gcc48-O3} & 4.37\% & 6.77\% & 0.96\% \\ \hline
\textbf{gcc49-O2} & -0.49\% & -0.54\% & -0.42\% \\ \hline
\textbf{gcc49-O3} & 4.60\% & 6.85\% & 1.41\% \\ \hline
\textbf{gcc49-O2-LTO} & 1.41\% & 1.54\% & 1.23\% \\ \hline
\textbf{gcc49-O3-LTO} & 8.02\% & 9.82\% & 5.62\% \\ \hline
\textbf{gcc49-O3-LTO-UG5} & 6.93\% & 9.29\% & 3.77\% \\ \hline
\textbf{gcc49-O3-PGO} & 8.29\% & 8.35\% & 8.21\% \\ \hline
\textbf{gcc49-O3-LTO-PGO} & 12.30\% & 12.41\% & 12.16\% \\ \hline
	\end{tabular}
	\caption{SPEC CPU2006 speed-up summary.}
	\label{fix:Spec2006SpeedupSummary}
\end{table}

\begin{figure}[htbp]
	\begin{center}
		\includegraphics[width=140mm]{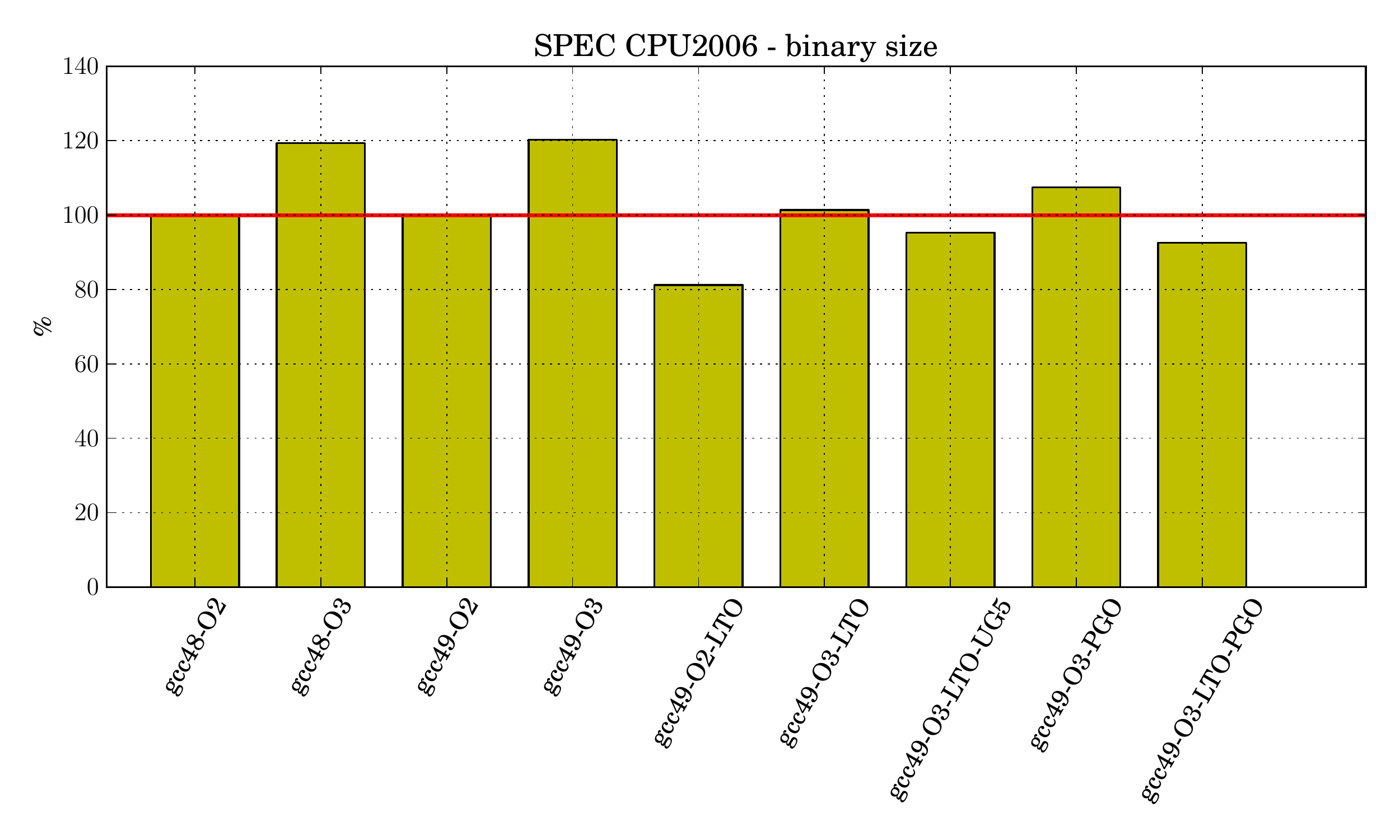}
		\vspace{-10pt}
	\end{center}
	\caption{SPEC CPU2006 binary size graph.}
\end{figure}

\begin{table}[htbp]
	\centering
	\begin{tabular}{|l|r|r|r|}
	\hline
\textbf{} & \textbf{Size } & \textbf{Size (FP)} & \textbf{Size (INT)} \\ \hline
\textbf{gcc48-O2} & 0.00\% & 0.00\% & 0.00\% \\ \hline
\textbf{gcc48-O3} & 19.32\% & 22.92\% & 14.20\% \\ \hline
\textbf{gcc49-O2} & -0.06\% & -0.09\% & -0.01\% \\ \hline
\textbf{gcc49-O3} & 20.25\% & 24.30\% & 14.52\% \\ \hline
\textbf{gcc49-O2-LTO} & -18.81\% & -19.76\% & -17.46\% \\ \hline
\textbf{gcc49-O3-LTO} & 1.35\% & 3.47\% & -1.65\% \\ \hline
\textbf{gcc49-O3-LTO-UG5} & -4.72\% & -1.23\% & -9.67\% \\ \hline
\textbf{gcc49-O3-PGO} & 7.43\% & 11.75\% & 1.33\% \\ \hline
\textbf{gcc49-O3-LTO-PGO} & -7.46\% & -4.68\% & -11.39\% \\ \hline
	\end{tabular}
	\caption{SPEC CPU2006 binary reduction.}
	\label{fix:Spec2006BinarySizeReduction}
\end{table}

\newpage

\section{Comparison of GNU ld and gold}

\texttt{gold} linker was one of the first system tools written in C++ language and the statistics presented at the end of the section proof that C++ is fast enough even for core tools. Compared to \texttt{GNU ld}, the new linker is custom-fitted just to ELF format, where data structure for a symbol table entry is almost half size for x86-64 (80\,B) and the symbol table is traversed just three times (the old linker walks the table thirteen times). Additionally, the linker script language is not presented at heart of the linker, although linker script are supported. There is an exception for the Linux kernel, where the linker script facility is extremely utilized even in undocumented manner. That is the reason why \texttt{gold} is not mature enough to build the Linux kernel. Despite the fact that \texttt{gold} is multi-threaded, in practice there is no significant speed improvement, presented in Figure \ref{fix:LinkerSpeedComparison}.

\begin{table}[htbp]
	\centering
	\begin{tabular}{|l|r|r|r|r|r|}
	\hline
	& \textbf{GNU ld} & \textbf{gold} & \textbf{gold (2T)} & \textbf{gold (3T)} & \textbf{gold (4T)} \\ \hline
	\texttt{libxul.so} & \textbf{9.887\,s} & \textbf{4.860\,s} & 4.742\,s & 4.742\,s & 4.796\,s \\ \hline
	\texttt{libxul.so} (comp.) & 100\,\% & 49.16\,\% & 47.96\,\% & 47.96\,\% & 48.50\,\% \\ \hline
	\texttt{chrome} & \textbf{8.738\,s} & \textbf{3.573\,s} & 3.401\,s & 3.406\,s & 3.363\,s \\ \hline
	\texttt{chrome} (comp.) & 100\,\% & 40.89\,\% & 38.92\,\% & 38.98\,\% & 38.49\,\% \\ \hline
	Inkscape & \textbf{1.061\,s} & \textbf{0.370\,s} & 0.365\,s & 0.372\,s & 0.375\,s \\ \hline
	Inkscape (comp.) & 100\,\% & 34.87\,\% & 34.40\% & 35.06\,\% & 35.34\,\% \\ \hline
	\end{tabular}
	\caption{Linker speed comparison.}
	\label{fix:LinkerSpeedComparison}
\end{table}

According to collected data, \texttt{gold} is really fast for small to large applications. Firefox main library \texttt{libxul.so} is linked twice faster with \texttt{gold} and uses just a third of the time for smaller applications as Inkscape. Both linkers were taken from official stable release of \texttt{binutils} (version 2.23.52.20130526).

Apart from speed improvement, \texttt{gold} offers additional new features not present in the old linker:

\begin{itemize}
	\item \textbf{One Definition Rule (ODR) detection} (\texttt{--detect-odr-violations})
	
	ODR says that any translation unit, template, type, function, or object must not have more than one definition. In a large and complex application, it is hard to detect violation of the rule. The linker uses heuristics, in which the size of symbols is compared and eventually, debugging information is compared. In fact, there are many false positives and false negatives.

	\item \textbf{Compression of debugging information} (\texttt{--compress-debug-sections})
	
	Using \texttt{zlib} library, the linker compresses debugging information, which can save up to a half of the size.
	
	\item \textbf{Incremental Linking}
	
	In an application compound of many objects, incremental linking is a technique which accelerates the software build after a single file (or small amount of files) is modified. The linker modifies existing binary, either by replacing existing code (if the newly created object is smaller), or by finding a new location in ELF binary. The linker falls back to normal linkage if one of the three happens: a command line changes, a linker script changes or if there is a symbol which migrates from one object file to another. If required, conditions are satisfied, \texttt{gold} traverses all object and archive files included on command line, checks them for timestamp changes and build a list of symbol changes. After changes are applied to appropriate sections, the dynamic relocation is applied to all changed files and all files referring to a symbol that changes. To support incremental linking, the linker creates auxiliary ELF sections like \texttt{.gnu_incremental_inputs}, \texttt{.gnu_incremental_symtab}, \texttt{.gnu_incremental_relocs}, \texttt{.gnu_incremental_got_plt} and \texttt{.gnu_incremental_strtab}. In case of the Chromium application, these sections cover about 80\,MB (36\% of the binary) and incremental link takes 1.837\,s (51.41\% of the time spent by a full link).

	\item \textbf{Concurrent Linking}	
	
	Concurrent linking is still under development. Initial implementation was presented in Sander Mathijs van Veen's work~\cite{ConcurrentGoldLinking}. In the compilation cluster, object files are created in parallel and it is possible to execute a linker in the same way. Every time a object file is generated, the linker can read the file, allocate sections in the output file and process relocation of already known symbols. As more files are finished, less of relocation work is remaining.
	
	\item \textbf{Identical Code Folding (ICF)}	 (\texttt{--icf=safe/all})
	
	Optimization technique will be presented in Subsection \ref{subsec:ICF}.
	
\end{itemize}

\section{Conclusion}

As shown in aforementioned sections, a lot of problems are connected to large shared libraries. These libraries contain position independent code, which runs fast on x86-64 architecture. Majority of large programs are organized as an executable that loads a large shared library. Maintainers often forget to mark internal symbols of the shared library as hidden and dynamic symbol resolution process is hurt by a huge number of seek operations. As we shown in Section \ref{sec:ElfHack}, ElfHack is a generic tool that should be either separated from Mozilla source base or the idea should be implemented naturally in a dynamic linker.

\chapter{Profile-Guided Reordering Pass}
\label{chap:ProfileGuidedReorderingPass}

Function reordering was introduced in~\cite{Muchnick}. This standard algorithm reorders functions in way that all functions that call each other frequent reside in similar location. The algorithm is implemented by the Google plug-in that was presented at the end of Subsection \ref{subsec:ReorderingInLinker}.

The \textit{profile-guided reordering}, proposed in this chapter, aims to track the call order of functions during start-up of an application. We are primarily focused on minimizing the number of disk pages needed to be read by a hard drive of any speed. In order to reduce the disk misses, we put all these called functions next to each other in \texttt{.text} section. Important to notice, out scope of control is reduced by generating the ELF section \texttt{.text}, where the order of function can be rearranged by the linker. Apart from that, there are also special function sections shown in~\cite{GlekHubickaLTO}. The pass utilizes value profiling infrastructure, described in~\cite{PDOInGCC}.

The reordering optimization is divided into the following stages:

\begin{enumerate}
	\item \textit{Value profile instrumentation}. During this phase, the compiler examines all functions and splits the edge entering to the first basic block (BB). We must be sure that the first BB will be called just once. The existing BB can be visited in the function more than once. We utilize existing time profiling infrastructure with an exception that our newly created histogram does not belong to any GIMPLE statement. Nevertheless, the histogram is always appended and read as a first the histogram in the function. Histogram-type property prevents any confusion of the existing types. Function call \texttt{__gcov_time_profiler} (implemented in \texttt{libgcov} library) is added to every newly created basic block.
	
	We add histogram counter with following data members:
	
	\begin{itemize}	
		\item First visit of a function. The counter is set just once we first time visit the function.
		\item Last visit of a function. We set the counter every time the function is called.
		\item Program run counter. If a function is visited we set the counter to one. In merge phase, this counter indicates how many time profiles do we have.
	\end{itemize}
	
	\item \textit{Time profile evaluation}. Profile instrumented binary should be run in a common way. Every time we first run a function, \texttt{__gcov_time_profiler} fills current value of the global function counter (\texttt{function_counter}). After that, the value of the counter is incremented. Furthermore, we increment the counter only if a function has not been run, thus function numbers are distributed continuously.
	
	\item \textit{Time profile merge}. At the end of execution, with values collected from the previous run, the profiling runtime (\texttt{libgcov}) merges histograms. It would make sense to introduce a counter for last call of the function and build a heuristics which will make a decision whether the symbol is actually a start-up function. When the distance between first and last run is small and the function is called relatively few times, we can mark the function.
	
	\item \textit{Time profile read}. Time profile is read  from a file (with \texttt{.gcov} extension) produced by runtime and stored to the call graph.
	
	\item \textit{LTO partitioning reorder}. To utilize full capacity of parallelism in LTO, in WHOPR mode~\cite{WHOPR}, we sort all functions with collected time profile in ascending order. After that, LTO partitioning algorithm distributes intended symbols to first \texttt{K} partitions. The functions with lowest time profile will reside in the first partition, followed by symbols in the second partition, etc.
	
	\item \textit{LTO streaming} Time profile, stored in \texttt{struct cgraph_node}, is serialized and deserialized during \textit{Write optimization} and \textit{Read optimization summary} phase of LTO respectively.
	
	\item \textit{LTRANS function reorder}. With enabled \texttt{-ftoplevel-reoder} option, the gcc compiler sorts all functions with profile in ascending order.
	
	\item \textit{ELF section creation}. Created \texttt{.text} sections of the ELF files generated by \textit{Local transformation} are concatenated and final \texttt{.text} section preserves the given order.
	
\end{enumerate}	
	
Assume a function \texttt{foo} which contains a loop comprising entry basic block. To guarantee single execution of profiling code, we split the entering edge. At the same time, the newly created basic block does the call to \mbox{\texttt{__gcov_time_profile}}.	
	
\section{Results}

We did an experiment of medium-sized applications, where the \texttt{.text} section is in amount of megabytes. In Table \ref{fix:DiskPagesStatistics}, we decrease the number of disk page misses by about 25\% in case of default settings of a hard drive. Even though the speed-up is negligible for \textit{GIMP}, for \textit{Inkscape} we are able to measure slight speed improvement.

\begin{figure}[htbp]
	\includegraphics[width=130mm]{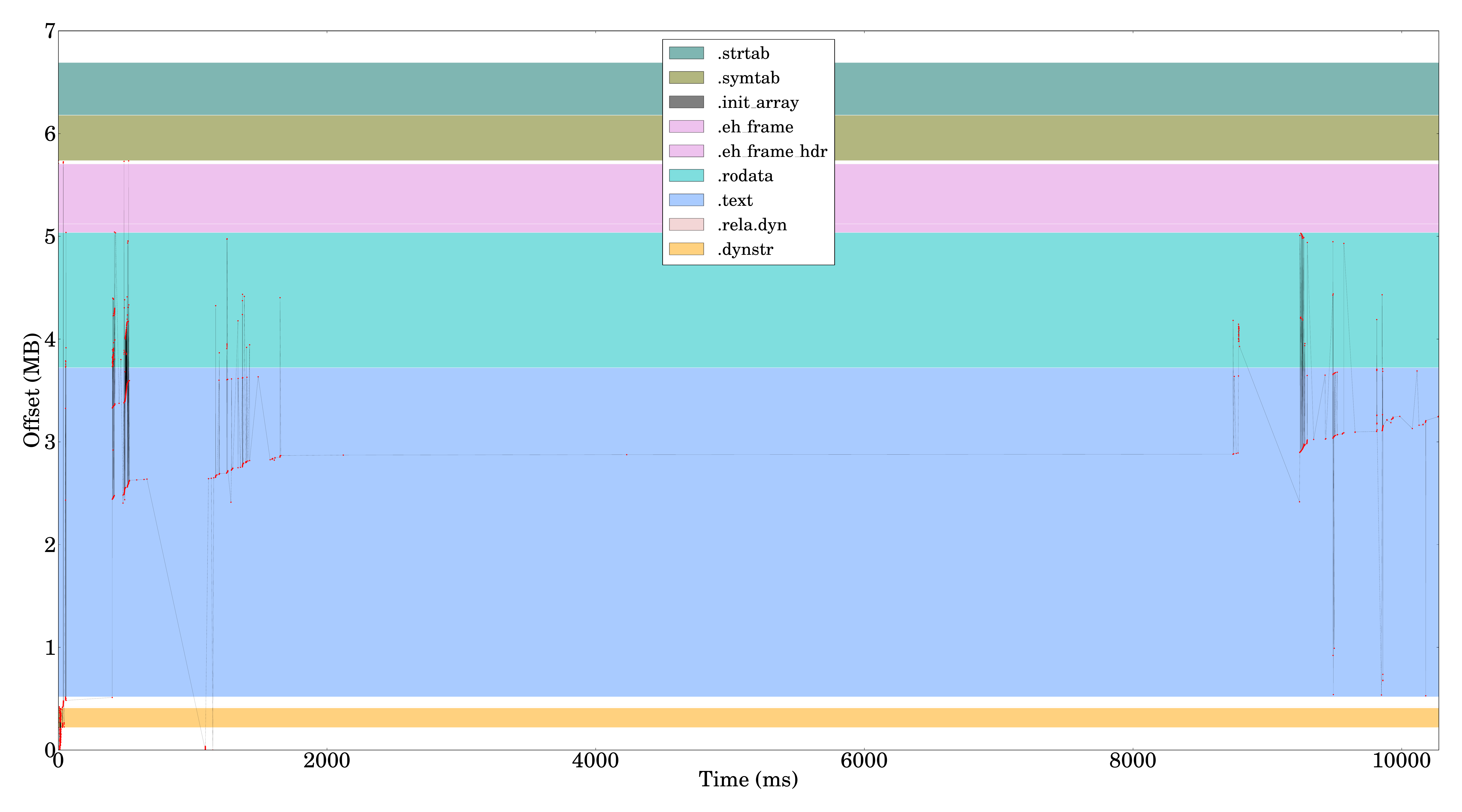}
	\caption{Disk page faults for reordered \textit{GIMP}.}
	\label{fig:DiskSeekGimpReordered}
	\vspace{-10pt}
\end{figure}

Every time the binary touches a single byte of a disk page which the Linux kernel does not include in its cache, the disk will most probably  need to read more pages (16 as the maximum). If we disable the kernel read-ahead feature, we are able to save almost one half of read pages. Page faults of \textit{Inkscape} are demonstrated in Figure \ref{fig:DiskSeekGimpReordered}.

Blog posts that deal with function reordering in Mozilla Firefox can be seen in~\cite{GlandiumFaultyLib} and~\cite{GlandiumBinaryLayout}.

\begin{table}[htbp]
	\centering
	\begin{tabular}{|l|r|r|r|}
	\hline
	& Disk pages read & Saved read pages & Start speed-up \\ \hline
	GIMP & \num{1106} & & \\ \hline
	GIMP (reordered) & \num{850} & -23.15\% & 0\,\% \\ \hline
	GIMP w/o disk cache (r.) & \num{629} & -43.13\% & N/A \\ \hline
	Inkscape & \num{2500} & & \\ \hline
	Inkscape (reordered) & \num{1428} & -23.15\% & 10.74\% \\ \hline
	\end{tabular}
	\caption{Disk page faults during start-up.}
	\label{fix:DiskPagesStatistics}
\end{table}

\section{Conclusion}

	Still, there is no native support by a linker (both \texttt{GNU ld} and \texttt{gold}) to propagate the order information in the form of section name convention. We are in contact with \texttt{gold} developers in mailing list at \url{http://sourceware.org/ml/binutils/2012-12/msg00227.html}. We discussed new section names like \texttt{.text[bucket](function_name)}, the bucket of which is a number the linker will respect.
	
At present time, GCC mainline has a problem with PGO, because collected profile validation checksum values do not correspond to functions seen in the second phase.

The pass is planned to be merged to GCC 4.9 and we want to combine the pass with an IPA pass that groups functions according to hot edges. To introduce a cooperation of these passes, we have to find a threshold in the time profile after that functions are no more classified as start-up.

\chapter{Semantic Function Equality}
\label{chap:SemanticFunctionEquality}

\section{Introduction}

As programming in languages with higher level of abstraction (e.g. C++) is getting more popular, heavy use of class hierarchy and template utilization brings many specialized functions. It is quite common that arguments of these functions are pointer types with e.g. a different class type. In fact, these functions are binary equivalents and are candidates to be merged. Apart from that, languages like C++ generate large amount of constructors and destructors that lead to generation of equivalent classes. These is an existing function merging solution at the level of linker and we describe this \texttt{gold} feature in subsection \ref{subsec:ICF}.

We implemented a new GCC inter-procedural pass called \texttt{ipa-sem-equality} which tries to proof semantic equivalence of two functions. Generally, we compare all function attributes, arguments, their types, and we test if all GIMPLE statements correlate. If so, our new pass can either create an alias instead or an equivalent function body is removed and function call wrapper is generated (also called \texttt{thunk}). The new pass is a simple IPA pass that is primarily intended for new passes and further pass development will be discussed later on.

With growing program visibility, the pass is getting more efficient. Although the pass is designed as a simple IPA pass and we are unable to see function bodies during whole program analysis (WPA), the pass can be still be successful. One can either run the pass with a single LTO partition, for small to medium applications, or LTRANS partitions generated for programs like Firefox are still large enough to cover almost all equivalents.

As a small demonstration, a dump of top ten semantically equal functions can be in Firefox. We try to comment on inclusion of the following functions and the reason why they are equal from semantic point of view. The following listing presents ten top-most merged functions in Firefox:

\footnotesize
\begin{lstlisting}[basicstyle=\ttfamily]
   1397 AddRef
    744 Release
    175 s_HashKey
    164 s_MatchEntry
    132 UnmarkPurple
    131 s_InitEntry
    131 GetNodeValue
    130 SetNodeValue
    130 RemoveChild
    130 LookupPrefix
\end{lstlisting}
\normalsize

First two most often used symbols do support implement custom reference counting in Firefox. This function has following GIMPLE representation:

\begin{minipage}{\linewidth}
\footnotesize
\begin{lstlisting}[basicstyle=\ttfamily]

AddRef (struct ProfileChangeStatusImpl * const this)
{
  <bb 2>:
  _4 = MEM[(struct nsAutoRefCnt *)this_1(D) + 8B].mValue;
  _5 = _4 + 1;
  MEM[(struct nsAutoRefCnt *)this_1(D) + 8B].mValue = _5;
  return _5;
}
\end{lstlisting}
\normalsize
\end{minipage}

\vspace{5mm}

The function \texttt{AddRef} is passed \texttt{this} pointer for a class struct. Structure pointer is cast to a corresponding \texttt{nsAutoRefCnt} and counter property \texttt{mValue} is read. All classes inherited from the reference counting class have the structure on the same offset. As a result, class pointers passed as the argument are commutable. The rest of the function's body just increments the counter and returns current value.

Common denominator for the rest of the listed functions is HTML element class hierarchy. HTML\footnote{\url{https://en.wikipedia.org/wiki/HTML}} as a language introducing large variety of elements that are located in the HTML tree. Functions like \texttt{RemoveChild}, \texttt{SetNodeValue} or \texttt{GetNodeValue} manipulates the tree and create big groups of semantically equivalent functions.

\section{Pass Implementation}

The pass is implemented as \texttt{SIMPLE_IPA_PASS} and located in \texttt{ipa-sem-equality.c} source file. The gcc compiler's pass manager launches the pass as the last one for every LTRANS partition. That is, after all normal IPA passes execute their local transformations. The pass is divided to following stages:

\begin{enumerate}
	\item \textit{Preparation phase}. During the phase, various data structures are allocated and initialized. The gcc compiler examines every functions, one at a time, and fills basic information about the functions to the corresponding structure.
	
	\item \textit{Congruent groups creation}. We compute an essential hash value for all functions and the hash is used as a congruent class sorter.
	
	\item \textit{Congruent groups resolution phase}. According to function calls, congruent resolution algorithm splits all congruent classes where we proof the functions cannot be potentially equivalent.
	
	\item \textit{Semantic function comparison phase}. All congruent groups with more than one member are iterated and the pass tries to proof that the functions are really equivalent. We examine function's bodies and all necessary correlations the functions must fulfil.
	
	\item \textit{Function merging phase}. When we encounter a pair of functions which can be merged, the pass chooses the best way of replacement, or in special cases we are not allowed to do any kind of merging operation.
	
	\item \textit{Cleanup}. The pass releases all idle memory.
\end{enumerate}

\section{Core Data Structures}

To each function of the source file that the pass analyzes we assign a single instance of \texttt{struct sem_func_t}, the data structure created in \textit{Preparation stage}. The structure encompasses the following items: a pointer to call graph node \texttt{node}, a tree with the declaration (func_decl), tree nodes for all function's arguments and a result type (\texttt{arg_types} and \texttt{result_type}). Moreover, we compute elementary statistics about the number of basic blocks, edges and SSA names that are used by the symbol. For congruence group resolution phase, we also comprise a list of all symbols that the function calls. Finally, a list of \texttt{sem_bb_t} structures is shown.

\footnotesize
\begin{lstlisting}[label={lst:SemFuncStruct},basicstyle=\ttfamily,caption=The data structure associated with each call graph node (function).]
typedef struct sem_func
{
  /* Global unique function index.  */
  unsigned int index;
  /* Call graph structure reference.  */
  struct cgraph_node *node;
  /* Function declaration tree node.  */
  tree func_decl;
  /* Exception handling region tree.  */
  eh_region region_tree;
  /* Result type tree node.  */
  tree result_type;
  /* Array of argument tree types.  */
  tree *arg_types;
  /* Number of function arguments.  */
  unsigned int arg_count;
  /* Basic block count.  */
  unsigned int bb_count;
  /* Total amount of edges in the function.  */
  unsigned int edge_count;
  /* Array of sizes of all basic blocks.  */
  unsigned int *bb_sizes;
  /* Control flow graph checksum.  */
  hashval_t cfg_checksum;
  /* Total number of SSA names used in the function.  */
  unsigned ssa_names_size;
  /* Array of structures for all basic blocks.  */
  sem_bb_t **bb_sorted;
  /* Vector for all calls done by the function.  */
  vec<tree> called_functions;
  /* Computed semantic function hash value.  */
  hashval_t hash;
} sem_func_t;
\end{lstlisting}
\normalsize

The last mentioned structure \texttt{sem_bb_t}, presented in Listing \ref{lst:SemBBStruct}, aggregates statistics about the count of edges and non-debug statements. Lastly, a hash based on kind of statements resides in the structure as well.

\begin{minipage}{\linewidth}
\footnotesize
\begin{lstlisting}[label={lst:SemBBStruct},basicstyle=\ttfamily,caption=The data structure associated with each basic block.]
/* Basic block struct for sematic equality pass.  */
typedef struct sem_bb
{
  /* Basic block the structure belongs to.  */
  basic_block bb;
  /* Reference to the semantic function this BB belongs to.  */
  sem_func_t *func;
  /* Number of non-debug statements in the basic block.  */
  unsigned nondbg_stmt_count;
  /* Number of edges connected to the block.  */
  unsigned edge_count;
  /* Computed hash value for basic block.  */
  hashval_t hash;
} sem_bb_t;
\end{lstlisting}
\normalsize
\end{minipage}

We would like to enumerate one important data structured named \texttt{func_dict_t}, shown in Listing \ref{lst:FuncDictStruct}. The structure is created for every single comparison of a pair of functions. First two members represent bidirectional mapping for SSA names. If  \textit{i}th SSA name in the source function corresponds to \textit{j}th in the second one, \texttt{source[i]} hash assigns \texttt{j} and similarly the value \texttt{target[j]} is equal to \textit{i}. Remaining hash tables are used in the same manner for all kind of declarations, and edges respectively.

\footnotesize
\begin{lstlisting}[label={lst:FuncDictStruct},basicstyle=\ttfamily,caption=The data structure for a mapping of correspondent items used during comparison of a pair of functions.]
/* Struct used for all kind of function dictionaries like
   SSA names, call graph edges and all kind of declarations.  */
typedef struct func_dict
{
  /* Source mapping of SSA names.  */
  vec<int> source;
  /* Target mapping of SSA names.  */
  vec<int> target;
  /* Hash table for correspondence declarations.  */
  hash_table <decl_var_hash> decl_hash;
  /* Hash table for correspondence of edges.  */
  hash_table <edge_var_hash> edge_hash;
} func_dict_t;
\end{lstlisting}
\normalsize

\section{Algorithm}

\subsection{Preparation Stage}

In preparation stage, we visit all functions and fill all data structure members of \texttt{sem_func_t}. We are primarily motivated to collect as many hashable items as possible. Control flow checksum of a function is given by \texttt{coverage_compute_cfg_checksum}, the function used for validation of profile information read from a \texttt{gcov} file. To improve the granularity of function groups, the pass iterates all basic blocks and each non-debug statement which resides in the block. Apart from that, the algorithm stores all function calls to their callees.
Every parsed function is pushed to \texttt{semantic_functions} vector.

\subsection{Congruent Groups Creation}

\begin{figure}[htbp]
	\begin{center}
		\includegraphics[width=130mm]{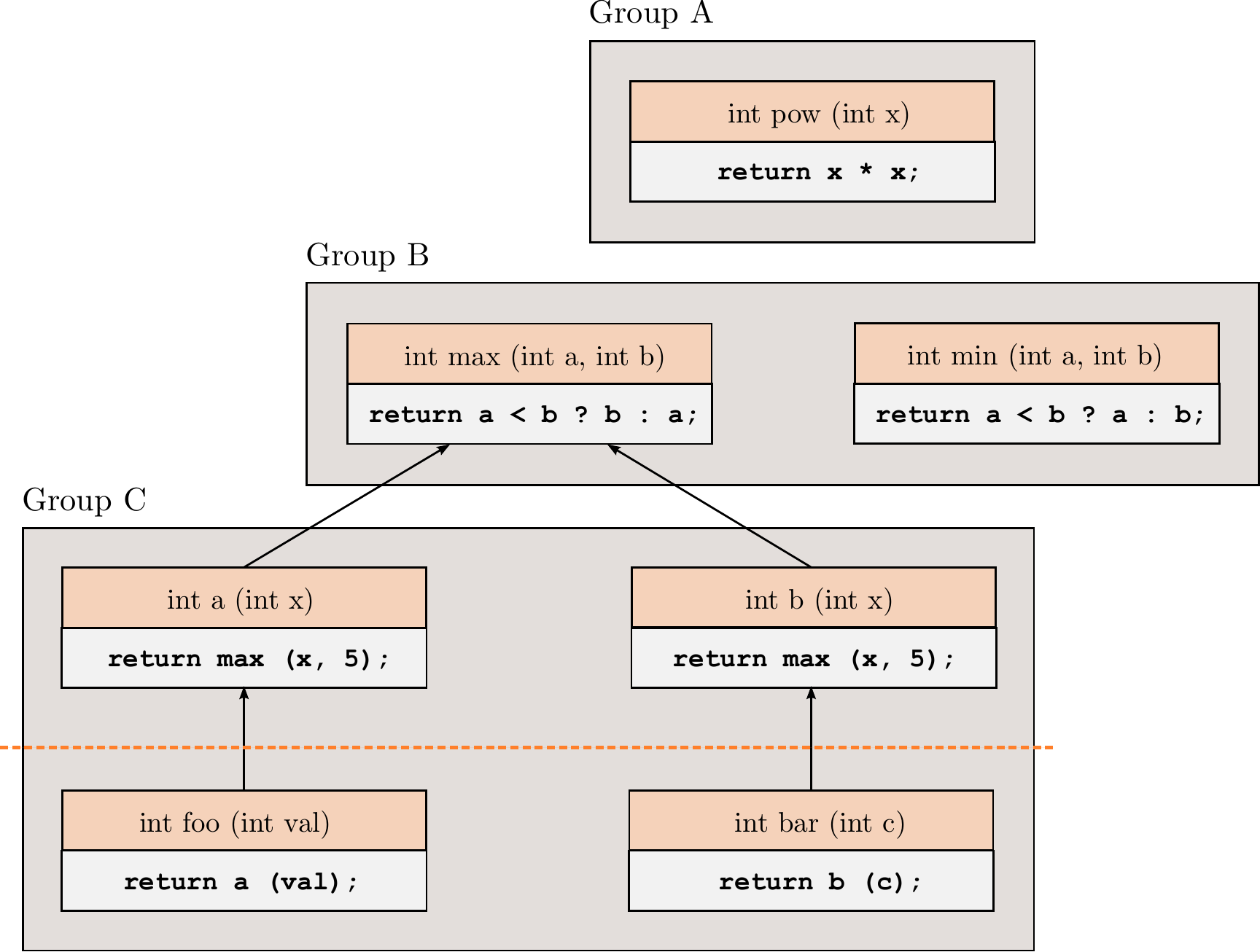}
		\vspace{-10pt}
	\end{center}
	\caption{Congruent classes at the beginning of resolution process.}
	\label{fig:CongruentFunctionClasses}
\end{figure}

We define a \textit{congruent group} as a set of functions that are candidates for function equality. Figure \ref{fig:CongruentFunctionClasses} presents the example of functions in a program and their separation to these groups. 

To create congruent groups, we build a compound hash value for each previously parsed function. The hash consists of following components:

\begin{itemize}
	\item The number of arguments, basic blocks and edges.
	\item The control flow graph checksum.
	\item The number of non-debug statements in each basic block.
	\item The compound has for every basic block which aggregates types of GIMPLE statements.
\end{itemize}

After that, the pass sorts each function to a congruent class according to its hash value and the vector of classes is allocated. Finally, function calls in \texttt{called_functions} property are mapped to newly created classes of symbols.

\subsection{Congruent Groups Resolution}

We chose an algorithm used for Value Numbering in~\cite{Alpern1992Value} with having O($N log N$) worst case running time, where N is the number of functions in a program. Originally, the algorithm was based on finite-state machine minimization.

The algorithm starts from the assumption that all functions with equal hash reside in the same congruent class. During the algorithm we compare functions in these groups and new groups can be created. We never merge these groups and the algorithm finishes after no additional split is needed.

All functions in the example return an integer as a result and accept either one or two arguments. \textit{Group A} contains just a function \texttt{pow} and its body contains a simple assignment statement. Even thought functions in the \textit{Group C} have the same number of arguments, call statements in their bodies do distinction. And finally, \textit{Group B} is made of symbols with exactly two arguments. The resolution, as described in previous sentences, is done according to value of the hash which encompasses the aforementioned differences.

With $P\backslash_{i}Q$, we denote all functions from group \textit{P} that call a function from group \textit{Q} as \textit{i}-th function call. Symmetrically, $P/_{i}Q$ contains functions which do not call as \textit{i}-th call any function from group \textit{Q}. \textit{Q} \textit{properly splits} \textit{P} for some \textit{i} if neither $P\backslash_{i}Q$ nor $P/_{i}Q$ is empty. A naive implementation of the algorithm can lead to quadratic complexity, but as described in the paper, there is an algorithm that can be O($N log N$). There are two important observations leading to much faster approach:

\begin{itemize}
	\item By always picking the class with fewer functions, only O($N log N$) congruent classes need to be used to split the others.
	\item By iterating a class \textit{Q} we can simultaneously build new classes $P\backslash_{i}Q$ for every class \textit{P}.
\end{itemize}

In the presented example, all functions have at maximum one call and function \texttt{foo} calls symbol \texttt{a} from the same \textit{Group C}. On the contrary, \texttt{a} does a call from \textit{Group B}. Therefore, the groups muse be split and we mark new classes with dashed orange line. These newly created classes are pushed to a work list and the algorithm converges when the work list is empty.

\subsection{Semantic Function Comparison Phase}

After the previous step is finished, all candidates in a group must be proved to be really semantically equivalent. Assuming a pair of functions from a congruent group, the following list of comparisons is examined to proof the functions are equal:

\begin{itemize}
	\item The number of arguments, non-debug basic blocks, edges and control flow checksum must be equal.
	\item Argument and result type is compared with predicate \texttt{types_compatible_p} which returns true if a conversion to one of types is not necessary.
	\item Function attributes are compared, e.g. \texttt{nothrow}, \texttt{noreturn} or \texttt{constructor}.
	\item Exception handling regions must correspond, function \texttt{compare_eh_regions} traverses the EH tree and checks all key data structures which must be identical.
	\item All basic blocks are collated in \texttt{compare_bb}. This comparison function covers majority of the pass and is responsible for particular equality of every GIMPLE statement in the basic block. We handle these GIMPLE statements for equality:	
	\begin{itemize}
		\item \texttt{GIMPLE_CALL}. If the calls have a different number of arguments, false is returned. The pass does not support the call based on two different variables (e.g. \texttt{func_ptr(1)}). If the call refers to a function that is not visible to the pass, we require symbol pointer identity. In addition, the arguments and result type are validated with predicate \mbox{\texttt{check_operand}}.
		\item \texttt{GIMPLE_ASSIGN}. For an assignment statement, we check an expression computed by the statement and the code of the expression and iteration of all operands is validated with \mbox{\texttt{check_operand}}.
		\item \texttt{GIMPLE_COND}. Condition assignment similarly holds expression code and left-hand and right-hand side operands are passed to the same predicate as in the previous statements.
		\item \texttt{GIMPLE_SWITCH}. Switch statements must have equal number of labels in both functions and be derived from a SSA name. Label constant ranges are accessed with function \texttt{CASE_LOW}, respectively \texttt{CASE_HIGH}.
		\item \texttt{GIMPLE_DEBUG} and \texttt{GIMPLE_EH_DISPATCH}. Both these GIMPLE codes are just skipped and do not affect equality in a negative way.
		\item \texttt{GIMPLE_RESX}. This statement is a placeholder for \textit{_Unwind_Resume} and we compare regions in both functions with \texttt{gimple_resx_region}.
		\item \texttt{GIMPLE_LABEL}. For labels, we are interested in label declaration check performed by \texttt{check_operand}.
		\item \texttt{GIMPLE_RETURN}. Return statement can either return \texttt{void} type or \texttt{check_operand} returns true if operands are equal.
		\item \texttt{GIMPLE_GOTO}. We require both target destinations to be driven by SSA names and the validation goes via \texttt{check_operand} function.
		\item \texttt{GIMPLE_ASM}. Assembler statements are not supported yet and prevent function folding.
	\end{itemize}

	\item Edges in basic blocks are validated for correct indices and flags. We build a hast table of appropriate edges, used by further analyses.
	
	\item Finally, PHI nodes verifier iterates all non-virtual nodes to test if the source and the destination edges correlate. If so, the pass validates PHI arguments with \texttt{check_operand}.
\end{itemize}

This stage of the algorithm intensively uses the following set of functions:

\begin{itemize}
	\item 	\texttt{check_operand} - the function tests for given \texttt{tree} arguments if they are equivalent. Declaration tree types like \texttt{VAR_DECL}, \texttt{PARM_DECL} and \texttt{LABEL_DECL} are handled by \texttt{check_declaration}. Constant types are supported by either \texttt{operand_equal_p} predicate or compared by value in case of an integer constant. Moreover, the function is capable of working with SSA names (\texttt{check_ssa_names} and handled components. Handled components are more complex types as e.g. \texttt{ARRAY_REF}, \texttt{COMPONENT_REF} and require more detail scan, done in \texttt{compare_handled_component}.

	\item \texttt{check_declaration} - for a given pair of \texttt{tree} types, we do a hash table lookup, if the source declaration is presented in the table. If this is the case, types are compatible if the target declaration is presented in the lookup result. Otherwise, we add this new pair to the hash table.

	\item \texttt{check_ssa_names} - similarly to previous function, with use of \text{check_dict_ssa_lookup}, we check if the passed SSA names \texttt{t1} and \texttt{t2} do correspond in both functions. If so, and if the SSA names are default definitions, we must validate the declaration as well.
	
	\item \texttt{compare_handled_component} - for a complex type, we unwind tree operands the function is called recursively, until we reach elementary TREE code.

\end{itemize}

\subsection{Function Merging Phase}

If we encounter a pair of functions as a candidate for merge, we can create either an \textit{alias} or \textit{thunk} (function wrapper). We prefer to do an alias, which is cheaper and merging is applied according to following rules:

\begin{enumerate}
	\item If the address of at least one function is not taken, alias can be used.
	\item But if the function is part of COMDAT section that can be replaced, we must use thunk.
	\item If we create a thunk and none of functions is writeable, we can redirect calls instead.
\end{enumerate}

\section{Related Work}

As described in the aforementioned chapter, \texttt{gold} linker comes with interesting feature called \textit{Identical Code Folding}. It is actually very similar to our pass, but unlike the pass, merging is applied at the level of assembler and linker. 

Microsoft compiler offers a compiler option (/Gy) that places each function to a separate COMDAT section. Microsoft linker does a similar algorithm named \textit{Identical COMDAT Folding}. Users are not allowed to combine this feature with debugging symbols and profiling. Because the toolchain is closed, we have not done any deeper analysis.

\subsection{Identical Code Folding}
\label{subsec:ICF}

Identical Code Folding in the \texttt{gold} linker is quite a new feature that folds bit-identical function, each residing in a separate ELF section. More precisely, a pair of functions is equal if and only if their \texttt{.text} sections are bit-identical and the relocations point to sections (functions) which are identical as such. We must provide object files compiled with \texttt{-ffunction-sections} options, because the linker's basic working unit is a section, not a function.

In order to detect equal functions, the linker splits content of all functions to \textit{variable} and \textit{constant}. Constant in this context means that content will not be changed through the analysis. On the other hand, variable parts refer to sections that might be possibly folded. After the stage, the constant part sorts functions to groups and a function relocation in each section is replaced by an identifier of the group. Then, the checksum based on a constant and a variable part is computed and groups are sorted more sensitively. The rest of algorithm is performed in the following steps:

\begin{enumerate}
	\item We replace all variable content of the folding candidate function with a corresponding group identifier.	
	\item Function checksum is recomputed.	
	\item New groups identifier is determined by a lookup to a hash table with group to identifier mapping.
	\item Repeat previous steps 1 to 3 until the convergence of function groups is reached.
\end{enumerate}

ICF can operate in following two modes: \texttt{safe} and \textit{all} (linker option:\\\texttt{--icf=all/safe}). Folding can be unsafe if an application compares a pair of pointers referring to a function. Thus, the algorithm merges two functions with a different address and the comparison assumptions are no longer valid. The safe mode operates pessimistically and turns off all functions with a taken address as not foldable. In many cases, none of these pointers are used for comparison.

The method introduces some kind of unwinding across merged functions and a new table to DWARF debugging information is added. This kind of information is capable to disambiguate PC counter by examining the call chain. Measured statistics about the ICF will be presented in the following section together with our inter-procedural pass.

\section{Results}

As seen in Table \ref{fix:Spec2006BinarySizeReductionSE}, the number of semantically equal function for SPEC benchmark is very small. The pass gives very similar results to ICF, which was enabled with more aggressive level of code folding (\texttt{--icf=all}).

% 1) performance
\begin{table}[htbp]
	\centering
	\begin{tabular}{|l|r|r|r|}
	\hline
\textbf{} & \textbf{Speed-up} & \textbf{Speed-up (FP)} & \textbf{Speed-up (INT)} \\ \hline
\textbf{gcc49} & 0.00\% & 0.00\% & 0.00\% \\ \hline
\textbf{gcc49-ICF} & -0.05\% & 0.11\% & -0.26\% \\ \hline
\textbf{gcc49-SE} & -0.04\% & 0.30\% & -0.46\% \\ \hline
	\end{tabular}
	\caption{SPEC CPU2006 speed-up summary for semantically equal functions.}
	\label{fix:Spec2006SpeedupSummarySE}
\end{table}

% 2) size
\begin{table}[htbp]
	\centering
	\begin{tabular}{|l|r|r|r|}
	\hline
\textbf{} & \textbf{Size } & \textbf{Size (FP)} & \textbf{Size (INT)} \\ \hline
\textbf{gcc49} & 0.00\% & 0.00\% & 0.00\% \\ \hline
\textbf{gcc49-ICF} & -0.53\% & -0.00\% & -1.19\% \\ \hline
\textbf{gcc49-SE} & -0.58\% & -0.30\% & -0.93\% \\ \hline
	\end{tabular}
	\caption{SPEC CPU2006 binary reduction for semantically equal functions.}
	\label{fix:Spec2006BinarySizeReductionSE}
\end{table}

More interesting data are presented in the following table. All these tests were run with LTO and we compare our pass with existing code folding, implemented in \texttt{gold} linker. The compiler pass saves about 2.6\% in Firefox that is even more than ICF in safe mode. On the other hand, as we do not cover semantic equality of variables, on folding possibilities are motived. If we combine both techniques, the result is more improved. This fact shows that the implementation in the GCC compiler merges functions that are not proofed to be equivalent in the linker.

\begin{table}[htbp]
	\centering
	\begin{tabular}{|l|r|r|r|r|r|}
	\hline
	& \textbf{LTO build} & \textbf{IPA sem eq.} & \textbf{ICF safe} & \textbf{ICF all} & \textbf{SE \& ICF} \\ \hline
	GIMP & \num{3129999}\,B & \num{3126800}\,B & \num{3127600}\,B & \num{3116411}\,B & \num{3117347}\,B \\ \hline
	GIMP (p.) & 100\% & 99.89\% & 99.92\% & 99.57\% & 99.59\% \\ \hline
	Inkscape & \num{6960970}\,B & \num{6877078}\,B & \num{6889249}\,B & \num{6870837}\,B & \num{6803162}\,B \\ \hline
	Inkscape (p.) & 100\% & 98.79\% & 98.97\% & 98.71\% & 97.73\% \\ \hline
	Firefox & \num{18735539}\,B & \num{18243965}\,B & \num{18125214}\,B & \num{17721274}\,B & N/A \\ \hline
	Firefox (p.) & 100\% & 97.38\% & 96.74\% & 94.59\% & N/A \\ \hline
	\end{tabular}
	\caption{Binary size reduction affected by function folding techniques}
	\label{fix:FunctionFoldingStatistics}
\end{table}

\section{Conclusion}

Both solutions, inter-procedural pass and ICF are not a subset of the second technique. The pass achieves interesting results on larger applications and there is still place for improvements. Starting to support variable merging and better understanding of specific structures like virtual tables, the pass can reduce the number of functions even more. We plan to integrate the pass to WHOPR mode of link-time optimization.

\chapter{Conclusion}
\label{chap:Conclusion}

\section{Summary}

After a short introduction in Chapter \ref{chapt:Introduction}, in Chapter \ref{chapt:ProgrammersGuide} we described a number of aspects concerning problematic of large applications. We have provided a guide for programmers that are not familiar with link-time optimization and want to speed-up a large application. We have showed differences between build systems and we have advised programmer, how to integrate these techniques to an existing project. The second half of the chapter has analysed the representation of executables and shared libraries and we have also briefly discussed bottlenecks of the mostly used format.

Then we have moved to look on complexity of large applications from perspective of compilation and growth of the source code. We have shown real world usage of link-time optimization and we have presented the long process of bug issues we have dealt with. Moreover, we have done deeper analysis for optimization techniques used in dynamic linker and we have introduced why is start-up of large application slow. At the end of the chapter we have performed SPEC benchmarks of aforementioned build systems with a various levels of optimization.

In Chapter \ref{chap:ProfileGuidedReorderingPass}, we have presented the first inter-procedural optimization done as a part of the thesis. We were inspired by similar solutions done e.g. in linker and we have implemented an approach that utilizes profiling infrastructure in the GCC compiler. In addition, the optimization is combined with link-time optimization as well.

Chapter \ref{chap:SemanticFunctionEquality} is the most significant one. It describes the implementation of \textit{semantic function equality} we have implemented as a part of the thesis. In order to demonstrate usefulness, we have done comparison to the existing solution in \texttt{gold} linker. Apart from that, we have defined all core data structures and all phases of the algorithm in detail.

\section{Future Work}

There are still several tasks that could be addressed in the future. First of all, the semantic function equality pass should be enhanced for a support to merge variables. This can bring further improvements for large applications, as well as integration of the pass to link-time optimization framework.

As was mentioned above, the function reordering pass has been implemented as the independent optimization and the function placement should be done more precisely. That means, the function reordering should be done more sensitively in different contexts.

\appendix

\chapter{SPEC CPU2006 Results}
\label{app:SpecCPU2006Results}

% 3) int_performance
\begin{table}[htbp]
	\centering
	\begin{tabular}{|l|r|r|r|r|r|r|}
	\hline
\textbf{} & \textbf{48-O2} & \textbf{48-O3} & \textbf{O2} & \textbf{O3} & \textbf{O2-LTO} & \textbf{O3-LTO} \\ \hline
\textbf{400.perlbench} & 0.00\% & 1.06\% & 0.38\% & -3.39\% & \cellcolor{SpecBad}-8.31\% & -0.40\% \\ \hline
\textbf{401.bzip2} & 0.00\% & -0.26\% & \cellcolor{SpecBad}-6.49\% & -2.50\% & -2.17\% & -4.88\% \\ \hline
\textbf{403.gcc} & 0.00\% & -2.54\% & 0.34\% & -0.32\% & -0.86\% & 0.81\% \\ \hline
\textbf{429.mcf} & 0.00\% & -4.06\% & 1.35\% & -1.46\% & 0.15\% & 0.77\% \\ \hline
\textbf{445.gobmk} & 0.00\% & -0.58\% & 0.70\% & 0.45\% & -1.12\% & 0.93\% \\ \hline
\textbf{456.hmmer} & 0.00\% & \cellcolor{SpecGood}5.75\% & 0.07\% & \cellcolor{SpecGood}6.94\% & 0.11\% & \cellcolor{SpecGood}7.41\% \\ \hline
\textbf{458.sjeng} & 0.00\% & 1.81\% & 0.09\% & 4.64\% & -0.93\% & 4.11\% \\ \hline
\textbf{462.libquantum} & 0.00\% & 0.05\% & 0.16\% & 0.76\% & -0.18\% & \cellcolor{SpecGood}6.90\% \\ \hline
\textbf{464.h264ref} & 0.00\% & 1.75\% & -2.78\% & 2.41\% & 1.96\% & 2.89\% \\ \hline
\textbf{471.omnetpp} & 0.00\% & \cellcolor{SpecGood}5.36\% & 2.95\% & 3.33\% & \cellcolor{SpecGood}9.27\% & \cellcolor{SpecBetter}18.39\% \\ \hline
\textbf{473.astar} & 0.00\% & 4.92\% & -1.74\% & 4.57\% & \cellcolor{SpecGood}6.18\% & \cellcolor{SpecGood}6.58\% \\ \hline
\textbf{483.xalancbmk} & 0.00\% & -1.73\% & -0.09\% & 1.46\% & \cellcolor{SpecBetter}10.70\% & \cellcolor{SpecBetter}23.94\% \\ \hline
\hline
\textbf{average} & \textbf{0.00\%} & \textbf{0.96\%} & \textbf{-0.42\%} & \textbf{1.41\%} & \textbf{1.23\%} & \textbf{\cellcolor{SpecGood}5.62\%} \\ \hline
	\end{tabular}
	\caption{SPEC CPU2006 INT speed-up, part I.}
	\label{fix:Spec2006IntSpeedupPart1}
\end{table}

\begin{table}[htbp]
	\centering
	\begin{tabular}{|l|r|r|r|r|r|}
	\hline
\textbf{} & \textbf{O3-LTO-UG5} & \textbf{O3-P} & \textbf{O3-LTO-P} & \textbf{O2-L} & \textbf{O3-L} \\ \hline
\textbf{400.perlbench} & -0.97\% & \cellcolor{SpecBetter}11.56\% & \cellcolor{SpecBetter}10.09\% & N/A & N/A \\ \hline
\textbf{401.bzip2} & -4.63\% & 4.86\% & \cellcolor{SpecGood}5.20\% & N/A & N/A \\ \hline
\textbf{403.gcc} & 0.72\% & \cellcolor{SpecBetter}10.14\% & \cellcolor{SpecBetter}12.10\% & N/A & N/A \\ \hline
\textbf{429.mcf} & 1.44\% & 1.06\% & 0.60\% & -1.95\% & -0.93\% \\ \hline
\textbf{445.gobmk} & -0.80\% & \cellcolor{SpecGood}6.58\% & \cellcolor{SpecGood}7.07\% & 2.59\% & 2.77\% \\ \hline
\textbf{456.hmmer} & \cellcolor{SpecGood}7.20\% & \cellcolor{SpecGood}5.55\% & 3.69\% & 2.33\% & 2.57\% \\ \hline
\textbf{458.sjeng} & 2.16\% & \cellcolor{SpecGood}8.46\% & \cellcolor{SpecGood}9.09\% & \cellcolor{SpecGood}9.01\% & \cellcolor{SpecGood}9.63\% \\ \hline
\textbf{462.libquantum} & \cellcolor{SpecGood}7.90\% & \cellcolor{SpecGood}9.54\% & \cellcolor{SpecBetter}10.03\% & \cellcolor{SpecGood}6.20\% & \cellcolor{SpecGood}8.31\% \\ \hline
\textbf{464.h264ref} & 2.43\% & 3.91\% & \cellcolor{SpecGood}5.43\% & \cellcolor{SpecGood}9.24\% & \cellcolor{SpecBetter}10.94\% \\ \hline
\textbf{471.omnetpp} & \cellcolor{SpecGood}8.51\% & \cellcolor{SpecGood}7.83\% & \cellcolor{SpecBetter}15.73\% & N/A & N/A \\ \hline
\textbf{473.astar} & \cellcolor{SpecGood}6.69\% & \cellcolor{SpecBetter}16.03\% & \cellcolor{SpecBetter}14.89\% & \cellcolor{SpecBetter}10.60\% & \cellcolor{SpecBetter}10.95\% \\ \hline
\textbf{483.xalancbmk} & \cellcolor{SpecBetter}14.64\% & \cellcolor{SpecBetter}12.97\% & \cellcolor{SpecBetter}51.99\% & N/A & N/A \\ \hline
\hline
\textbf{average} & \textbf{3.77\%} & \textbf{\cellcolor{SpecGood}8.21\%} & \textbf{\cellcolor{SpecBetter}12.16\%} & \textbf{\cellcolor{SpecGood}5.43\%} & \textbf{\cellcolor{SpecGood}6.32\%} \\ \hline
	\end{tabular}
	\caption{SPEC CPU2006 INT speed-up, part II.}
	\label{fix:Spec2006IntSpeedupPart2}
\end{table}

% 4) fp_performance
\begin{table}[htbp]
	\centering
	\begin{tabular}{|l|r|r|r|r|r|r|}
	\hline
\textbf{} & \textbf{48-O2} & \textbf{48-O3} & \textbf{O2} & \textbf{O3} & \textbf{O2-LTO} & \textbf{O3-LTO} \\ \hline
\textbf{410.bwaves} & 0.00\% & -2.00\% & 0.96\% & -0.37\% & \cellcolor{SpecGood}5.06\% & \cellcolor{SpecGood}7.15\% \\ \hline
\textbf{433.milc} & 0.00\% & -0.99\% & -1.59\% & -0.87\% & 2.18\% & \cellcolor{SpecGood}5.61\% \\ \hline
\textbf{434.zeusmp} & 0.00\% & \cellcolor{SpecBetter}11.55\% & 0.02\% & \cellcolor{SpecBetter}11.75\% & 0.63\% & \cellcolor{SpecBetter}14.62\% \\ \hline
\textbf{435.gromacs} & 0.00\% & 3.68\% & 4.18\% & \cellcolor{SpecGood}7.87\% & 4.69\% & \cellcolor{SpecGood}8.94\% \\ \hline
\textbf{436.cactusADM} & 0.00\% & \cellcolor{SpecBetter}30.47\% & \cellcolor{SpecBad}-5.42\% & \cellcolor{SpecBetter}26.60\% & -1.81\% & \cellcolor{SpecBetter}31.08\% \\ \hline
\textbf{437.leslie3d} & 0.00\% & \cellcolor{SpecBetter}22.21\% & 0.08\% & \cellcolor{SpecBetter}22.97\% & -0.02\% & \cellcolor{SpecBetter}24.72\% \\ \hline
\textbf{444.namd} & 0.00\% & 0.03\% & -0.03\% & -0.11\% & 0.42\% & -0.09\% \\ \hline
\textbf{447.dealII} & 0.00\% & 1.61\% & -1.09\% & 2.24\% & \cellcolor{SpecBetter}10.82\% & \cellcolor{SpecBetter}12.23\% \\ \hline
\textbf{450.soplex} & 0.00\% & 1.10\% & -2.20\% & 2.11\% & -0.09\% & -1.43\% \\ \hline
\textbf{453.povray} & 0.00\% & 1.26\% & -0.32\% & 0.84\% & 0.78\% & \cellcolor{SpecGood}5.22\% \\ \hline
\textbf{454.calculix} & 0.00\% & \cellcolor{SpecGood}6.88\% & -0.94\% & \cellcolor{SpecGood}6.74\% & 0.49\% & \cellcolor{SpecGood}7.95\% \\ \hline
\textbf{459.GemsFDTD} & 0.00\% & 1.74\% & -1.06\% & -0.74\% & 1.19\% & -0.43\% \\ \hline
\textbf{465.tonto} & 0.00\% & 0.12\% & -1.17\% & 0.29\% & -0.65\% & 1.13\% \\ \hline
\textbf{470.lbm} & 0.00\% & -0.16\% & 0.40\% & 0.28\% & -0.19\% & 0.30\% \\ \hline
\textbf{481.wrf} & 0.00\% & \cellcolor{SpecBetter}21.46\% & 0.45\% & \cellcolor{SpecBetter}20.88\% & 0.44\% & \cellcolor{SpecBetter}20.99\% \\ \hline
\textbf{482.sphinx3} & 0.00\% & \cellcolor{SpecBetter}17.42\% & -2.30\% & \cellcolor{SpecBetter}16.04\% & 0.71\% & \cellcolor{SpecBetter}19.21\% \\ \hline
\hline
\textbf{average} & \textbf{0.00\%} & \textbf{\cellcolor{SpecGood}6.77\%} & \textbf{-0.54\%} & \textbf{\cellcolor{SpecGood}6.85\%} & \textbf{1.54\%} & \textbf{\cellcolor{SpecGood}9.82\%} \\ \hline
	\end{tabular}
	\caption{SPEC CPU2006 FP speed-up, part I.}
	\label{fix:Spec2006FpSpeedupPart1}
\end{table}

\begin{table}[htbp]
	\centering
	\begin{tabular}{|l|r|r|r|r|r|}
	\hline
\textbf{} & \textbf{O3-LTO-UG5} & \textbf{O3-P} & \textbf{O3-LTO-P} & \textbf{O2-L} & \textbf{O3-L} \\ \hline
\textbf{410.bwaves} & \cellcolor{SpecGood}6.95\% & 4.39\% & \cellcolor{SpecGood}8.30\% & N/A & N/A \\ \hline
\textbf{433.milc} & 4.20\% & -2.49\% & 4.22\% & -0.00\% & \cellcolor{SpecGood}6.34\% \\ \hline
\textbf{434.zeusmp} & \cellcolor{SpecBetter}14.26\% & \cellcolor{SpecBetter}14.09\% & \cellcolor{SpecBetter}16.83\% & N/A & N/A \\ \hline
\textbf{435.gromacs} & \cellcolor{SpecGood}8.95\% & \cellcolor{SpecGood}5.97\% & \cellcolor{SpecGood}5.82\% & N/A & N/A \\ \hline
\textbf{436.cactusADM} & \cellcolor{SpecBetter}28.80\% & \cellcolor{SpecBetter}30.52\% & \cellcolor{SpecBetter}28.60\% & N/A & N/A \\ \hline
\textbf{437.leslie3d} & \cellcolor{SpecBetter}23.68\% & \cellcolor{SpecBetter}26.63\% & \cellcolor{SpecBetter}25.73\% & N/A & N/A \\ \hline
\textbf{444.namd} & 0.28\% & 0.40\% & 0.32\% & N/A & N/A \\ \hline
\textbf{447.dealII} & \cellcolor{SpecBetter}12.13\% & \cellcolor{SpecGood}7.33\% & \cellcolor{SpecBetter}19.75\% & N/A & N/A \\ \hline
\textbf{450.soplex} & 0.09\% & 2.70\% & 4.39\% & N/A & N/A \\ \hline
\textbf{453.povray} & 1.79\% & \cellcolor{SpecGood}6.97\% & \cellcolor{SpecBetter}29.16\% & N/A & N/A \\ \hline
\textbf{454.calculix} & \cellcolor{SpecGood}7.82\% & \cellcolor{SpecGood}8.62\% & \cellcolor{SpecGood}9.06\% & N/A & N/A \\ \hline
\textbf{459.GemsFDTD} & -0.16\% & -0.13\% & -0.44\% & N/A & N/A \\ \hline
\textbf{465.tonto} & 1.11\% & 4.92\% & \cellcolor{SpecGood}5.22\% & N/A & N/A \\ \hline
\textbf{470.lbm} & 0.34\% & -0.04\% & 0.23\% & -0.88\% & -1.18\% \\ \hline
\textbf{481.wrf} & \cellcolor{SpecBetter}20.97\% & \cellcolor{SpecBetter}23.58\% & \cellcolor{SpecBetter}24.15\% & N/A & N/A \\ \hline
\textbf{482.sphinx3} & \cellcolor{SpecBetter}17.48\% & \cellcolor{SpecBetter}12.21\% & \cellcolor{SpecBetter}17.21\% & \cellcolor{SpecGood}6.54\% & \cellcolor{SpecGood}7.30\% \\ \hline
\hline
\textbf{average} & \textbf{\cellcolor{SpecGood}9.29\%} & \textbf{\cellcolor{SpecGood}8.35\%} & \textbf{\cellcolor{SpecBetter}12.41\%} & \textbf{1.89\%} & \textbf{4.15\%} \\ \hline
	\end{tabular}
	\caption{SPEC CPU2006 FP speed-up, part II.}
	\label{fix:Spec2006FpSpeedupPart2}
\end{table}

% 5) int_size
\begin{table}[htbp]
	\centering
	\begin{tabular}{|l|r|r|r|r|r|r|}
	\hline
\textbf{} & \textbf{48-O2} & \textbf{48-O3} & \textbf{O2} & \textbf{O3} & \textbf{O2-LTO} & \textbf{O3-LTO} \\ \hline
\textbf{400.perlbench} & 0.00\% & 21.13\% & -0.09\% & 22.09\% & -3.33\% & 23.75\% \\ \hline
\textbf{401.bzip2} & 0.00\% & \cellcolor{SpecBad}26.72\% & 0.00\% & \cellcolor{SpecBad}29.55\% & -11.67\% & 15.96\% \\ \hline
\textbf{403.gcc} & 0.00\% & 19.33\% & -0.05\% & 19.55\% & -7.46\% & 20.62\% \\ \hline
\textbf{429.mcf} & 0.00\% & 9.91\% & 0.45\% & 9.73\% & -16.65\% & -8.23\% \\ \hline
\textbf{445.gobmk} & 0.00\% & 3.26\% & -0.01\% & 3.21\% & -0.78\% & 6.77\% \\ \hline
\textbf{456.hmmer} & 0.00\% & 15.08\% & 0.00\% & 15.94\% & \cellcolor{SpecBetter}-51.60\% & \cellcolor{SpecGood}-36.47\% \\ \hline
\textbf{458.sjeng} & 0.00\% & 18.48\% & 0.11\% & 18.80\% & -4.94\% & 20.53\% \\ \hline
\textbf{462.libquantum} & 0.00\% & 8.45\% & 0.00\% & 6.51\% & \cellcolor{SpecGood}-44.00\% & \cellcolor{SpecGood}-41.40\% \\ \hline
\textbf{464.h264ref} & 0.00\% & 13.72\% & 0.42\% & 14.17\% & -8.53\% & 8.11\% \\ \hline
\textbf{471.omnetpp} & 0.00\% & 10.58\% & -0.54\% & 11.56\% & -15.62\% & 2.76\% \\ \hline
\textbf{473.astar} & 0.00\% & 14.31\% & 0.12\% & 14.23\% & -20.69\% & -12.15\% \\ \hline
\textbf{483.xalancbmk} & 0.00\% & 9.46\% & -0.54\% & 8.91\% & -24.25\% & -20.05\% \\ \hline
\hline
\textbf{average} & \textbf{0.00\%} & \textbf{14.20\%} & \textbf{-0.01\%} & \textbf{14.52\%} & \textbf{-17.46\%} & \textbf{-1.65\%} \\ \hline
	\end{tabular}
	\caption{SPEC CPU2006 INT size reduction, part I.}
	\label{fix:Spec2006IntSizeReductionPart1}
\end{table}

\begin{table}[htbp]
	\centering
	\begin{tabular}{|l|r|r|r|r|r|}
	\hline
\textbf{} & \textbf{O3-LTO-UG5} & \textbf{O3-P} & \textbf{O3-LTO-P} & \textbf{O2-L} & \textbf{O3-L} \\ \hline
\textbf{400.perlbench} & 6.13\% & 9.45\% & 18.71\% & N/A & N/A \\ \hline
\textbf{401.bzip2} & 6.09\% & 16.36\% & 4.97\% & N/A & N/A \\ \hline
\textbf{403.gcc} & 0.77\% & 7.19\% & 17.14\% & N/A & N/A \\ \hline
\textbf{429.mcf} & -8.23\% & \cellcolor{SpecBad}28.13\% & 14.33\% & \cellcolor{SpecWorse}52.86\% & \cellcolor{SpecWorse}52.86\% \\ \hline
\textbf{445.gobmk} & 2.35\% & -0.00\% & 0.40\% & 4.77\% & 4.40\% \\ \hline
\textbf{456.hmmer} & \cellcolor{SpecGood}-44.28\% & -13.29\% & \cellcolor{SpecBetter}-55.44\% & -20.74\% & -20.74\% \\ \hline
\textbf{458.sjeng} & 8.27\% & -9.15\% & -14.12\% & 5.08\% & 5.08\% \\ \hline
\textbf{462.libquantum} & \cellcolor{SpecGood}-41.40\% & -5.21\% & \cellcolor{SpecGood}-41.41\% & -15.78\% & -15.78\% \\ \hline
\textbf{464.h264ref} & 4.71\% & -6.58\% & -14.94\% & -2.86\% & -2.16\% \\ \hline
\textbf{471.omnetpp} & -12.21\% & -5.21\% & -17.78\% & \cellcolor{SpecWorse}53.42\% & \cellcolor{SpecWorse}53.42\% \\ \hline
\textbf{473.astar} & -12.15\% & -0.13\% & -23.67\% & 23.66\% & 24.01\% \\ \hline
\textbf{483.xalancbmk} & \cellcolor{SpecGood}-26.03\% & -5.61\% & -24.91\% & \cellcolor{SpecWorse}73.15\% & \cellcolor{SpecWorse}73.22\% \\ \hline
\hline
\textbf{average} & \textbf{-9.67\%} & \textbf{1.33\%} & \textbf{-11.39\%} & \textbf{19.28\%} & \textbf{19.37\%} \\ \hline
	\end{tabular}
	\caption{SPEC CPU2006 INT size reduction, part II.}
	\label{fix:Spec2006IntSizeReductionPart2}
\end{table}

% 6) fp_size
\begin{table}[htbp]
	\centering
	\begin{tabular}{|l|r|r|r|r|r|r|}
	\hline
\textbf{} & \textbf{48-O2} & \textbf{48-O3} & \textbf{O2} & \textbf{O3} & \textbf{O2-LTO} & \textbf{O3-LTO} \\ \hline
\textbf{410.bwaves} & 0.00\% & \cellcolor{SpecBad}28.74\% & 0.00\% & \cellcolor{SpecBad}31.26\% & 0.14\% & 21.36\% \\ \hline
\textbf{416.gamess} & 0.00\% & 22.02\% & -0.16\% & 22.73\% & -6.43\% & 11.12\% \\ \hline
\textbf{433.milc} & 0.00\% & 15.70\% & -0.02\% & 16.09\% & \cellcolor{SpecGood}-31.82\% & -7.97\% \\ \hline
\textbf{434.zeusmp} & 0.00\% & \cellcolor{SpecWorse}57.49\% & -0.42\% & \cellcolor{SpecWorse}61.80\% & -5.54\% & \cellcolor{SpecBad}49.71\% \\ \hline
\textbf{435.gromacs} & 0.00\% & 12.53\% & 0.08\% & 12.65\% & -21.91\% & -10.34\% \\ \hline
\textbf{436.cactusADM} & 0.00\% & \cellcolor{SpecBad}25.20\% & 0.14\% & \cellcolor{SpecBad}25.58\% & \cellcolor{SpecGood}-36.38\% & -8.95\% \\ \hline
\textbf{437.leslie3d} & 0.00\% & \cellcolor{SpecWorse}57.83\% & -0.15\% & \cellcolor{SpecWorse}65.42\% & -1.59\% & \cellcolor{SpecBad}47.17\% \\ \hline
\textbf{444.namd} & 0.00\% & 3.25\% & 0.31\% & 3.57\% & 0.74\% & 4.06\% \\ \hline
\textbf{447.dealII} & 0.00\% & 2.74\% & -0.19\% & 2.86\% & \cellcolor{SpecBetter}-84.01\% & \cellcolor{SpecBetter}-83.41\% \\ \hline
\textbf{450.soplex} & 0.00\% & 12.52\% & -0.87\% & 11.71\% & -20.35\% & -14.47\% \\ \hline
\textbf{453.povray} & 0.00\% & 24.55\% & -0.01\% & \cellcolor{SpecBad}26.11\% & -6.27\% & 20.83\% \\ \hline
\textbf{454.calculix} & 0.00\% & 10.02\% & -0.20\% & 10.66\% & \cellcolor{SpecGood}-28.59\% & -15.53\% \\ \hline
\textbf{459.GemsFDTD} & 0.00\% & \cellcolor{SpecBad}42.78\% & -0.12\% & \cellcolor{SpecBad}45.84\% & -5.47\% & \cellcolor{SpecBad}49.86\% \\ \hline
\textbf{465.tonto} & 0.00\% & 12.04\% & 0.09\% & 12.50\% & \cellcolor{SpecGood}-29.46\% & -12.15\% \\ \hline
\textbf{470.lbm} & 0.00\% & 2.74\% & 0.00\% & 2.74\% & -6.62\% & -3.92\% \\ \hline
\textbf{481.wrf} & 0.00\% & \cellcolor{SpecBad}37.96\% & -0.01\% & \cellcolor{SpecBad}40.01\% & -24.60\% & 13.62\% \\ \hline
\textbf{482.sphinx3} & 0.00\% & 21.63\% & -0.02\% & 21.60\% & \cellcolor{SpecGood}-27.69\% & -2.06\% \\ \hline
\hline
\textbf{average} & \textbf{0.00\%} & \textbf{22.92\%} & \textbf{-0.09\%} & \textbf{24.30\%} & \textbf{-19.76\%} & \textbf{3.47\%} \\ \hline
	\end{tabular}
	\caption{SPEC CPU2006 FP size reduction, part I.}
	\label{fix:Spec2006FpSizeReductionPart1}
\end{table}

\begin{table}[htbp]
	\centering
	\begin{tabular}{|l|r|r|r|r|r|}
	\hline
\textbf{} & \textbf{O3-LTO-UG5} & \textbf{O3-P} & \textbf{O3-LTO-P} & \textbf{O2-L} & \textbf{O3-L} \\ \hline
\textbf{410.bwaves} & 21.36\% & \cellcolor{SpecBad}32.87\% & \cellcolor{SpecBad}34.72\% & N/A & N/A \\ \hline
\textbf{416.gamess} & 11.12\% & -8.31\% & -15.39\% & N/A & N/A \\ \hline
\textbf{433.milc} & -21.67\% & -5.75\% & -23.68\% & 18.13\% & 13.79\% \\ \hline
\textbf{434.zeusmp} & \cellcolor{SpecBad}49.71\% & \cellcolor{SpecWorse}83.22\% & \cellcolor{SpecWorse}65.77\% & N/A & N/A \\ \hline
\textbf{435.gromacs} & -14.34\% & -8.40\% & \cellcolor{SpecGood}-25.72\% & N/A & N/A \\ \hline
\textbf{436.cactusADM} & -15.90\% & -1.65\% & \cellcolor{SpecGood}-38.33\% & N/A & N/A \\ \hline
\textbf{437.leslie3d} & \cellcolor{SpecBad}47.17\% & \cellcolor{SpecWorse}72.23\% & \cellcolor{SpecWorse}72.73\% & N/A & N/A \\ \hline
\textbf{444.namd} & 4.06\% & -7.55\% & -7.34\% & \cellcolor{SpecBad}44.14\% & \cellcolor{SpecBad}44.14\% \\ \hline
\textbf{447.dealII} & \cellcolor{SpecBetter}-84.91\% & -17.46\% & \cellcolor{SpecBetter}-84.13\% & \cellcolor{SpecWorse}74.71\% & \cellcolor{SpecWorse}73.94\% \\ \hline
\textbf{450.soplex} & -24.79\% & -5.67\% & -19.87\% & \cellcolor{SpecWorse}111.13\% & \cellcolor{SpecWorse}111.13\% \\ \hline
\textbf{453.povray} & 2.35\% & -5.56\% & -9.43\% & \cellcolor{SpecWorse}125.08\% & \cellcolor{SpecWorse}125.08\% \\ \hline
\textbf{454.calculix} & -18.80\% & -6.51\% & \cellcolor{SpecGood}-28.82\% & N/A & N/A \\ \hline
\textbf{459.GemsFDTD} & \cellcolor{SpecBad}49.86\% & \cellcolor{SpecWorse}59.44\% & \cellcolor{SpecWorse}53.52\% & N/A & N/A \\ \hline
\textbf{465.tonto} & -22.10\% & -5.12\% & \cellcolor{SpecGood}-30.77\% & N/A & N/A \\ \hline
\textbf{470.lbm} & -3.92\% & 7.03\% & 0.66\% & 2.71\% & 2.71\% \\ \hline
\textbf{481.wrf} & 13.62\% & 7.09\% & -13.57\% & N/A & N/A \\ \hline
\textbf{482.sphinx3} & -13.77\% & 9.79\% & -9.96\% & 7.96\% & 8.70\% \\ \hline
\hline
\textbf{average} & \textbf{-1.23\%} & \textbf{11.75\%} & \textbf{-4.68\%} & \textbf{\cellcolor{SpecWorse}54.84\%} & \textbf{\cellcolor{SpecWorse}54.21\%} \\ \hline
	\end{tabular}
	\caption{SPEC CPU2006 FP size reduction, part II.}
	\label{fix:Spec2006IntSizeReductionPart2}
\end{table}

\chapter{SPEC CPU2006 Results for Inlining}

The presented collection of SPEC benchmarks results was run with enabled LTO optimization and different degree of optimization. We compare default -O2 and -O3 configuration and we added -O3 profiles with set \texttt{unit-growth} equal to 100\%, respectively 200\%.

% 3) int_performance
\begin{table}[htbp]
	\begin{tabular}{|l|r|r|r|r|}
	\hline
\textbf{} & \textbf{O2} & \textbf{O3} & \textbf{O3-UG100} & \textbf{O3-UG200} \\ \hline
\textbf{403.gcc} & 0.00\% & 2.60\% & 2.02\% & 2.32\% \\ \hline
\textbf{429.mcf} & 0.00\% & -0.92\% & 3.00\% & 0.83\% \\ \hline
\textbf{445.gobmk} & 0.00\% & 1.05\% & 3.87\% & 2.87\% \\ \hline
\textbf{456.hmmer} & 0.00\% & \cellcolor{SpecGood}5.28\% & \cellcolor{SpecGood}6.00\% & \cellcolor{SpecGood}5.55\% \\ \hline
\textbf{458.sjeng} & 0.00\% & \cellcolor{SpecGood}5.57\% & \cellcolor{SpecGood}5.53\% & 4.80\% \\ \hline
\textbf{462.libquantum} & 0.00\% & 1.17\% & 1.95\% & 0.93\% \\ \hline
\textbf{464.h264ref} & 0.00\% & 3.76\% & 1.80\% & -1.99\% \\ \hline
\textbf{471.omnetpp} & 0.00\% & \cellcolor{SpecGood}6.80\% & \cellcolor{SpecGood}5.26\% & \cellcolor{SpecGood}5.06\% \\ \hline
\textbf{473.astar} & 0.00\% & 0.38\% & 0.79\% & -0.35\% \\ \hline
\textbf{483.xalancbmk} & 0.00\% & \cellcolor{SpecBetter}12.35\% & \cellcolor{SpecBetter}20.89\% & \cellcolor{SpecBetter}20.09\% \\ \hline
\hline
\textbf{average} & \textbf{0.00\%} & \textbf{3.80\%} & \textbf{4.29\%} & \textbf{4.01\%} \\ \hline
	\end{tabular}
	\caption{SPEC CPU2006 INT speed-up for inlining, part I}
\end{table}

% 4) fp_performance
\begin{table}[htbp]
	\begin{tabular}{|l|r|r|r|r|}
	\hline
\textbf{} & \textbf{O2} & \textbf{O3} & \textbf{O3-UG100} & \textbf{O3-UG200} \\ \hline
\textbf{410.bwaves} & 0.00\% & 2.51\% & 1.21\% & 3.18\% \\ \hline
\textbf{416.gamess} & 0.00\% & -0.52\% & -1.95\% & -1.70\% \\ \hline
\textbf{433.milc} & 0.00\% & \cellcolor{SpecGood}5.74\% & 4.31\% & 4.27\% \\ \hline
\textbf{434.zeusmp} & 0.00\% & \cellcolor{SpecBetter}16.10\% & \cellcolor{SpecBetter}13.56\% & \cellcolor{SpecBetter}13.19\% \\ \hline
\textbf{435.gromacs} & 0.00\% & 1.84\% & 3.14\% & 3.80\% \\ \hline
\textbf{436.cactusADM} & 0.00\% & \cellcolor{SpecBetter}30.17\% & \cellcolor{SpecBetter}32.76\% & \cellcolor{SpecBetter}29.20\% \\ \hline
\textbf{437.leslie3d} & 0.00\% & \cellcolor{SpecBetter}24.28\% & \cellcolor{SpecBetter}23.73\% & \cellcolor{SpecBetter}21.79\% \\ \hline
\textbf{444.namd} & 0.00\% & 1.66\% & -0.59\% & 1.53\% \\ \hline
\textbf{447.dealII} & 0.00\% & 1.27\% & 2.49\% & 1.59\% \\ \hline
\textbf{450.soplex} & 0.00\% & 2.84\% & 3.17\% & 3.04\% \\ \hline
\textbf{453.povray} & 0.00\% & 4.67\% & 3.09\% & \cellcolor{SpecGood}5.58\% \\ \hline
\textbf{454.calculix} & 0.00\% & \cellcolor{SpecGood}8.02\% & \cellcolor{SpecGood}8.02\% & \cellcolor{SpecGood}6.20\% \\ \hline
\textbf{459.GemsFDTD} & 0.00\% & -1.02\% & -0.31\% & 0.42\% \\ \hline
\textbf{465.tonto} & 0.00\% & 1.31\% & 0.39\% & 0.40\% \\ \hline
\textbf{470.lbm} & 0.00\% & 1.04\% & 0.99\% & 0.47\% \\ \hline
\textbf{481.wrf} & 0.00\% & \cellcolor{SpecBetter}21.22\% & \cellcolor{SpecBetter}20.21\% & \cellcolor{SpecBetter}19.85\% \\ \hline
\textbf{482.sphinx3} & 0.00\% & \cellcolor{SpecBetter}18.36\% & \cellcolor{SpecBetter}18.88\% & \cellcolor{SpecBetter}16.70\% \\ \hline
\hline
\textbf{average} & \textbf{0.00\%} & \textbf{\cellcolor{SpecGood}8.21\%} & \textbf{\cellcolor{SpecGood}7.83\%} & \textbf{\cellcolor{SpecGood}7.62\%} \\ \hline
	\end{tabular}
	\caption{SPEC CPU2006 FP speed-up for inlining, part I}
\end{table}

% 5) int_size
\begin{table}[htbp]
	\begin{tabular}{|l|r|r|r|r|}
	\hline
\textbf{} & \textbf{O2} & \textbf{O3} & \textbf{O3-UG100} & \textbf{O3-UG200} \\ \hline
\textbf{400.perlbench} & 0.00\% & N/A & \cellcolor{SpecWorse}60.03\% & N/A \\ \hline
\textbf{401.bzip2} & 0.00\% & N/A & \cellcolor{SpecBad}26.06\% & N/A \\ \hline
\textbf{403.gcc} & 0.00\% & \cellcolor{SpecBad}30.34\% & \cellcolor{SpecWorse}94.63\% & \cellcolor{SpecWorse}155.08\% \\ \hline
\textbf{429.mcf} & 0.00\% & 9.70\% & 9.70\% & 9.70\% \\ \hline
\textbf{445.gobmk} & 0.00\% & 7.45\% & 14.13\% & 14.13\% \\ \hline
\textbf{456.hmmer} & 0.00\% & \cellcolor{SpecBad}30.65\% & \cellcolor{SpecBad}30.65\% & \cellcolor{SpecBad}30.65\% \\ \hline
\textbf{458.sjeng} & 0.00\% & \cellcolor{SpecBad}26.74\% & \cellcolor{SpecBad}26.74\% & \cellcolor{SpecBad}26.74\% \\ \hline
\textbf{462.libquantum} & 0.00\% & 4.32\% & 4.32\% & 4.32\% \\ \hline
\textbf{464.h264ref} & 0.00\% & 17.55\% & 17.55\% & 17.55\% \\ \hline
\textbf{471.omnetpp} & 0.00\% & 21.65\% & \cellcolor{SpecBad}45.16\% & \cellcolor{SpecBad}45.16\% \\ \hline
\textbf{473.astar} & 0.00\% & 9.31\% & 9.31\% & 9.31\% \\ \hline
\textbf{483.xalancbmk} & 0.00\% & 5.47\% & \cellcolor{SpecBad}26.71\% & \cellcolor{SpecBad}26.71\% \\ \hline
\hline
\textbf{average} & \textbf{0.00\%} & \textbf{16.32\%} & \textbf{\cellcolor{SpecBad}30.42\%} & \textbf{\cellcolor{SpecBad}33.93\%} \\ \hline
	\end{tabular}
	\caption{SPEC CPU2006 INT size reduction for inlining, part I}
\end{table}

% 6) fp_size
\begin{table}[htbp]
	\begin{tabular}{|l|r|r|r|r|}
	\hline
\textbf{} & \textbf{O2} & \textbf{O3} & \textbf{O3-UG100} & \textbf{O3-UG200} \\ \hline
\textbf{410.bwaves} & 0.00\% & 20.29\% & 20.29\% & 20.29\% \\ \hline
\textbf{416.gamess} & 0.00\% & 20.17\% & 20.17\% & 20.17\% \\ \hline
\textbf{433.milc} & 0.00\% & \cellcolor{SpecBad}34.28\% & \cellcolor{SpecBad}34.28\% & \cellcolor{SpecBad}34.28\% \\ \hline
\textbf{434.zeusmp} & 0.00\% & \cellcolor{SpecWorse}58.56\% & \cellcolor{SpecWorse}58.56\% & \cellcolor{SpecWorse}58.56\% \\ \hline
\textbf{435.gromacs} & 0.00\% & 14.13\% & 14.13\% & 14.13\% \\ \hline
\textbf{436.cactusADM} & 0.00\% & \cellcolor{SpecBad}42.09\% & \cellcolor{SpecBad}42.09\% & \cellcolor{SpecBad}42.09\% \\ \hline
\textbf{437.leslie3d} & 0.00\% & \cellcolor{SpecBad}49.33\% & \cellcolor{SpecBad}49.33\% & \cellcolor{SpecBad}49.33\% \\ \hline
\textbf{444.namd} & 0.00\% & 3.21\% & 3.21\% & 3.21\% \\ \hline
\textbf{447.dealII} & 0.00\% & 3.72\% & 17.87\% & 17.87\% \\ \hline
\textbf{450.soplex} & 0.00\% & 7.12\% & 19.30\% & 19.30\% \\ \hline
\textbf{453.povray} & 0.00\% & \cellcolor{SpecBad}28.98\% & \cellcolor{SpecWorse}77.00\% & \cellcolor{SpecWorse}80.35\% \\ \hline
\textbf{454.calculix} & 0.00\% & 17.92\% & 17.92\% & 17.92\% \\ \hline
\textbf{459.GemsFDTD} & 0.00\% & \cellcolor{SpecWorse}58.58\% & \cellcolor{SpecWorse}58.58\% & \cellcolor{SpecWorse}58.58\% \\ \hline
\textbf{465.tonto} & 0.00\% & 23.56\% & 23.56\% & 23.56\% \\ \hline
\textbf{470.lbm} & 0.00\% & 2.69\% & 2.69\% & 2.69\% \\ \hline
\textbf{481.wrf} & 0.00\% & \cellcolor{SpecWorse}50.88\% & \cellcolor{SpecWorse}50.88\% & \cellcolor{SpecWorse}50.88\% \\ \hline
\textbf{482.sphinx3} & 0.00\% & \cellcolor{SpecBad}35.36\% & \cellcolor{SpecBad}35.42\% & \cellcolor{SpecBad}35.42\% \\ \hline
\hline
\textbf{average} & \textbf{0.00\%} & \textbf{\cellcolor{SpecBad}27.70\%} & \textbf{\cellcolor{SpecBad}32.08\%} & \textbf{\cellcolor{SpecBad}32.27\%} \\ \hline
	\end{tabular}
	\caption{SPEC CPU2006 FP size reduction for inlining, part I}
\end{table}

\chapter{Attached CD's Content}
\label{app:ContentsOfCD}

The thesis is accompanied by a CD with the following items:

\begin{itemize}
	\item Patch for function reordering applicable to gcc main development tree (made on revision 201397) that itself contains entire implementation.
	\item Patch for inter-procedural semantic function equality pass applicable to gcc main development tree (made on revision 201397) that itself contains entire implementation.
	\item Patch for more flexible unit-growth that can be applicable to gcc main development tree (made on revision 201397).
	\item PDF version of this thesis.
	\item A rich variety of Python scripts related to ELF format, SPEC benchmarks and statistic data aggregation. Collection of scripts is also a Github project located at \url{https://github.com/marxin/script-misc}.
\end{itemize}

%%% Seznam použité literatury je zpracován podle platných standardů. Povinnou citační
%%% normou pro diplomovou práci je ISO 690. Jména časopisů lze uvádět zkráceně, ale jen
%%% v kodifikované podobě. Všechny použité zdroje a prameny musí být řádně citovány.

\def\bibname{Bibliography}

%%% Přílohy k diplomové práci, existují-li (různé dodatky jako výpisy programů,
%%% diagramy apod.). Každá příloha musí být alespoň jednou odkazována z vlastního
%%% textu práce. Přílohy se číslují.
% \chapwithtoc{Attachments}

\openright

\begin{thebibliography}{99}
\addcontentsline{toc}{chapter}{\bibname}

\bibitem{lipo}
	\textsc{Li}, Xinliang, \textsc{Ashok}, Raksit, \textsc{Hundt}, Robert.
	\emph{LIPO - Profile Feedback Based Lightweight IPO} [online].
	2009 [cit. July 5, 2013].
	Available from: \url{http://gcc.gnu.org/wiki/LightweightIpo}.

\bibitem{gcc-lto}
	\textsc{GCC}.
	\emph{Link Time Optimization, GCC Internals} [online].
	[cit. July 5, 2013].
	Available from: \url{http://gcc.gnu.org/onlinedocs/gccint/LTO.html}.

\bibitem{FirefoxTelemetry}
	\textsc{Glek}, Taras.
	\emph{Firefox Telemetry} [online].
	2011 [cit. July 30, 2013].
	Available from \url{https://blog.mozilla.org/tglek/2011/05/13/firefox-telemetry/}.

\bibitem{glandium-elfhack}
	\textsc{Hommey}, Mike.	
	\emph{Improving libxul start-up I/O by hacking the ELF format} [online].
	2010 [cit. July 2, 2013].
	Available from: \url{http://glandium.org/blog/?p=1177}.

\bibitem{glandium-preload}
	\textsc{Hommey}, Mike.
	\emph{Preloading for dummies} [online].
	2011 [cit. July 15, 2013].
	Available from: \url{http://glandium.org/blog/?p=1746}.

\bibitem{GlandiumSystemtap}
	\textsc{Hommey}, Mike.
	\emph{Attempting to track I/O with systemtap} [online].
	2011 [cit. July 15, 2013].
	Available from: \url{http://glandium.org/blog/?p=1476}.
		
\bibitem{GlandiumFaultyLib}
	\textsc{Hommey}, Mike.
	\emph{faulty.lib vs. Firefox for Android} [online].
	2012 [cit. July 15, 2013].
	Available from: \url{http://glandium.org/blog/?p=2467}.

\bibitem{GlandiumBinaryLayout}
	\textsc{Hommey}, Mike.
	\emph{Improving binary layout for progressive decompression} [online].
	2011 [cit. July 15, 2013].
	Available from: \url{http://glandium.org/blog/?p=2320}.
	
\bibitem{Drepper}
	\textsc{Drepper}, Ulrich.
	\emph{How to write shared libraries} [online].
	2011 [cit. July 9, 2013].
	Available from: \url{http://www.akkadia.org/drepper/dsohowto.pdf}.

\bibitem{Prelink}
	\textsc{Jelínek} Jakub.
	\emph{Prelink} [online].
	2004 [cit. July 11, 2013].
	Available from: \url{http://people.redhat.com/jakub/prelink.pdf}.

\bibitem{AndiKleenKernelLTO}
	\textsc{Kleen}, Andi.
	\emph{GCC link time optimization and the Linux kernel} [online].
	2013 [cit. July 12, 2013].
	Available from: \url{http://halobates.de/kernel-lto.pdf}.

\bibitem{LightweightFDOIPO}
	\textsc{Li}, David Xinliang, \textsc{Ashok}, Raksit, \textsc{Hundt}, Robert.
	Lightweight feedback-directed cross-module optimization.
	In: \emph{Proceedings of the 8th annual IEEE/ACM international symposium on Code generation and optimization}.
	New York, NY, USA: ACM, 2010, pages 53-61.
	ISBN 978-1-60558-635-9.
	Available from: \url{research.google.com/pubs/archive/36355.pdf}.

\bibitem{WHOPR}
	\textsc{Briggs} P., \textsc{Evans} D., \textsc{Grant}, B., \textsc{Hundt}, R., \textsc{Maddox}, W., \textsc{Novillo}, D., \textsc{Park}, S., \textsc{Sehr}, D., \textsc{Taylor}, I., \textsc{Wild}, O.
	\emph{Fast and Scalable Whole Program Optimizations in GCC}.
	2007.
	Available from: \url{http://gcc.gnu.org/projects/lto/whopr.pdf}.

\bibitem{GlekHubickaLTO}
	\textsc{Glek}, Taras, \textsc{Hubička}, Jan.
	Optimizing real-world applications with GCC Link Time Optimization.
	In: \emph{Proceedings of the GCC Developers Summit}.
	Ottawa, Ontario, Canada: 2010, pages 107-113.
	Available from: \url{http://gcc.gnu.org/wiki/summit2010?action=AttachFile&do=get&target=2010-GCC-Summit-Proceedings.pdf}.

\bibitem{PDOInGCC}
	\textsc{Dvořák}, Zdeněk, \textsc{Hubička}, Jan, \textsc{Nejedlý}, Pavel, \textsc{Zlomek}, Josef.
	Infrastructure for Profile Driven Optimizations in GCC Compiler.
	2002.
	Available from: \url{http://www.ucw.cz/~hubicka/papers/proj.ps.gz}.

\bibitem{SafeICF}
	\textsc{Tallam}, Sriraman, \textsc{Coutant}, Cary, \textsc{Taylor}, Ian Lance, \textsc{Li}, David Xinliang, \textsc{Demetriou} Chris .
	Safe ICF: Pointer Safe and Unwinding Aware Identical Code Folding in Gold.
	In: \emph{Proceedings of the GCC Developers Summit}.
	Ottawa, Ontario, Canada: 2010, pages 107-113.
	Available from: \url{http://gcc.gnu.org/wiki/summit2010?action=AttachFile&do=get&target=2010-GCC-Summit-Proceedings.pdf}.

\bibitem{DynamicCodeManagement}
	\textsc{Huang}, Xianglong, \textsc{Lewis}, Brian T, \textsc{McKinley}, Kathryn S.
	Dynamic code management: improving whole program code locality in managed runtimes.
	In: \emph{Proceedings of the 2nd international conference on Virtual execution environments}.
	Ottawa, Ontario, Canada: ACM, 2006, pages 133-143.
	ISBN 1-59593-332-8.
	Available from \url{https://www.usenix.org/legacy/events/vee06/full_papers/p133-huang.pdf}.

\bibitem{Ramirez99softwaretrace}
	\textsc{Ramírez}, Alex, \textsc{Larriba-pey}, Josep-l., \textsc{Navarro}, Carlos, \textsc{Torrellas}, Josep, \textsc{Valero}, Mateo.
	Software Trace Cache.
	In: \emph{Proceedings of the 13th Intl. Conference on Supercomputing}.
	1999: pages: 119-126.
	Available from: \url{http://www.ac.upc.es/homes/aramirez/papers/ieee-toc05.pdf}.
	
\bibitem{ConcurrentGoldLinking}
	\textsc{Mathijs van Veen}, Sander.
	\emph{Concurrent linking in GNU Gold}.
	Amsterdam, 2013.
	Bachelor thesis.
	University of Amsterdam.
	Available from: \url{http://smvv.kompiler.org/gold.pdf}.

\bibitem{Alpern1992Value}
	\textsc{Alpern}, Bowen, \textsc{Zadeck}, K.
	Value Numbering.
	In: \emph{Optimization in Compilers, edited by F. Allen, B. Rosen, and K. Zadeck}.	
	To be published by ACM Press, 1992.	
	Available from: \url{http://www.cs.ucr.edu/~gupta/teaching/553-07/Papers/value.pdf}.

\bibitem{Gold}
	\textsc{Taylor}, Ian Lance.
	A New ELF Linker.
	In: \emph{Proceedings of the GCC Developers Summit}.	
	Ottawa, Ontario, Canada: 2008, pages 129-136.	
	Available from: \url{http://ols.fedoraproject.org/GCC/Reprints-2008/taylor-reprint.pdf}.

\bibitem{CompaqAlphaLTO}
	\textsc{Muth}, R., \textsc{Debray}, S. K., \textsc{Watterson}, S. and \textsc{De Bosschere}, K.
	alto: a link-time optimizer for the Compaq Alpha.
	In: \emph{In: Software: Practice and Experience}.
	2001: pages 67-71.

\bibitem{Muchnick}
	\textsc{Muchnick}, Steven S.
	Advanced compiler design and implementation.
	an Francisco, CA, USA: Morgan Kaufmann Publishers Inc., 1997.
	ISBN:1-55860-320-4.	

\bibitem{Automake}
	\emph{Automake: Documentation}.
	2013 [cit. 27.6.2013].
	Available from: \url{http://www.gnu.org/software/automake/manual/automake.pdf}.

\bibitem{GNUMake}
	\emph{GNU Make: Documentation}.
	2013 [cit. 21.6.2013].
	Available from: \url{http://www.gnu.org/software/make/manual/make.html}

\bibitem{GCC}
	\emph{GNU GCC: Manual}.
	2013 [cit. 27.6.2013].
	Available from: \url{http://gcc.gnu.org/onlinedocs/gcc-4.8.1/gcc/}.

\bibitem{LVVM}
	\emph{LVVM: Manual}.
	2013 [cit. 29.6.2013].
	Available from: \url{http://llvm.org/docs/}.

\bibitem{Open64}
	\emph{Using the x86 Open64 Compiler Suite}
	2013 [cit. 27.6.2013].
	Available from: \url{http://developer.amd.com/wordpress/media/2012/10/open64.pdf}

\bibitem{HPCMO1}
	\textsc{Hewlett-Packard}.
	Compiler architecture for cross-module optimization.
	USA.
	United States Patent 5375242.
	
\bibitem{HPCMO2}
	\textsc{Hewlett-Packard}.
	Compiler architecture for cross-module optimization.
	USA.
	United States Patent 5920723.

\bibitem{OhlohFirefox}	
	\textsc{BLACK DUCK}.
	Ohloh: Project Summary for Mozilla Firefox.
	2013 [cit. 29.6.2013].
	Available from: \url{http://www.ohloh.net/p/firefox}.

\end{thebibliography}
\end{document}